\documentclass[aps]{revtex4}
\bibpunct{(}{)}{;}{a}{,}{,}%

\usepackage{graphicx}
\usepackage{subfigure}
\usepackage{amsmath}
\renewcommand{\vec}[1]
{
  \textbf{#1}
}
\newcommand{\Abs}[1]
{
  \left\vert {#1} \right\vert
}
\renewcommand{\(} 
{
  \left(
}
\renewcommand{\)} 
{
  \right)
}
\newcommand{\w}
{
  \omega
}
\newcommand{\wpe}
{
  \w_{\mathrm{pe}}
}
\newcommand{\wpi}
{
  \w_{\mathrm{pi}}
}
\newcommand{\wce}
{
  \w_{\mathrm{ce}}
}
\newcommand{\wci}
{
  \w_{\mathrm{ci}}
}
\newcommand{\wuh}
{
  \w_{\mathrm{uh}}
}
\newcommand{\wlh}
{
  \w_{\mathrm{lh}}
}
\renewcommand{\a}
{
 \mathrm{\alpha}
}
\newcommand{\f}
{
 \widehat{f}
}
\newcommand{\Fi}
{
 \f_{\mathrm{i}}
}
\newcommand{\Fe}
{
 \f_{\mathrm{e}}
}

\renewcommand{\i} {\mathrm{i}}

\newcommand{\E}
{
  {\bf E}
}
\newcommand {\B}
{
  {\bf B}
}
\newcommand {\Bext}
{
  {\bf B}_{\mathrm{ext}}
}
\newcommand {\A}
{
  {\bf A}
}
\renewcommand {\v}
{
  \vec{v}
}
\newcommand {\x}
{
  \vec{x}
}
\newcommand {\Eta}
{
  \boldsymbol{\eta}
}
\newcommand {\vGamma}
{
  \boldsymbol{\Gamma}
}
\renewcommand{\d}
{
  \partial
}
\newcommand {\D}
{
  \boldsymbol{\nabla}
}
\newcommand{\grad}
{
  \boldsymbol{\nabla}
}
\renewcommand{\div}
{
  \boldsymbol{\nabla}\cdot
}
\newcommand{\rot}
{
  \boldsymbol{\nabla}\times
}

\newcommand {\Dt}[1]
{
  \frac{\d #1}{\d t}
}
\newcommand {\Dtt}[1]
{
  \frac{\d^2 #1}{\d t^2}
}
\newcommand {\X}
{
  \times
}
\renewcommand {\epsilon}
{
  \varepsilon
}
\newcommand{\diff}
{
  \mathrm{d}
}

\newcommand {\F}
{
  \mathrm{F}
}
\newcommand {\Ti}
{
  T_\mathrm{i}
}
\newcommand {\mi}
{
  m_\mathrm{i}
}

\newcommand{\Te}
{
  T_\mathrm{e}
}
\newcommand{\me}
{
  m_\mathrm{e}
}

\newcommand{\vthe}
{
  v_\mathrm{th,e}
}
\newcommand{\vthi}
{
  v_\mathrm{th,i}
}
\newcommand{\rD}
{
  r_\mathrm{D}
}

\newcommand {\etamax}
{
  \eta_{\mathrm{max}}
}
\newcommand {\Part} [2]
{
  \frac{\d #1}{\d #2}
}
\newcommand {\dfdxdeta}
{
  \frac{\d^2\f}{\d x \d \eta}
}

\newcommand {\intInfty}
{
  \int_{-\infty}^{\infty}
}

\newcommand{\veff}  { v_{\mathrm{eff}} }       

\begin{document}

\title{Numerical simulations of the Fourier transformed Vlasov-Maxwell
system in higher dimensions --- Theory and applications}

\author{Bengt Eliasson}    
\affiliation{Institut f\"ur Theoretische Physik IV,
Fakult\"at f\"ur Physik und Astronomie,
Ruhr--Universit\"at Bochum, D-44780 Bochum, Germany}
\affiliation{Department of Physics, Ume{\aa} University,
SE-901 87 Ume{\aa}, Sweden}

\begin{abstract}
We present a review of recent developments of simulations of the Vlasov-Maxwell system of equations using a Fourier transform method in velocity space. In this method, the distribution functions for electrons and ions are Fourier transformed in velocity space, and the resulting set of equations are solved numerically. In the original Vlasov equation, phase mixing may lead to an oscillatory behavior and sharp gradients of the distribution function in velocity space, which is problematic in simulations where it can lead to unphysical electric fields and instabilities and to the recurrence effect where parts of the initial condition recur in the simulation. The particle distribution function is in general smoother in the Fourier transformed velocity space, which is desirable for the numerical approximations. By designing outflow boundary conditions in the Fourier transformed velocity space, the highest oscillating terms are allowed to propagate out through the boundary and are removed from the calculations, thereby strongly reducing the numerical recurrence effect. The outflow boundary conditions in higher dimensions including electromagnetic effects are discussed. The Fourier transform method is also suitable to solve the Fourier transformed Wigner equation, which is the quantum mechanical analogue of the Vlasov equation for classical particles.
\end{abstract}

\received{8 March 2010}
\revised{17 July 2010}

\maketitle

\tableofcontents
%
\section{Introduction}

The Vlasov equation governs the dynamics of the distribution function of charged particles (electrons, ions) in the six-dimensional phase space, consisting of 3 velocity (or momentum) dimensions and 3 position dimensions, plus time. It offers an accurate description of a plasma in the collisionless limit, i.e., when the particles are affected by long-range electric and magnetic fields only, and when short-range fields from its nearest neighbors can be neglected.

The most common method to solve the Vlasov equation numerically is the Particle-In-Cell (PIC) method \citep{Birdsall91,Matsumoto93}, where the Vlasov equation is solved by following the trajectories
of a set of statistically distributed super-particles, which resolves the particle distribution functions in phase space. Each super-particle represents a large number of real particles. PIC simulations have proven to be extremely successful due to their relative simplicity and adaptivity. However, the statistical noise of
PIC simulations sometimes overshadows the physical results, and for some problems, the low-density velocity tail of the particle distribution cannot be resolved with high enough accuracy by the super-particles.

Grid-based Vlasov solvers treat the particle distribution function as a phase fluid that is represented on a grid in both position and velocity (or momentum) space. The advantage with grid-based Vlasov solvers is that they do not give rise to statistical noise in the simulations, and that the dynamical range is larger than for PIC methods, so that the low-density velocity tail of the particle distribution can be resolved much more accurately. A disadvantage with grid-based Vlasov solvers in higher dimensions is that the full phase-space has to be represented on a numerical grid grid, which makes both the storage of the data in the computer's memory, and the numerical calculations, extremely demanding. Another problem is the tendency of the distribution function to become oscillatory in velocity space due to phase mixing. This is the cause of Landau damping and other kinetic effects, but may lead to unphysical noise and recurrence effects in the numerical solution. This was recognized in early Vlasov simulations and special methods were devised to resolve this problem, including the Fourier-Fourier and Hermite expansion methods \citep{Armstrong_70}. Various other methods have also been developed for solving the Vlasov equation, the classical and widely used time-spitting method \citep{Knorr_76,Cheng_77} where a smoothing operator was applied to remove the highest oscillations in velocity space, a Van Leer dissipative scheme \citep{Mangeney_02}, the finite volume method \citep{Elkina_06}, a back-substitution method for magnetized plasma \citep{Schmitz_06}, etc. Several Eulerian grid-based solvers are reviewed and compared by \citet{Arber_02} and \citet{Filbet_03}.

One special category of methods are transform methods \citep{Armstrong_70}, where the particle distribution function is expressed as a sum or integral of basis functions in velocity space. The structuring of velocity space in this case leads to the excitation of higher and higher modes in the transformed velocity space, and special care must be taken when these excitations reach the highest mode represented in the numerical simulation. For example, for methods using Hermite polynomials to resolve the velocity space, the highest-order Hermite polynomials is designed to absorb the oscillations in velocity space \citep{Knorr_74,Gibelli_06}, thereby reducing numerical recurrence effects strongly. \citet{Klimas_87} and \citet{Klimas_94} devised filtering methods for the Fourier-Fourier method to remove the highest oscillations in velocity space. (In the Fourier-Fourier method, the Vlasov equation is Fourier transformed both in velocity space and position space, and the resulting equation is solved numerically.) For the Fourier method in one, two and three dimensions, \citet{Eliasson_01,Eliasson_02,Eliasson_03,Eliasson_07} designed absorbing boundary conditions at the largest Fourier mode in velocity space so that the highest oscillations in velocity space were removed from the solution. In this method, the Vlasov equation is Fourier transformed only in velocity space and not in position space, so that the electromagnetic fields enter by multiplications instead of convolutions in the transformed Vlasov equation. The Fourier transformed Vlasov equation is also interesting to study in its own respect. Some mathematical aspects of the Fourier transformed Vlasov equation are given by \citet{Neunzert71,Neunzert81}, and in their analytic and and simulation studies, \citet{Sedlacec92,Sedlacec02} presented interpretations of the Landau damping, the time echo phenomenon, etc., in terms of imperfectly trapped (leaking) waves in the Fourier transformed velocity space.

The main topic of this paper is a the properties of the Vlasov equation and the Fourier transformed Vlasov equation. The full Vlasov-Maxwell system, and the Vlasov-Maxwell system Fourier transformed in velocity space are presented in section II. In section III, we discuss the properties of the Vlasov equation that makes simulations a challenging task. We note that oscillations of the particle distribution function in velocity space corresponds to wave packets in the Fourier transformed velocity space. Special attention is put on the absorbing artificial boundary conditions in the Fourier transformed velocity space, where the highest Fourier modes are absorbed and removed from the calculations. These boundary conditions have the following attractive features: (i) They substantially reduce the unphysical reflections at the artificial boundaries, thereby reducing unphysical noise and recurrence effects in simulations. (ii) The boundary conditions are local in time and involve only boundary points (but they are non-local along the boundary). (iii) The boundary conditions together with the interior differential equation defines a well-posed problem. Some examples of simulations of one-dimensional kinetic structures, electron and ion holes, are also discussed in section III.
In section IV, we will briefly mention the discrete approximations used in the numerical simulations of the Fourier transformed Vlasov equation. Especially the important topics of the representation of particle velocities by the numerical scheme and how to choose domain and grid sizes in the Fourier transformed velocity space are discussed. In sections V and VI, we present the generalizations of the Fourier technique to two and three dimensions, respectively. Here the treatment of the spatially varying magnetic field has to be treated carefully in the design of absorbing boundary conditions int the Fourier transformed velocity space. We mention here that the well-posedness of the boundary conditions for the one-, two- and three-dimensional Fourier transformed Vlasov equation have been proved by energy estimates \citep{Eliasson_01,Eliasson_02,Eliasson_07}, where an energy norm is non-increasing in time. Finally, in chapter VII, we discuss extensions of the Vlasov equation to incorporate quantum and relativistic effects. The quantum analogue to the Vlasov equation is the Wigner equation, and we will see that the absorbing boundary conditions used for the Vlasov equation can be applied unchanged to the Wigner equation. For the relativistic Vlasov equation, the relativistic gamma factor enters into the Vlasov equation and leads to a convolution in the Fourier transformed velocity space. This convolution operator is non-local in space and may lead to non-local absorbing boundary conditions in space and time.
\section{The Vlasov-Maxwell system of equations}
The non-relativistic Vlasov equation
\begin{align}
  \label{3D_vlasov}
  \Dt{f_{\a}}+\v\cdot\D_x f_{\a}+\frac{\vec{F}_\a}{m_{\a}}\cdot \D_v f_{\a}=0
\end{align}
where the Lorentz force is
\begin{equation}
  \label{LorentzForce}
  \vec{F}_\a=q_{\a}[\E+\v\X(\B+\Bext)]
\end{equation}
describes the evolution of the distribution function $f_{\a}$ of electrically charged particles
of type $\a$ (e.g., ``electrons'' or ``singly ionized
oxygen ions''), each particle having the electric charge $q_{\a}$ and mass $m_{\a}$.
Here, the magnetic field is separated into two parts, where $\Bext$ is an external
magnetic field (e.g., the Earth's geomagnetic field), and $\B$ is the self-consistent
part of the magnetic field, created by the plasma.
One Vlasov equation
is needed for each species of particles.

The particles interact via the electromagnetic field.
The charge and current densities, $\rho$ and $\vec{j}$, act as sources
of self-consistent electromagnetic fields according to the Maxwell equations
\begin{align}
  \label{3D_Gauss}
  \div \E &= \frac{\rho}{\epsilon_0} \\
  \label{3D_divB}
  \div \B &= 0 \\
  \label{3D_Ampere}
  \rot \E &= -\Dt{\B}\\
  \label{3D_Maxwell}
  \rot \B &= \mu_0 \vec{j} +
  \frac{1}{c^2} \Dt{\E}
\end{align}
The charge and current densities are related to the particle
number densities $n_{\a}$ and mean velocities $\v_{\a}$ as
\begin{align}
\label{3D_rho}
\rho&=\sum_{\a}q_{\a}n_{\a}
\intertext{and}
\label{3D_j}
\vec{j}&=\sum_{\a}q_{\a}n_{\a}\v_{\a}
\end{align}
respectively,
where the particle number densities and mean velocities are obtained
as moments of the particle distribution functions, as
\begin{align}
 \label{3D_na}
  n_{\a}(\x,t)&=\intInfty f_{\a}(\x,\v,t)\,\diff^3v \\
 \intertext{and}
 \label{3D_va}
  \v_{\a}\(\x,t\)&=\frac{1}{n_{\a}\(\x,t\)}
  \intInfty \v f_{\a}\(\x,\v,t\)\,\diff^3 v
\end{align}
respectively.

The Vlasov equation (\ref{3D_vlasov}) together with the Maxwell equations
(\ref{3D_Gauss})--(\ref{3D_Maxwell})
and the constitutive
equations (\ref{3D_rho})--(\ref{3D_va}) form
a closed, coupled system of nonlinear partial differential equations
and integral equations.
\subsection{The Fourier transformed Vlasov equation}
By using the Fourier transform pair
\begin{align}
  f_{\a}(\x,\v,t)&=\intInfty \f_{\a}(\x,\Eta,t)
  e^{-\i\Eta\cdot\v}\,\diff^3\eta \\
  \label{3D_F3D2}
  \f_{\a}(\x,\Eta,t)&=\frac{1}{(2\pi)^3}\intInfty f_{\a}(\x,\v,t)
  e^{\i\Eta\cdot\v}\,\diff^3\eta
\end{align}
the velocity variable $\v$ is transformed into a new variable $\Eta$
and the distribution function $f(\x,\v,t)$ is changed to
a new, complex valued, function $\f(\x,\Eta,t)$, which obeys the
transformed Vlasov equation
\begin{align}
 \label{3D_eta1}
 \frac{\d \f_{\a}}{\d t}-\i\D_x\cdot\D_{\eta}\f_{\a}
 -\frac{q_{\a}}{m_{\a}}\big(\i\E\cdot\Eta\f_{\a}+
 \D_{\eta}\cdot\{[(\B+\Bext)\X\Eta]\f_{\a}\}\big) =0
\end{align}
The nabla operators $\D_x$ and $\D_{\eta}$ denote differentiation
with respect to $\x$ and $\Eta$, respectively.

Equation (\ref{3D_eta1}) together with
the Maxwell equations (\ref{3D_Gauss})--(\ref{3D_Maxwell})
and the constitutive equations (\ref{3D_rho})--(\ref{3D_j}) where the
particle number densities and mean velocities are obtained as
\begin{align}
  \label{3D_density}
  n_\a(\x,t)&=(2\pi)^3\f_{\a}(\x,\vec{0},t)
  \intertext{and}
  \label{3D_velocity}
  \v_\a(\x,t)&=-\i\frac{(2\pi)^3}{n_{\a}(\x,t)}\left[\D_{\eta}\f_{\a}
  (\x,\Eta,t)\right]_{\Eta=\vec{0}}
\end{align}
respectively, form a new closed set of equations. One can note that the integrals over infinite
$\v$ space in Eqs.~(\ref{3D_na}) and (\ref{3D_va}) have been converted to evaluations in $\Eta$ space
in Eqs.~(\ref{3D_density} and \ref{3D_velocity}).
The factor $(2\pi)^3$ in Eqs.~(\ref{3D_F3D2}), (\ref{3D_density}) and (\ref{3D_velocity})
is valid for three velocity dimensions. For $n$ velocity dimensions
this factor is $(2\pi)^n$.

\subsection{Invariants of the Vlasov-Maxwell system}
The Vlasov equation (\ref{3D_vlasov}) coupled with (\ref{3D_Gauss})--(\ref{3D_Maxwell})
conserves the energy norm
\begin{align}
  \| f_\alpha \|^2&=
  \int
  \int
     f_\alpha^2
     \,\diff^3 v
     \,\diff^3 x,
     \label{norm1}
  \intertext{the total number of particles}
  N_\alpha &=
  \int
  \int
     f_\alpha
     \,\diff^3v \diff^3 x,\\
     \intertext{the total linear momentum}
  \mathbf{p}&=
  \int
  \left[\int
     {\bf v} (m_i f_i+m_e f_e)
     \,\diff^3v +\varepsilon_0{\E}\times {\B}\right]
     \,\diff^3x,
     \intertext{and the total energy}
     \label{energy}
  W&=
  \int
  \left[
  \int
     \frac{1}{2} {\v}^2 (m_i f_i+m_e f_e)
     \,\diff^3v
    +\frac{1}{2}\left(\varepsilon_0{\E}^2 +\frac{{\B}^2}{\mu_0}\right)\right]
    \,\diff^3x.
\end{align}
The corresponding invariants for the Fourier-transformed
Vlasov-Maxwell system (\ref{3D_eta1}) and (\ref{3D_Gauss})--(\ref{3D_Maxwell}) are:
\begin{align}
  \|\f_\alpha\|^2&=\frac{1}{(2\pi)^3}
  \int
  \int
    |\f_\alpha|^2
    \,\diff^3\eta
    \,\diff^3 x,
    \label{energy_norm}
    \\
    \label{particles}
  N_\alpha&=\int
    (2\pi)^3 (\f_\alpha)_{{\Eta}=0}\,\diff^3 x, \\
    \label{momentum_eta}
  \mathbf{p}&=\int
    [-\i (2\pi)^3 \D_{\Eta} (m_{\rm i} f_{\rm i} + m_{\rm e} f_{\rm e})_{\Eta=0}
    +\varepsilon_0 {\E}\times {\B}
    ]\,\diff^3x,
    \intertext{and}
    \label{energy_eta}
  W&=\int \bigg[-\frac{1}{2}(2\pi)^3
     \nabla_{\Eta}^2 (m_{\rm i} f_{\rm i} + m_{\rm e} f_{\rm e})_{\Eta=0}
     +\frac{1}{2}\left(\varepsilon_0{\E}^2 +\frac{{\B}^2}{\mu_0}\right) \bigg]\,\diff^3x,
\end{align}
respectively, where the norm (\ref{energy_norm}) follows from
(\ref{norm1}) via the Parseval relation. These invariants can be
used to check the accuracy of the numerical scheme. When the system
is restricted to a bounded domain in $\Eta$ space with appropriate
boundary conditions (discussed in Section \ref{sec:wellposed}
below), the norm $\|\f_\alpha\|^2$ will be a non-increasing,
positive function of time, while the other three quantities will
still be conserved.
%

\subsection{The electrodynamic scalar and vector potentials}

The Maxwell equations (\ref{3D_Gauss})--(\ref{3D_Maxwell}) can be written in terms of the
scalar and vector potentials $\Phi$ and $\A$, which are related to the
electromagnetic field as
\begin{align}
  \label{pot_E}
  \E&=-\D \Phi-\Dt{\A} \\
  \label{pot_B}
  \B&={\bf B}_0+\rot\A,
\end{align}
where ${\bf B}_0$ is the external magnetic field.
Introducing these expressions into Eqs. (\ref{3D_Gauss})--(\ref{3D_Maxwell}), and choosing the Lorentz condition,
\begin{align}
  \div\A+\frac{1}{c^2}\Dt{\Phi}=0
\end{align}
yields the electrodynamic waves equations
\begin{align}
  \label{Lorentz1}
  \frac{1}{c^2}\Dtt{\Phi}-\nabla^2\Phi&=\frac{\rho}{\varepsilon_0}\\
  \label{Lorentz2}
  \frac{1}{c^2}\Dtt{\A}-\nabla^2\A&=\mu_0\vec{j}.
\end{align}
In this description, the divergence of the magnetic field is zero, since the divergence of the right hand
side of Eq.~(\ref{pot_B})
is zero by the vector relation $\div(\rot \A)=0$.
The divergence of the electric field can be set to the correct value by
using the Maxwell equation for the divergence of the electric field (\ref{3D_Gauss})
together with Eq.~(\ref{pot_E}), yielding
\begin{align}
  \div\E&=-\nabla^2 \Phi-\div \Dt{\A}=\frac{\rho}{\varepsilon_0}
  \intertext{or}
  \label{nablaphi}
  -\nabla^2 \Phi&=\frac{\rho}{\varepsilon_0} + \div \Dt{\A}
\end{align}
This equation for $\Phi$ conserves the divergence of the electric field.

By introducing a separate variable $\vGamma$ for the time derivative of
the vector potential $\A$,
the wave equations (\ref{Lorentz2}) and (\ref{nablaphi}) can be rewritten in a first-order system with respect to time,
\begin{align}
  \label{Lorentz3}
  \Dt{\A}&=\vGamma & \Dt{\vGamma}&=c^2(\nabla^2\A+\mu_0\vec{j}) \\
  \label{Lorentz3_2}
  -\nabla^2\Phi&=\frac{\rho}{\varepsilon_0}+\div\vGamma &&
\end{align}
and the electric and magnetic fields are calculated as
\begin{align}
  \E&=-\grad\Phi-\vGamma
  \label{E_grad}
  \intertext{and}
  \label{B_rot}
  \B&=\rot\A
\end{align}
respectively, in the new variables.

The system (\ref{Lorentz3})--(\ref{B_rot})
produces physical electric and magnetic fields regardless of the initial
conditions on $\A$ and $\vGamma$, in the sense that the first two
Maxwell equations for the divergences (\ref{3D_Gauss})--(\ref{3D_divB})
are fulfilled.
Therefore, a consistent numerical scheme will
also produce physical solutions, up to the local truncation error of
the numerical scheme, even after a long time; no
artificial electric and magnetic charges are
created and accumulated by the numerical scheme, which could be the case if
the two last Maxwell equations (\ref{3D_Ampere})--(\ref{3D_Maxwell})
are integrated numerically in time.
This general property of the system is an advantage, since
it is not necessary to use special, divergence-conserving schemes \cite{Wagner_98}
to solve these equations, and
it therefore opens up the possibility to switch between different numerical methods
without having to pay too much attention to the divergences of the electromagnetic field.
In complicated geometries, it may be a disadvantage that an elliptic
equation (\ref{Lorentz3_2}) has to be solved numerically to obtain the potential $\Phi$,
while in the simple geometries considered here, with periodic boundary conditions,
Eq. (\ref{Lorentz3_2}) is efficiently solved by means of Fourier transform techniques.

\subsection{The reduction of spatial and velocity dimensions}
In the study of problems with certain symmetries, it is sometimes
possible to make a choice of the coordinate system so that
the the problem can be analyzed in a smaller number of
dimensions. Numerically, this is very convenient because
unnecessary information is removed
from the problem and a smaller number of sampling points
is needed to represent the solution on a numerical grid.

One such assumption is that the problem is homogeneous
in one or more dimensions, in which case the derivatives in these
dimensions vanish. In the study of plane waves in
plasma, the number of dimensions in $\x=(x_1,x_2,x_3)$ space can be reduced
to one dimension, $\x=(x_1,0,0)$, so that only
derivatives with respect to $x_1$ (and not $x_2$ and $x_3$) remain.
In this manner, the Vlasov equation can be reduced from three
spatial and velocity dimensions, to one spatial and three
velocity dimensions, plus time.

For the non-relativistic Vlasov equation, it turns out that it
is also possible to reduce the number of velocity dimensions, but in a different
manner than for the spatial dimensions.
For electrostatic electron waves in an unmagnetized plasma, the reduction
to one spatial dimension $x_1$ also leads to that terms containing
factors of $v_2$ and $v_3$, and derivatives with respect to
$v_2$ and $v_3$, also vanish, giving rise to the system
\begin{align}
  \label{red1}
  &\Dt{f}+v_1\frac{\d f}{\d x_1}-
  \frac{e E_1}{\me} \frac{\d f}{\d v_1}=0
  \\
  \label{red2}
  &\frac{\d E_1}{\d x_1} =
  \frac{e}{\epsilon_0}\bigg[n_0-
  \intInfty\intInfty\intInfty f(x_1,v_1,v_2,v_3,t) \,\diff v_1\,\diff v_2\,\diff v_3 \bigg]
\end{align}
The dependence on $(v_2,v_3)$ only appears in the integral over all
velocity space for calculating the electric field $E_1$.
Similarly, for waves in a magnetized plasma, propagating in the
$(x_1,x_2)$ plane perpendicular to the magnetic field directed in the
$x_3$ direction and with the electric field directed in the $(x_1,x_2)$ plane
perpendicular to the magnetic field, any dependencies on the distribution in $v_3$
vanish in the Vlasov equation, and one has one Vlasov equation
in $(x_1,x_2,v_1,v_2,t)$ space for each value on $v_3$.
In these cases, it is possible (and convenient) to reduce the
number of dimensions also in $\v$ space.

In the study of collective phenomena in plasma,
the electromagnetic fields do not depend explicitly
on the exact velocity distribution of particles but on the
charge and current densities in $\x$ space, calculated as
integrals (moments) of the distribution function.
This makes it possible to reduce the number of velocity dimensions
in the Vlasov equation. For the case of electrostatic waves
in an unmagnetized plasma discussed above, it is simple to show that linear
combinations of distribution functions with different $(v_2,v_3)$
are solutions to the one-dimensional Vlasov equation (\ref{red1}), because
these distribution functions separately are solutions to the same Vlasov
equation. In particular, taking the limit of a continuous
``linear combination,'' the one-dimensional distribution function
\begin{align}
  \label{f1D}
  f^{1\mathrm{D}}(x_1,v_1,t)=\intInfty\intInfty f(x_1,v_1,v_2,v_3,t)\,\diff v_2\,\diff v_3
\end{align}
is a solution to the one-dimensional Vlasov equation (\ref{red1}) because the
function $f(x_1,v_1,v_2,v_3,t)$ is a solution to the the Vlasov equation (\ref{red1})
for each value on $(v_2,v_3)$.
The electric field is calculated from Eq.~(\ref{red2}) where, by Eq~(\ref{f1D}),
\begin{align}
  &\intInfty\intInfty\intInfty f \,\diff v_1\,\diff v_2\,\diff v_3
  =\intInfty f^{1\mathrm{D}}\,\diff v_1
\end{align}
and the resulting one-dimensional Vlasov equation coupled with Gauss' law is
\begin{align}
  &\Dt{f^{1\mathrm{D}}}+v_1\frac{\d f^{1\mathrm{D}}}{\d x_1}-
  \frac{e E_1}{\me} \frac{\d f^{1\mathrm{D}}}{\d v_1}=0
  \\
  &\frac{\d E_1}{\d x_1} =
  \frac{e}{\epsilon_0}\bigg[n_0-\intInfty f^{1\mathrm{D}}(x_1,v_1,t) \,\diff v_1 \bigg]
\end{align}
for the unknown function $f^{1\mathrm{D}}$.

For waves propagating perpendicularly to a magnetic field, mentioned above,
it is possible to derive the two-dimensional Vlasov-Maxwell system, which
depends on the distribution functions in the form
\begin{align}
  f^{2\mathrm{D}}(x_1,x_2,v_1,v_2,t)=\intInfty f(x_1,x_2,v_1,v_2,v_3,t)\,\diff v_3
\end{align}
For waves propagating with some angle to the magnetic field, it is more
 difficult to reduce the number of velocity dimensions
in the manner described above, since all three velocity components
will appear explicitly in the resulting Vlasov equation. Even so,
the reduction of the number of velocity dimensions for the Vlasov
equation has been done also for this case, in which a full Vlasov kinetic
description is maintained only along one ``dominant'' spatial coordinate,
and with the perpendicular dimensions modeled by reduced moment-based methods
\citep{Newmann_04}.

\section{Properties of the Vlasov equation} \label{chap:vlasov1D}

A well-known property of the Vlasov equation is that an initially smooth distribution function which evolves in time may become increasingly oscillatory in velocity space due to phase mixing of the distribution function. This leads to Landau damping and other kinetic effects, but it also makes the numerical solution of the Vlasov equation a challenging task \citep{Armstrong_70,Knorr_76}. We here discuss the main features of the Vlasov equation and the numerical difficulties arising from the phase mixing effects. The phase mixing and oscillatory behavior of the Vlasov equation in velocity space leads to wave packets in the Fourier transformed velocity space, which points to the idea to absorb the highest Fourier modes via boundary conditions in the Fourier transformed Vlasov equation.

\subsection{The filamentation in velocity space and the numerical recurrence effect}

The behavior of the Vlasov equation is illustrated by the one-dimensional
Vlasov equation for electrons with stationary, singly charged ions, coupled with Gauss' law for the
electrostatic field,
\begin{align}
  &\Dt{f_e}+v\frac{\d f_e}{\d x}-\frac{eE}{m_e} \frac{\d f_e}{\d v}=0,
  \label{vlasov_PDE1}
  \\
  &\frac{\d E}{\d x} = \frac{e}{\varepsilon_0}\bigg[n_0-\intInfty f_e(x,v,t) \,\diff v\bigg],
  \label{vlasov_PDE_E1}
\end{align}
describing the evolution of the electron distribution function $f_e$ in a
self-consistent electric field $E$, where $n_0$ is the equilibrium electron (and ion) number density, $e$ is the magnitude of the electron charge, $m_e$ is the electron mass, and $\varepsilon_0$ is the electric vacuum permittivity. It can be cast into the dimensionless form
\begin{align}
  \label{vlasov_PDE}
  &\Dt{f_e}+v\frac{\d f_e}{\d x}-E \frac{\d f_e}{\d v}=0,
  \\
  \label{vlasov_PDE_E}
  &\frac{\d E}{\d x} = 1-\intInfty f_e(x,v,t) \,\diff v,
\end{align}
were the distribution function $f_e$ is normalized by $n_0/v_{th,e}$, time $t$ by $\omega_{pe}^{-1}$, space $x$ by $r_{De}$, velocity $v$ by $v_{th,e}$, and the electric field $E$ by
$v_{th,e}^2 m_e /e r_{De}$. Here $\omega_{pe}=(n_0 e^2/\varepsilon_0 m_e)^{1/2}$ is the electron plasma frequency, $r_{De}=(k_B T_e \varepsilon_0/n_0 e^2)^{1/2}$ is the electron Debye length, $v_{th,e}=(k_B T_e/m_e)^{1/2}$ is the electron thermal speed, $T_e$ is the electron temperature, and $k_B$ is Boltzmann's constant.

\begin{figure}[htb]
  \includegraphics[width=10cm]{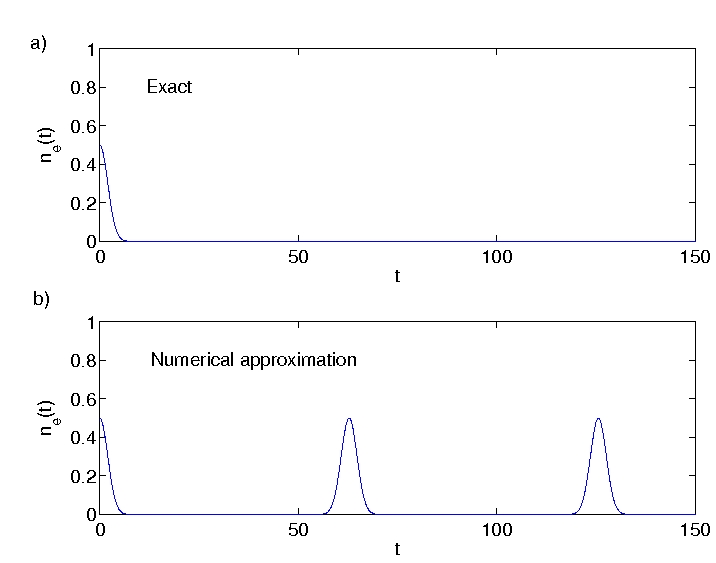}
  \caption{a) Exact and b) numerical approximation of the electron number density $n_e$. One can see a numerical recurrence effect with periodicity ${T_{\rm recurrence}=2\pi /(k_x\, \Delta v)\approx 62.8}$.}
  \label{fig:recurrence}
\end{figure}

To illustrate the main numerical difficulty of the Vlasov equation, we consider the
reduced, free-streaming problem
\begin{align}
  &\frac{\partial f}{\partial t}+v\frac{\partial f}{\partial x}=0,
  \qquad   f(x,v,t=0)=\frac{A}{\sqrt{2\pi}}\cos(k_x x)\exp(-\frac{v^2}{2})
\end{align}
which has the exact solution
\begin{equation}
f=(2\pi)^{-1/2}A\cos[{k_x} (x-{v t})]\exp(-v^2/2).
\label{f_exact}
\end{equation}
This solution becomes more and more oscillatory in the velocity space with increasing time due to the $kvt$ term inside the cosinus function. We note that the distribution function $f$ does not decay in time, but, due to phase mixing between positive and negative values of $f$, the number density
\begin{equation}
 n_e=\int f\,dv=A{\exp(-k_x^2 t^2/2)}\cos(k_x x),
\end{equation}
decays super-exponentially fast with time. Assume now that we have the exact solution $f$ of the electron distribution function, and want to calculate the electron number density numerically via a sum representation of the integral over velocity space. If we resolve velocity space with an equidistant grid as $v=v_j=j\Delta v$, $j=0$, $\pm 1$, $\pm 2$, ..., $\pm M$, where $M$ is a large integer, then a numerical approximation of the electron number density is
\begin{equation}
  n_e\approx \sum_{j=-M}^{M} (2\pi)^{-1/2}A\cos[{k_x} (x-{j\Delta v t})]\exp(-j^2 \Delta v^2/2)\,\Delta v,
\end{equation}
which turns out to be {\emph periodic} in time with periodicity ${T_{\rm recurrence}=2\pi /(k_x\, \Delta v)}$.
For example, for $k_x=0.5$ and $\Delta v=0.2$, we have ${T_{\rm recurrence}=2\pi /(k_x\, \Delta v)\approx 62.8}$. While the exact number density decays super-exponentially, we see in Fig \ref{fig:recurrence} that in the numerical approximation, the initial condition recurs periodically with periodicity ${T_{\rm recurrence}\approx 62.8}$. This is the recurrence effect. It is in fact impossible to represent the solution on the grid after a finite time due to the Nyquist-Shannon sampling theorem, which states that one needs more than two grid points per wavelength to represent the solution an equidistant grid. Here we see from Eq. (\ref{f_exact}) that the ``wavelength'' in velocity space is $\lambda_v=2\pi/(k_x t)$. Hence, the sampling theorem $\lambda_v/\Delta v>2$ for representing the distribution function on the grid is violated for times $t\geq\pi/(k_x \Delta v)=T_{\rm recurrence}/2$.

\begin{figure}[htb]
  \centering
  \subfigure[The distribution function $f(x,v,t)$]{
      \includegraphics[height=7cm]{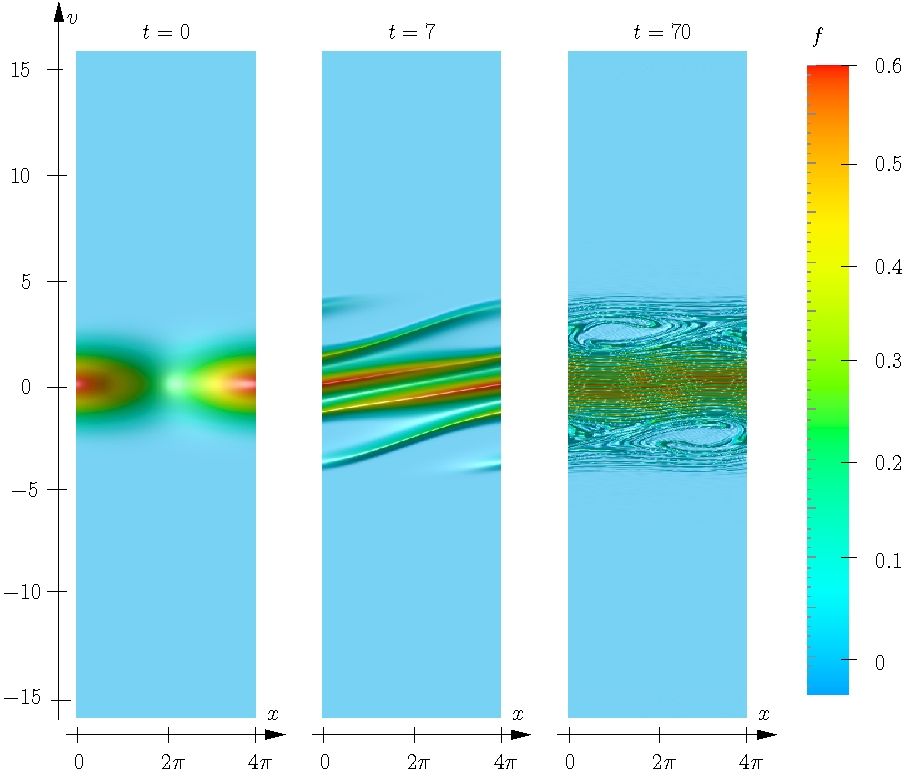}
    \label{fig:F_portrait}
  }
  \subfigure[The real part of the Fourier transformed distribution function $\f(x,\eta,t)$]{
      \includegraphics[height=7cm]{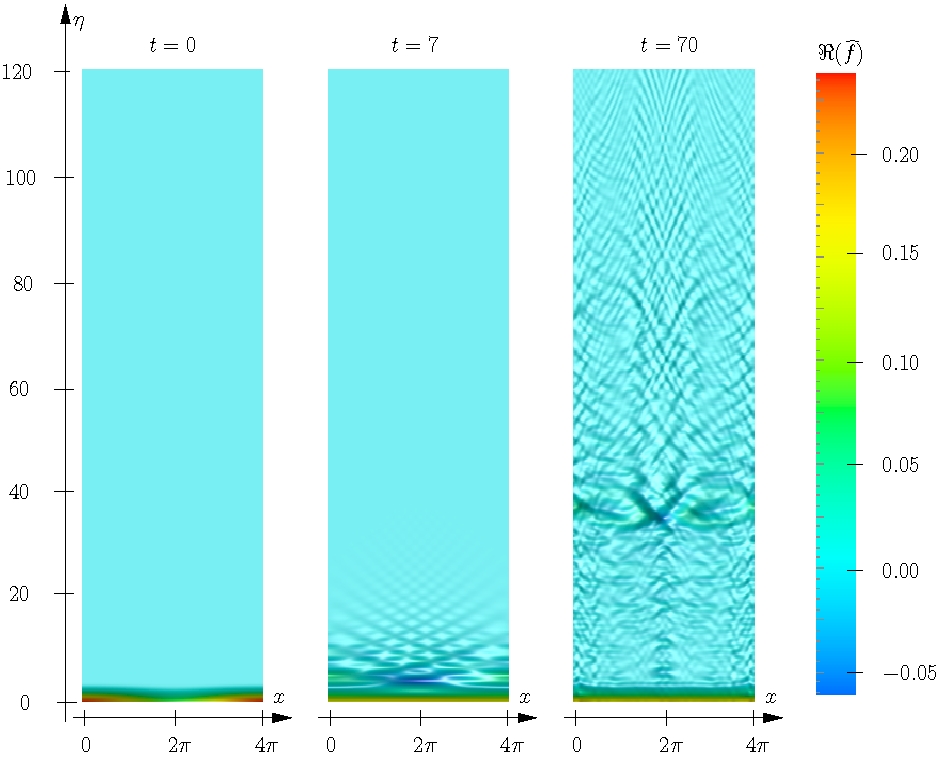}
    \label{fig:fre_portrait}
  }
  \caption{Phase space plots of the electron distribution function $f(x,v,t)$ and the real part of the Fourier transformed distribution function $\f(x,\eta,t)$ at different times. After \citet{Eliasson_06}.}
  \label{fig:phase_space}
\end{figure}



\begin{figure}[htb]
  \centering
    \includegraphics[width=10cm]{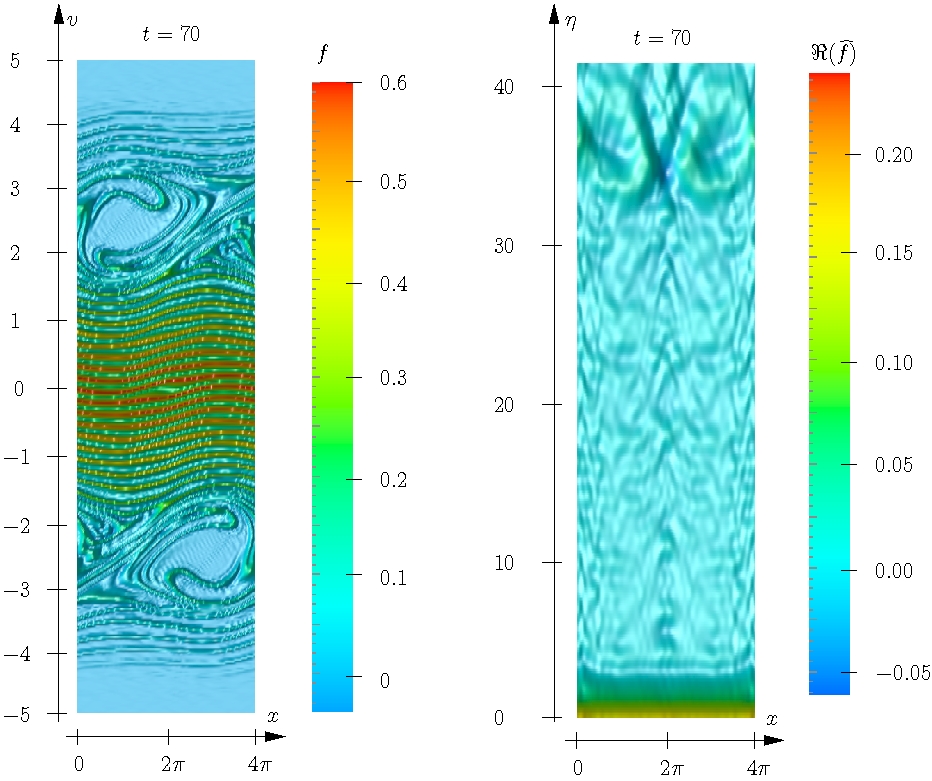}
  \caption{A closeup of the distribution function $f(x,v,t)$ and of the real part of the
  Fourier transformed distribution function $\f(x,\eta,t)$ at
  the time $t=70$. After \citet{Eliasson_06}.}
  \label{fig:close_portrait}
\end{figure}

We now return to the system (\ref{vlasov_PDE})--(\ref{vlasov_PDE_E}).
By employing the Fourier transform pair
\begin{align}
  f(x,v,t)&=\intInfty \f(x,\eta,t)\mathrm{e}^{-\i \eta v}\,
    \diff \eta \\
  \label{Fourier2}
  \f(x,\eta,t)&=\frac{1}{2\pi}\int_{-\infty}^{\infty}f(x,v,t)e^{\i\eta v}\,\diff v
\end{align}
the system (\ref{vlasov_PDE})--(\ref{vlasov_PDE_E}) is
transformed into a new set of equations
\begin{align}
  \label{TrVlasov}
  &\frac{\d \f}{\d t}-\i\dfdxdeta+
  \i E\eta \f=0
  \\
  \label{vlasov_PDE2_E}
  &\frac{\d E(x,t)}{\d x}=
  1-2 \pi \f(x,\eta,t)_{\eta=0}
\end{align}
for the Fourier transformed distribution function $\f(x,\eta,t)$. Here $\f$ is normalized by $n_0$ and the
Fourier transformed velocity variable $\eta$ is normalized by $v_{{\rm th,e}}^{-1}$, while
the other variables are normalized in the same manner as in Eqs. (\ref{vlasov_PDE})--(\ref{vlasov_PDE_E}).
As initial conditions for Eq. (\ref{vlasov_PDE}), we will use \citep{Armstrong_70,Knorr_76}
\begin{equation}
  \label{knorr0}
  f(x,v,0)=[1+A\cos(k_x x)]f_0(v),
\end{equation}
with $A=0.5$, $k_x=0.5$, and $f_0(v)=(2\pi)^{-1/2}
\exp\left(-v^2/2\right)$; see Fig.~\ref{fig:F_portrait} at $t=0$.
The corresponding initial condition for the Fourier transformed
Vlasov equation (\ref{TrVlasov}), shown at $t=0$ in Fig.~\ref{fig:fre_portrait}, is
\begin{equation}
  \label{knorr1}
  \f(x,\eta,0)=[1+A\cos(k_x x)]\f_0(\eta),
\end{equation}
with $\f_0(\eta)=(2\pi)^{-1}\exp(-\eta^2 /2)$.
We used the method of \citet{Eliasson_01} to solve the system (\ref{vlasov_PDE})--(\ref{vlasov_PDE_E}).
The simulation domain was
set to ${0 \le x\le 4 \pi}$ and $0\le \eta \le 120$ with $N_x=200$, $N_\eta=600$, and the time
domain was  $0 \le t \le 70$ with $N_t=35000$ and $\Delta t=0.002$. The numerical
dissipation parameter in $x$ space was set to ${\delta=0.001}$.

As can be seen in Fig.~\ref{fig:F_portrait}, the solution has
at $t=7$ formed filaments with large gradients in velocity
space. At $t=70$, the gradients have further steepened,
and two Bernstein-Green-Kruskal (BGK) waves
have been formed. A closeup of the solution is shown in the
left-hand panel of Fig.~\ref{fig:close_portrait}. The
initially smooth solution has evolved into an oscillatory solution with
steep gradients, primarily in $v$ space but also in $x$ space.
As a contrast, the Fourier transformed
solution $\f(x,\eta,t)$, displayed in Fig.~\ref{fig:fre_portrait}
for the same times as in Fig.~\ref{fig:F_portrait}, does not become oscillatory
in velocity space. Instead, wave packets are formed and are propagating away from the
origin $\eta=0$ in the Fourier transformed velocity space.

The smooth solution in the Fourier transformed velocity space is a desirable property
for numerical approximations of the Vlasov equation. It is possible to give an upper bound
on the derivatives in $\eta$ space and on numerical truncation errors for numerical schemes that
approximate the $\eta$ derivatives, which is not possible in the original velocity $v$ space.
By inspection of the three panels in Fig.~\ref{fig:F_portrait},
one can see that the solution has significantly non-zero values
only in the velocity interval $v=-5$ to $v=5$. This
suggests that the Vlasov equation in the Fourier transformed space
has a smooth solution. If one assumes that the solution in
Fig.~\ref{fig:F_portrait} for all times vanishes as a Gaussian function
for large $v$, with the estimate
\begin{equation}
  \label{f_estimate}
  \Abs{f(x,v,t)}<C\exp(-\gamma v^2)
\end{equation}
for some positive constants $C$ and $\gamma$, then the $\eta$
derivatives of the Fourier transformed solution are bounded as
\begin{equation}
\label{diff_estimate}
\begin{split}
  &\bigg|\frac{\d^n}{\d\eta^n}\f(x,\eta,t)\bigg| = \mbox{[Use Eq.~(\ref{Fourier2})]}
  =\bigg|\frac{1}{2\pi}\intInfty(\i v)^n e^{\i \eta v}f(x,v,t)\,\diff v\bigg| \\
  &<\mbox{[Use the triangle inequality]}<
  \frac{1}{2\pi}\intInfty|(\i v)^n e^{\i \eta v}f(x,v,t)|\,\diff v \\
  &=\frac{1}{2\pi}\intInfty|v|^n |f(x,v,t)|\,\diff v<\mbox{[By Eq.~(\ref{f_estimate})]} \\
  &<\frac{1}{2\pi}\intInfty|v|^n C\exp(-\gamma v^2)\,\diff v=
  \frac{1}{2\pi}\frac{C}{\gamma^{(n+1)/2}}a_n
\end{split}
\end{equation}
where the constant
\begin{equation}
  a_n=\left\{
  \begin{array}{ll}
    \sqrt{\pi}\, 2^{-n/2}(n-1)!!, &{\mbox{$n$ even}}
    \\
    $[(n-1)/2]!$ ,  &{\mbox{$n$ odd}}
  \end{array}
  \right.
\end{equation}
and where the symbols $!$ for the factorial and $!!$ for the
semi-factorial have their usual meaning. Thus, by the assumption
(\ref{f_estimate}) for $f(x,v,t)$ it follows that $\f(x,\eta,t)$
is infinitely differentiable with respect to $\eta$ with the
estimate (\ref{diff_estimate}) for the derivatives. It is
therefore possible to make an error estimate of the truncation
error of a difference scheme used to approximate the $\eta$
derivative in Eq.~(\ref{TrVlasov}). The 4th-order compact
Pad{\'e} difference scheme, which is used here to perform numerical
approximations of the first derivatives in $\Eta$ space (See Section \ref{Sec:Numerics} below)
has a truncation error of size
\begin{equation}
  |\epsilon|\leq\frac{1}{120}\Delta\eta^4 \max\bigg|\frac{\d^5\f}{\d\eta^5}\bigg|
\end{equation}
where the fifth derivative gives $n=5$ in
formula~(\ref{diff_estimate}). This gives the estimate
\begin{equation}
  \label{estimate}
  |\epsilon|<\frac{1}{2\pi}\frac{\Delta\eta^4}{60}\frac{C}{\gamma^3}
\end{equation}
for the truncation error. It is thus possible to make an error
estimate for the numerical differentiation in the Fourier
transformed velocity $\eta$ space in Eq.~(\ref{TrVlasov}), which
is not possible for a numerical differentiation in the original
velocity $v$ space in Eq.~(\ref{vlasov_PDE}).
In the closeup of the solution at time $t=70$, displayed in right panel of
Fig.~\ref{fig:close_portrait}, a wave packet can clearly be seen
at $\eta \approx 35$, which corresponds to the
``frequency'' of the oscillations in velocity space, seen in the
left panel of Fig.~\ref{fig:close_portrait}. Since the wave packet
is decoupled from the origin $\eta=0$ where the
electric field is calculated, it can be removed from the calculation
without immediately affecting the value of the electric field.
The wave packets eventually reach the artificial boundary at
$\eta=\etamax=120$; see the right panel of Fig. \ref{fig:fre_portrait}, where
they are absorbed, as described in the next section.

\begin{figure}
\centering
    \includegraphics[width=8cm]{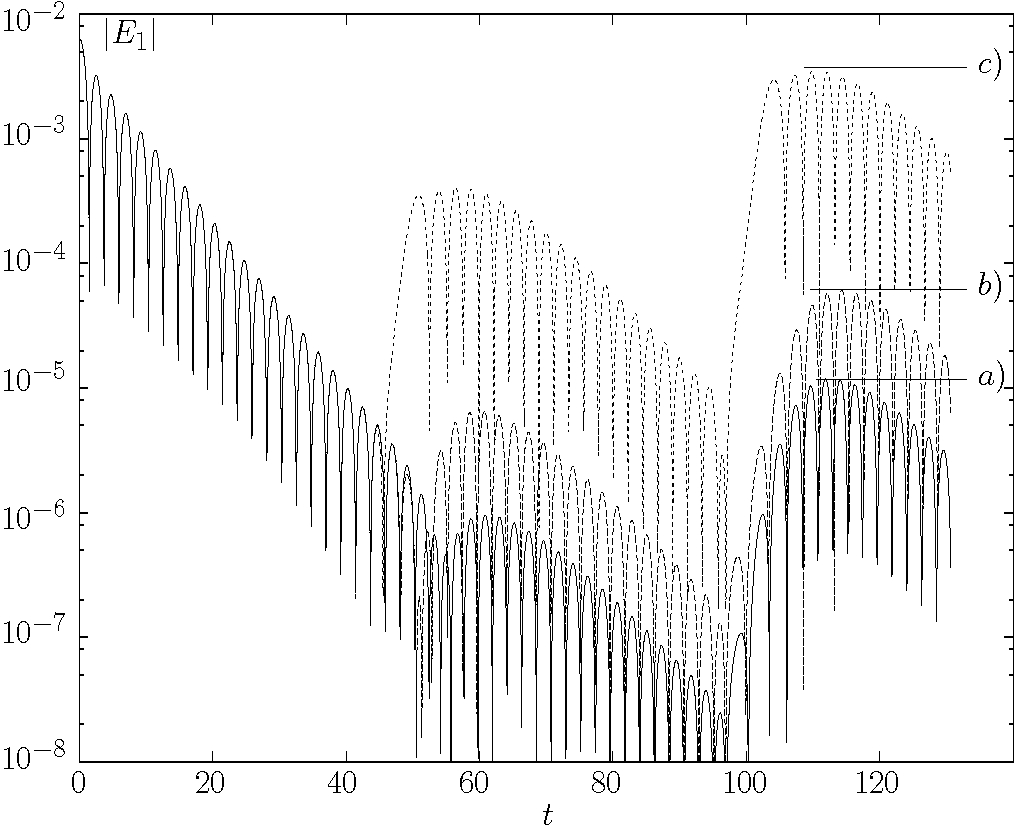}
\caption{The electric field for small-amplitude electrostatic waves. a) and b): outflow boundary conditions in $\eta$ space. c): $\widetilde{f}$ set to zero at boundary $\eta=\eta_{max}$. After \citet{Eliasson_01}.}
\label{reflections}
\end{figure}

\subsection{Outflow boundary conditions in Fourier transformed velocity space}
The idea developed in \citep{Eliasson_01,Eliasson_02,Eliasson_07} is to solve the
Vlasov equation in the Fourier transformed velocity space and to design absorbing
boundary conditions at the largest component; wave
packets that reach the artificial boundary in the Fourier transformed velocity space are
allowed to travel over the boundary and to be removed from the
solution, while incoming waves are set to zero at the boundary. In this manner a weakly
dissipative term is introduced in the Vlasov equations, which removes only the highest
oscillations in velocity space. By setting the artificial boundary further away from the
origin in the Fourier transformed velocity space, finer structures is resolved in velocity space.

The idea can be illustrated with the reduced problem
\begin{equation}
  \frac{\partial \widehat{f}}{\partial t}-i\frac{\partial^2 \widehat{f}}{\partial x\partial
  \eta}=0.
\end{equation}
A Fourier transform in space ($\partial/\partial x \rightarrow ik$)
gives the hyperbolic equation
\begin{equation}
  \frac{\partial \widetilde{f}}{\partial t}+k_x\frac{\partial \widetilde{f}}{\partial
  \eta}=0,
  \label{hyperbolic}
\end{equation}
which has solutions of the form $\widetilde{f}=\widetilde{f}_0(\eta-k_x t)$.
Well-posed outflow boundary conditions for Eq. (\ref{hyperbolic}) in $\eta$ space are found by
setting $\widehat{f}$ to zero at $\eta=\eta_{max}$ if $k_x<0$ and to zero at $\eta=-\eta_{max}$ if $k_x>0$.
It can be expressed as $\widetilde{f}=H(k_x)\widetilde{f}$ at $\eta=\eta_{max}$ and
$\widetilde{f}=H(-k_x)\widetilde{f}$ at $\eta=-\eta_{max}$
where $H$ is the Heaviside function, here defined as $H(k_x)=0$ for $k_x\leq 0$ and $H(k_x)=1$ for $k_x>0$.
Inverse Fourier transforming the boundary conditions, we obtain
well-posed outflow boundary conditions for $\widehat{f}$,
\begin{equation}
  \widehat{f}={\rm F}^{-1}[H(k_x){\rm F} \widehat{f}]\quad\mbox{at } \eta=\eta_{max}
  \label{outflow}
\end{equation}
and
\begin{equation}
  \widehat{f}={\rm F}^{-1}[H(-k_x){\rm F} \widehat{f}]\quad\mbox{at } \eta=-\eta_{max}
\end{equation}
where ${\rm F}$ and ${\rm F}^{-1}$ are the forward and inverse spatial Fourier transforms.

It turns out that the outflow boundary condition (\ref{outflow}) also works well for
the complete Fourier transformed system (\ref{TrVlasov})--(\ref{vlasov_PDE2_E}).
Figure \ref{reflections} shows a simulation with a small-amplitude wave in the
initial condition so that we have an
almost linear problem. The wavenumber is set to $k_x=0.5$ so that the wave is
strongly Landau damped. We are comparing cases where we have used the outflow
boundary condition (\ref{outflow}) at $\eta=\eta_{max}=30$ [panels a) and b) in Fig. \ref{reflections}] with a case where we simply set $\widetilde{f}$ to zero at the boundary [panel c)].
[A denser grid in $\eta$ space is used in a) compared to b).]
In the simulation in Fig. \ref{reflections}, the amplitude of the wave is initially decreasing
exponentially as expected from linear Vlasov theory. At $t\approx 60 \omega_{pe}^{-1}$, there
is a weak recurrence of the waves, and  at $t\approx 120\omega_{pe}^{-1}$ a stronger recurrence
takes place. We see that the recurrence phenomenon is much weaker for case a) and b)
where outflow boundary conditions were used in $\eta$ space,
while in case c), where the distribution function was set to zero at the $\eta$ boundary, the amplitude of the recurring wave at $t=120\,\omega_{pe}^{-1}$ is of the same order as in the initial condition. For the case c) the numerical results would be useless for a nonlinear problem after the recurrence has taken place at $t=120\,\omega_{pe}^{-1}$, while for the other cases the linear Landau damping effects are reasonably well represented and a nonlinear problem could be run beyond the recurrence time.

\begin{figure}[htb]
 \centering
    \includegraphics[width=8cm]{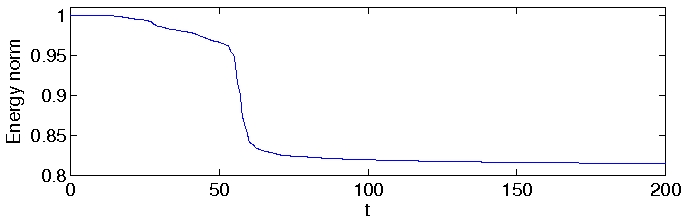}
\caption{Large amplitude case. Time development of the energy integral $||\widetilde{f}||_2$ in Eq. (\ref{Entropy}), normalized by its initial value.}
\label{fig:entropy}
\end{figure}

An interesting question is what is flowing out at the outflow boundary in $\eta$ space.
It is easy to show that the energy
(entropy) integrals $||f||^2$ and $||\widehat{f}||^2$ are conserved in time, where
\begin{align}
  ||f||^2=\int_0^L\!\!\int f^2 dv\,dx \mbox{ and } ||\widehat{f}||^2=\int_0^L\!\!\int |\widehat{f}|^2 d\eta\,dx.
\end{align}
if periodic boundary conditions are used in space.
They are related as $||f||^2=2\pi ||\widehat{f}||^2$ via Parseval's relation.
With the outflow boundary condition (\ref{outflow}) in $\eta$ space, it was shown by \citet{Eliasson_01} for the one-dimensional case that the energy integral
\begin{align}
  ||\widehat{f}||^2=\int_0^L \!\!\int_{-\eta_{max}}^{\eta_{max}} |\widehat{f}|^2 d\eta\,dx
  \label{Entropy}
\end{align}
is non-increasing in time, i.e. $d||\widehat{f}||^2/dt\leq 0$.
In fact, due to that $f$ is real valued, the symmetry condition $\widehat{f}(x,-\eta,t)=\widehat{f}^*(x,\eta,t)$ was used by \citet{Eliasson_01}, so that $\widehat{f}$ is real-valued at $\eta=0$.
On the other hand, the total number of particles,
\begin{align}
  N=\int_0^L 2\pi \widehat{f}_{\eta=0} \,dx
\end{align}
is always conserved, and so is the total (kinetic plus electrostatic) energy of the system,
\begin{align}
  W=\int_0^L \bigg(-\pi \frac{\partial^2\widehat{f}}{\partial\eta^2}\bigg|_{\eta=0} +\frac{E^2}{2}\bigg)\,dx.
\end{align}
The decrease of $||\widehat{f}||^2$ can be seen on as a loss of information about the finest details of
the distribution function. Hence, the outflow boundary conditions in $\eta$ space represents
a dissipative process in which the highest oscillations in velocity space are removed from the system and a partial thermalization is allowed.
In Fig. \ref{fig:entropy} we show the time evolution of the energy integral (\ref{Entropy})
relative to its initial value for a simulation of strongly
nonlinear electrostatic waves, with the initial conditions (\ref{knorr1}) with amplitude $A=0.5$ and
wavenumber $k_x=0.5$ (i.e. the same initial conditions as in Fig. \ref{fig:phase_space}), and using
the outflow boundary conditions (\ref{outflow}) at the boundary
$\eta=\eta_{max}=30$. Initially the energy integral decreases slowly, but at time
$t\approx 60$ it decreases sharply.
This time corresponds to $\eta_{max}/k$, i.e. the time when the wave packet reaches the boundary
$\eta=\eta_{max}$ as predicted by the solution of the reduced problem (\ref{hyperbolic}).
The simulation was continued till $t=7000$, and it could be observed that the energy integral decreased
to a value of about $0.8130$ after which it exhibited very small fluctuations \citep{Eliasson_01}.

\subsection{The relation between the outflow boundary condition and the Hilbert transform}
There is a simple relation between the outflow boundary
condition (\ref{outflow}) and the Hilbert transform, which in an infinite domain
is defined as
\begin{align}
  \mathcal{H}[f](x)=\frac{1}{\pi}\mathsf{p}\intInfty\frac{f(y)}{x-y}\diff y
  \label{Hilbert}
\end{align}
where $\mathsf{p}$ denotes the Cauchy principal value.
The outflow boundary condition (\ref{outflow}) at the
artificial boundary $\eta=\etamax$, extended to an infinite spatial domain, is
\begin{align}
  \f=\F^{-1}H(k_x)\F\f,
    \hspace{0.5cm} \eta=\etamax
\end{align}
where the spatial forward and inverse spatial Fourier transforms
are defined as
\begin{align}
  \F \phi &= \int_{-\infty}^{\infty} \phi(x) e^{-\i k_x x} \diff x
  \intertext{and}
  \F^{-1} \widetilde{\phi} &= \frac{1}{2\pi}
  \int_{-\infty}^{\infty} \widetilde{\phi}(k) e^{\i k_x x} \diff k,
\end{align}
and the Heaviside function as
\begin{align}
  H(k_x)=\left\{
  \begin{array}{l}
    1, \hspace{0.5cm} k_x>0
  \\
    0, \hspace{0.5cm} k_x \le 0
  \end{array}
  \right.
\end{align}

The projection operator ${\mathcal{G}}\equiv\F^{-1}H(k_x)\F$, acting on
$\f$ as
\begin{align}
  \mathcal{G}[f](x)=\F^{-1}H(k_x)\F f
\end{align}
projects the function $\f$ onto the space of functions with only
positive Fourier components in $x$ space. In other words, the
projection removes components with negative wavenumbers at the boundary
and leaves components with positive wavenumbers unchanged.
The Hilbert transform (\ref{Hilbert}) has the property that
\begin{align}
  \mathcal{H}[e^{\i k_x x}](x)=\mathrm{sign}(k_x) \i e^{\i k_x x}
\end{align}
and it follows that the boundary operator
$\mathcal{G}$ can be expressed in terms of the Hilbert transform as
\begin{align}
  \label{int_G}
  \mathcal{G}[f](x)=\frac{1}{2}\big[f(x)-\i \mathcal{H}[f](x)\big]=
  \frac{1}{2}\bigg[f(x)-\i\frac{1}{\pi}\mathsf{p}\intInfty\frac{f(y)}{x-y}\diff y\bigg]
\end{align}
i.e., as an operator in real $x$ space. Using Sohockij--Plemelj's formulas, we also have
\begin{align}
  \label{int_G2}
  \mathcal{G}[f](x)=-\frac{i}{2\pi} \lim_{\delta\rightarrow 0+} \intInfty\frac{f(y)}{x-y+i\delta}\diff y.
\end{align}
The integral formulations (\ref{int_G}) or (\ref{int_G2}) of the boundary operator may open up the possibility
to construct well-posed absorbing boundary conditions also for non-periodic problems,
where the integrals are restricted to bounded limits.

\section{The numerical aspects of the Fourier transformed Vlasov equation\label{Sec:Numerics}}
We here discuss the numerical approximations used to solve the Fourier transformed Vlasov equation.
As an example, we will discuss Fourier method for the one-dimensional
Vlasov equation coupled with Gauss' law \citep{Eliasson_01} in some detail.
Most of the results presented here carry over to the two- and three-dimensional cases \citep{Eliasson_02,Eliasson_03,Eliasson_07}, and we therefore
omit detailed discussions of the numerical methods in the multi-dimensional cases.
\subsection{Numerical approximations of the one-dimensional Vlasov equation}
We discretize the one-dimensional Vlasov equation on a
rectangular, equidistant grid with the numerical box $0\leq x < L$ in space,
and $0\leq\eta\leq\etamax$ in the Fourier transformed velocity space.
The discrete function values at the grid points are enumerated
such that
  $\f(x_i,\eta_j,t_k)\approx \f_{i,j}^k$
with the spatial variable $x_i=i \Delta x$, $i=0,\, 1,\,\ldots,\,N_x-1$,
and the Fourier transformed velocity variable
$\eta_j=j\Delta \eta$, $j=0,\, 1,\,\ldots,\,N_{\eta}$,
where $\Delta x = L/N_x$, $\Delta \eta = \etamax/N_{\eta}$.
The discrete time is obtained from the initial $t_0=0$ and then
$t_k=t_{k-1}+\Delta t_k$, $k=1,\, 2,\,\ldots,\,N_t$
The time step $\Delta t_k$ may be kept fixed or varied dynamically to maintain numerical
stability \citep{Eliasson_02}.

The system (\ref{TrVlasov})--(\ref{vlasov_PDE2_E}) restricted to a finite domain,
\begin{align}
  \label{alg1}
  &\Part{\f}{t}=\i\dfdxdeta-\i\eta E \f,
  \hspace{0.5cm} 0\le\eta<\etamax,\, 0\le x < L,
  \\
  \label{alg2}
  &\Part{E}{x}=1-2 \pi \f(x,0,t),
\end{align}
are supplemented by the outflow boundary condition (\ref{outflow}) at
$\eta=\eta_{\mathrm{max}}$,
\begin{align}
  \label{alg3}
  \f&=\F^{-1}H(k_x)\F\f,
  \quad \eta=\etamax,
  \intertext{and periodic boundary conditions in $x$ space,}
  \label{alg4}
  \f(x+L,\eta,t)&=\f(x,\eta,t).
\end{align}
Since $f$ is real valued, we have the symmetry $\f(x,-\eta,t)=\f^*(x,\eta,t)$,
and hence the real part of $\f$ is even and the imaginary part odd with respect to $\eta$ \citep{Armstrong_70, Eliasson_01}. At $\eta=0$ one can therefore use a symmetry boundary condition, as discussed below.
Equation (\ref{alg2}) is solved numerically to obtain $E$, which is
then inserted into right-hand side of Eq.~(\ref{alg1});
one can thus consider $E$ as a function of $f$.
The outflow boundary condition at $\eta=\etamax$ is
performed by the boundary operator $\F^{-1}H(k_x)\F$, which
removes all Fourier components with negative spatial
wavenumbers ($k_x<0$) at the boundary.
Discrete approximations are used
and for obtaining $E$ in Eq. (\ref{alg2}),
for the boundary operator in Eq. (\ref{alg3}), and
for the $\eta$ and $x$ derivatives the right-hand side
of Eq.~(\ref{alg1}). The semi-discretized system can then be written as
\begin{equation}
\label{alg5}
  \frac{d \f_{i,j}}{\d t}=P(\f)_{i,j}
\end{equation}
where $P$ is a grid function of all $\f_{i,j}$, representing the numerical
approximation of the right-hand side of Eqs. (\ref{alg1}). The
unknowns $\f_{i,j}$ are then advanced in time with the
4th-order Runge-Kutta algorithm:
\begin{quote}
\begin{enumerate}
\item
  $ F^{(1)}_{i,j} \leftarrow P(\f^k),\quad \forall\, i,j $
\item
  $ F^{(2)}_{i,j} \leftarrow P(\f^k+F^{(1)} \Delta t/2),\quad \forall\, i,j $
\item
  $ F^{(3)}_{i,j} \leftarrow P(\f^k+F^{(2)} \Delta t/2),\quad \forall\, i,j $
\item
  $ F^{(4)}_{i,j} \leftarrow P(\f^k+F^{(3)} \Delta t),\quad \forall\, i,j $
\item
  $ \f^{k+1}_{i,j} \leftarrow \f^{k}_{i,j}
  +\frac{\Delta t}{6}(F^{(1)}_{i,j}+2F^{(2)}_{i,j}+2F^{(3)}_{i,j}+F^{(4)}_{i,j}),\quad \forall\, i,j $
\end{enumerate}
\end{quote}
The steps needed for obtaining the approximation $P_{i,j}$ are:
\begin{quote}
\begin{enumerate}
\item
Calculate the electric field numerically from Eq.~(\ref{alg2}).
\item
Apply numerically the boundary operator on the right-hand side, according
to Eq.~(\ref{alg3}), for the points
along the boundary $\eta=\eta_\mathrm{max}$.
\item
Calculate a numerical approximation of the right-hand side of
Eq.~(\ref{alg1}), for all points including the points along the boundary
$\eta=\eta_\mathrm{max}$.
\end{enumerate}
\end{quote}

Pseudo-spectral methods are used to approximate $x$ derivatives
and to integrate the Poisson equation, using the
discrete Fourier transform (DFT) pair
\begin{align}
  \widehat{\phi}_\w &=\frac{1}{N_x}\sum_{j=0}^{N_x-1}
  \phi(x_j)\exp\(-\i 2\pi\w\frac{j}{N_x}\)\equiv\mathrm{DFT}\phi(x),\\
  \label{pseudo2}
  \phi(x_j)&=\sum_{\omega=-\left(N_x/2-1\right)}^{N_x/2}
  \phi_\w\exp\(\i 2 \pi\omega \frac{x_j}{L}\)\equiv\mathrm{DFT}^{-1}
  \widehat{\phi}_\w.
\end{align}
The $x$ derivatives are approximated in the pseudo-spectral method as
\begin{equation}
 \frac{\d \phi}{\d x}\approx\mathrm{DFT}^{-1}\left[\i k_x \mathrm{DFT}(\phi)\right],
\end{equation}
where $k_x=2\pi j\omega/L$,
and the integration of the electric field in (\ref{alg2}) is approximated by
\begin{equation}
  E\approx
  \mathrm{DFT}^{-1}\left[\frac{1}{\i k_x} \mathrm{DFT}(1-2\pi\f_{i,0}^k)\right],
\end{equation}
for all $k_x\neq 0$, while the component of $E$ corresponding to $k_x=0$ is
set equal to zero.

In $\eta$ space, the derivative $v=\d \f/\d \eta$
is calculated using the classical fourth order Pad\'e scheme
\citep{Lele_92}. For the inner points, the implicit
approximation
\begin{equation}
\label{eqsys1}
  v_{i,j-1}+4v_{i,j}+v_{i,j+1}=\frac{3}{\Delta\eta}
  \(\f_{i,j+1}-\f_{i,j-1}\),
  \hspace{0.5cm} j=1,\,2,\,\ldots,\,N_{\eta}-1
\end{equation}
is used. At the boundary $\eta=0$, the same approximation of the $\eta$
derivative is used as for the inner points, taking into
account the symmetry relations
$\f_{i,-1}=\f_{i,1}^*$ and $v_{i,-1}=-v_{i,1}^*$,
\begin{equation}
\label{eqsys2}
  -v_{i,1}^*+4v_{i,0}+v_{i,1}=\frac{3}{\Delta\eta}
  \(\f_{i,1}-\f_{i,1}^*\)
\end{equation}
or, for the real and imaginary parts,
\begin{align}
  \label{eqsysreal}
  v_{i,0}^{(\mathrm{Re})} &= 0 \\
  \label{eqsysimag}
  2 v_{i,0}^{(\mathrm{Im})}+v_{i,1}^{(\mathrm{Im})}&=
  \frac{3}{\Delta\eta}\f_{i,1}^{(\mathrm{Im})}
\end{align}
respectively.
At the boundary $\eta=\etamax$, the one-sided approximation
\begin{equation}
\label{eqsys3}
  v_{i,N_{\eta}}+2v_{i,N_{\eta}-1}=
  -\frac{1}{2\Delta\eta}
  \(-5\f_{i,N_{\eta}}+4\f_{i,N_{\eta}-1}
  +\f_{i,N_{\eta}-2}\)
\end{equation}
is used. This gives a truncation error of order $\Delta\eta^3$
at the boundary.
Equations (\ref{eqsys1}), (\ref{eqsys2}) and (\ref{eqsys3}) form one
real valued and one imaginary valued tri-diagonal
equation system for each subscript
$i=0,\,1,\,\ldots,\,N_x$, each system having $N_\eta$ unknowns.

At the boundary $\eta=\etamax$, the boundary condition (\ref{alg3})
is applied with the approximation
\begin{equation}
  \F^{-1}H(k_x)\F\phi(x,\eta_\mathrm{max},t)\approx
  \mathrm{DFT}^{-1}\left[H(k_x) \mathrm{DFT}(\phi_{i,N_\eta}^k)\right]
\end{equation}
where $\phi(x,\eta_\mathrm{max},t)$ is the right-hand side of
(\ref{alg1}) along the boundary $\eta=\etamax$ and
$\phi_{i,N_\eta}^k$ its discrete approximation.

\subsection{The numerical representation of particle velocities} \label{sec:velocities}
\begin{figure}[htb]
  \centering
  \resizebox{10cm}{!}{
    \includegraphics[]{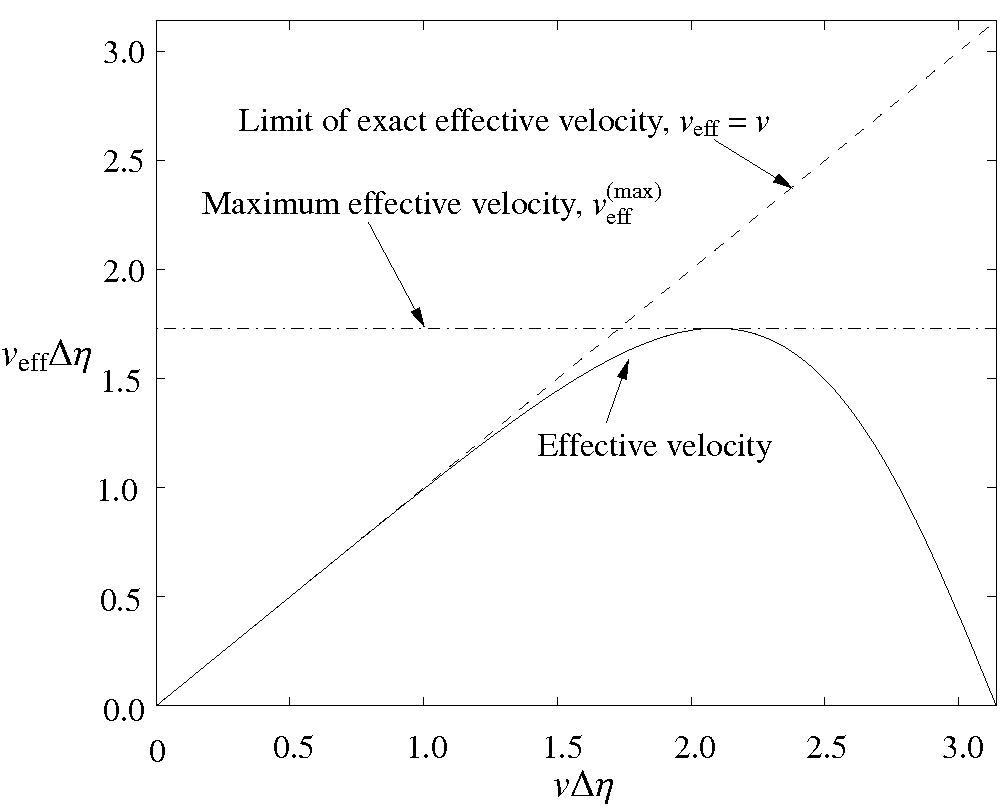}
  }
  \caption{The effective particle velocities produced by the 4th-order compact Pad{\'e} difference scheme
  in $\eta$ space. After \citet{Eliasson_06}.}
  \label{fig:veff}
\end{figure}
In order to investigate the impact of the difference scheme in the Fourier transformed
velocity space on effective particle velocities, we study an approximation of the
Fourier transformed Vlasov equation, in which the $x$ and $t$
derivatives are performed exactly, while the $\eta$ derivative is
approximated by Eq. (\ref{eqsys1}). The difference scheme can formally
be written as a difference operator $D_\eta$, giving rise to the
difference-differential equation
\begin{align}
  \label{DiffVlasov}
  \frac{\d \f}{\d t}-\i\frac{\d}{\d x}(D_\eta \f)+
  \i E\eta \f=0
\end{align}
In order to study the effect of the difference approximation
$D_\eta$ on the solution in $v$ space, the definition
(\ref{Fourier2}) is inserted into Eq.~(\ref{DiffVlasov}), giving
\begin{align}
  \label{DiffVlasov2}
  \frac{1}{2\pi}\intInfty\bigg[
  \Dt{f}e^{\i\eta v}-\i\frac{\d f}{\d x}D_\eta(e^{\i\eta v})+
   \i E\eta f e^{\i\eta v}
  \bigg]\,\diff v=0
\end{align}
where the difference operator gives $D_\eta(e^{\i\eta
v})=\i\veff(v)e^{\i\eta v}$ with the effective particle velocity $\i\veff(v)$, and
the term $\i E \eta f e^{\i\eta v}$ gives rise to the term $-e^{\i E \eta v}\d f/\d v$ by
an integration by parts, yielding
\begin{align}
  \label{DiffVlasov22}
  \frac{1}{2\pi}\intInfty\bigg[
  \Dt{f}+\veff(v)\frac{\d f}{\d x}-E \frac{\d f}{\d v}
  \bigg]e^{\i\eta v}\,\diff v=0,
\end{align}
and thus
\begin{align}
  \label{DiffVlasov3}
  \Dt{f}+\veff(v)\frac{\d f}{\d x}-E \frac{\d f}{\d v}=0,
\end{align}
which is an approximation of the Vlasov equation
(\ref{vlasov_PDE}), with the effective velocity $\veff(v)$
produced by the numerical differentiation in $\eta$ space. For the
Pad{\'e} scheme (\ref{eqsys1}) we have
\begin{equation}
  \veff(v)=\frac{3}{\Delta \eta}\frac{\sin(v\Delta \eta)}{[2+\cos(v\Delta \eta)]}.
\end{equation}
In the limit
$v\Delta\eta\rightarrow 0$, then $\veff\rightarrow v$; see
Fig.~\ref{fig:veff} where $\veff\Delta\eta$ has been plotted as a
function of $v\Delta \eta$. The equations for the particle trajectories
in $(x,v)$ space, produced by Eq.~(\ref{DiffVlasov3}), are
\begin{align}
  \label{charact_x2}
  \dot{x}(t)&=\veff[v(t)]=\frac{3}{\Delta \eta}\frac{\sin[v(t)\Delta \eta]}{\{2+\cos[v(t)\Delta \eta]\}} \\
  \label{charact_v2}
  \dot{v}(t)&=-\frac{e}{\me}E[x(t),t],
\end{align}
where the dots in the left-hand sides denote time derivatives $d/dt$.
Thus the particles are transported in $x$ space with the effective
velocity $\veff$, which is periodic in $v$.
If the product $v\Delta \eta$ is
too large, then the approximation $\veff\approx v$ breaks down;
see Fig.~\ref{fig:veff}. A maximum effective velocity,
$\veff^{(\mathrm{max})}\Delta\eta=\sqrt{3}\approx 1.73$, can be
found for $v\Delta \eta=2\pi/3\approx 2.09$. It means that even
though the largest \emph{represented} velocity is given by the
Nyquist limit $v_{\mathrm{max}}=\pi/\Delta \eta\approx 3.14/\Delta
\eta$, the highest effective velocity for transport of particles
in $x$ space is $\veff^{(\mathrm{max})}=\sqrt{3}/\Delta\eta$. In
numerical experiments, one has to choose a small enough
$\Delta\eta$, so that important phenomena in velocity space, for
example beams of particles, are well resolved, i.e., the
velocities of these particles must fulfil
$v<\veff^{(\mathrm{max})}$ with some margin.
In the simulation performed to produce the results in the present
section, particles were accelerated to velocities somewhat less
than $v=5$; see the left-hand panel of
Fig.~\ref{fig:close_portrait}. The grid size was
$\Delta\eta=(120/600)=0.2$, which gives
that $v\Delta\eta\approx 1.0$ for these particles. According to
the diagram in Fig.~\ref{fig:veff}, the effective velocity is
close to the limit of exact velocity for the value
$v\Delta\eta=1.0$, and thus the particle velocities for the
fastest particles are well resolved. The maximum effective
velocity produced in the simulation was
$\veff^{(\mathrm{max})}=\sqrt{3}/\Delta\eta\approx 8.6$.
\subsection{The choice of domain and grid sizes}
\label{sec:choice}
When using the numerical algorithm to solve physical problems, it is
important to know what is the computational domain and resolution in
the real velocity space, i.e., what is the maximum velocity
component $v_{{\rm max}}$, used in the real velocity space to resolve the particle
distribution function and what is the grid
size $\Delta v$ used in the numerical solution. For
given $\eta_{{\rm max}}$ and $\Delta \eta$, the maximum represented
velocity and the grid size in
velocity space are given by
\begin{equation}
  v_{{\rm max}}=\frac{\pi}{\Delta \eta}
\end{equation}
 and
\begin{equation}
  \Delta v=\frac{\pi}{\eta_{{\rm max}}},
\end{equation}
respectively. In view of the results in Fig.~\ref{fig:veff}, one should choose
$\Delta \eta$ small enough such that the maximum
velocity component $v_{{\rm max}}$ is more than twice the maximum particle velocity
one wants to resolve in the numerical solution, in order to avoid
dispersive errors on the particle velocities. One must also choose ${\eta_{{\rm max}}}$
large enough so that fine enough structures in the velocity distribution function
is resolved.

\subsection{One-dimensional simulations of electron and ion holes}
In simulations of processes with several timescales, it is important that the numerical scheme does
not introduce artificial heating of the electrons. Such effects can be a problem if a
diffusion operator is introduced in velocity space to minimize the filamentation effects in the Vlasov equation. In the Fourier method used here, there is minimal heating effects of the electrons since only the
highest Fourier modes in velocity space are absorbed by the outflow boundary condition in the Fourier transformed velocity space. As examples we will study the dynamics of electron and ion holes in an electron-ion plasma, which is governed by the Vlasov-Poisson system of equations for electrons and ions,
\begin{align}
  &\Dt{f_e}+v\frac{\d f_e}{\d x}+\frac{e}{m_e}\frac{\partial \Phi}{\partial x} \frac{\d f_e}{\d v}=0,
  \\
  &\Dt{f_i}+v\frac{\d f_i}{\d x}-\frac{e}{m_i}\frac{\partial \Phi}{\partial x} \frac{\d f_i}{\d v}=0,
  \\
  &-\frac{\d^2 \Phi}{\d x^2} = \frac{e}{\varepsilon_0}\bigg[\intInfty(f_i(x,v,t)-f_e(x,v,t))\,\diff v\bigg],
\end{align}
and where both the electron and ion dynamics are important. As initial conditions for the simulations, we have used Schamel's quasi-stationary solutions for electron and ion holes \citep{Schamel_79,Schamel_71,Schamel_81,Schamel_86}.

\begin{figure}[htb]
\includegraphics[width=8cm]{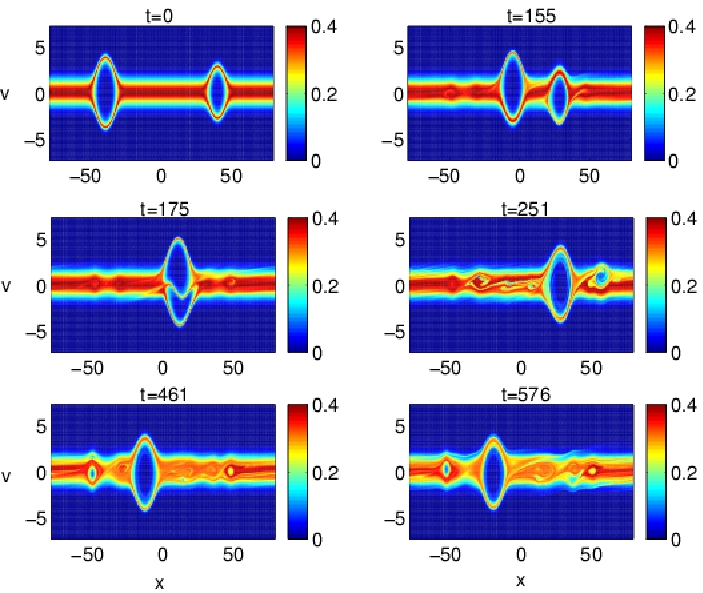}
\caption{
The electron distribution function for two electron holes
at time $t=0$ (upper left panel),
$t=155\,\omega_{pe}^{-1}$ (upper right panel), $t=175\,\omega_{pe}^{-1}$ (middle left panel),
$t=251\,\omega_{pe}^{-1}$ (middle right panel), $t=461\,\omega_{pe}^{-1}$ (lower left panel)
and $t=576$ (lower right panel).  The color bars go
from dark blue (small values) to dark red (large values).
After \citet{Eliasson_eh_04}.
}
\label{Fig6eh}
\end{figure}

\begin{figure}[htb]
\includegraphics[width=8cm]{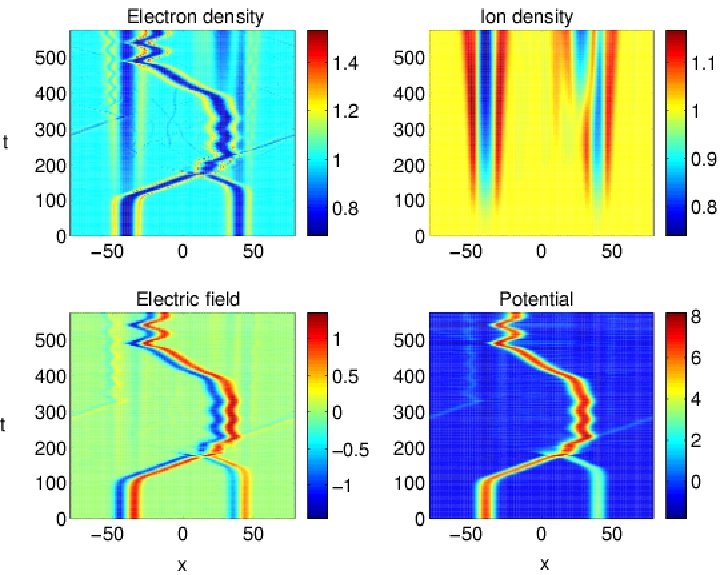}
\caption{The time and space evolution of
the electron density (upper left panel),
the ion density (upper right panel), the electric field (lower left panel) and
the potential (lower right panel) associated with the two
electron holes in Fig.~\ref{Fig6eh}.
After \citet{Eliasson_eh_04}.
}
\label{Fig5eh}
\end{figure}

Figures \ref{Fig5eh} and \ref{Fig6eh} show a simulation of interacting electron holes in
and electron-ion plasma with mobile ions and realistic ion to electron mass ratio of
29500 for oxygen ions \citep{Eliasson_eh_04}.
Two large-amplitude electron holes were initially placed at $x=-40\,r_{De}$ and $x=40\,r_{De}$, displayed at $t=0$ in the
 phase space number density plots in Fig. \ref{Fig6eh} (top left panel). A small electron density perturbation, centered at the two electron holes,
were introduced as a seed for any instability. The ion density was initially taken to be homogeneous.
We see in Fig. \ref{Fig6eh} that at $t=155\,\omega_{pe}^{-1}$ (top right panel), the two electron holes
having started moving towards each other, and that at $t=175\,\omega_{pe}^{-1}$ they are colliding and are merging to a new electron hole (middle left panel), and that the newly created electron hole at $t=461\,\omega_{pe}^{-1}$ has propagated to
$x=30\,r_{De}$ (middle right panel), and at a later stage it has propagated to $x=-30\,r_{De}$ where it remains throughout the simulation (lower panels). The reason for this complicated behavior of the electron holes can be understood by
studying the interaction with the ions in detail. Due to their positive potential, the electron
holes expel the ions and create local ion density cavities, which in turn eject and accelerate the
electron holes. This can clearly be seen in Fig. \ref{Fig5eh}, where the two electron holes start propagating
at $t\approx 100\,\omega_{pe}^{-1}$ and $t\approx 130\,\omega_{pe}^{-1}$, respectively.  At $t\approx 170\,\omega_{pe}^{-1}$, the
two electron holes collide and merge into a new electron hole with larger amplitude, which
propagates slightly in the positive $x$ direction,
and becomes trapped at a local ion density maximum at $x\approx 30\,r_{De}$;
see the upper right panel of Fig. \ref{Fig5eh} for the ion density and the lower
right panel for the potential.  After $t\approx 400\,\omega_{pe}^{-1}$, a new
ion density cavity is created where the electron hole is centered, and
at this time the electron hole is again ejected and accelerated in the negative $x$
direction. At $t\approx 480\,\omega_{pe}^{-1}$, the moving
electron hole again encounters an ion density maximum located at $x\approx-30\,r_{De}$,
where it is again trapped, performing large oscillations. We see that the electron holes
remain stable during the acceleration by ion density cavities
and survive on an ion time scale, much longer than the electron plasma period.

\begin{figure}[htb]
\includegraphics[width=8cm]{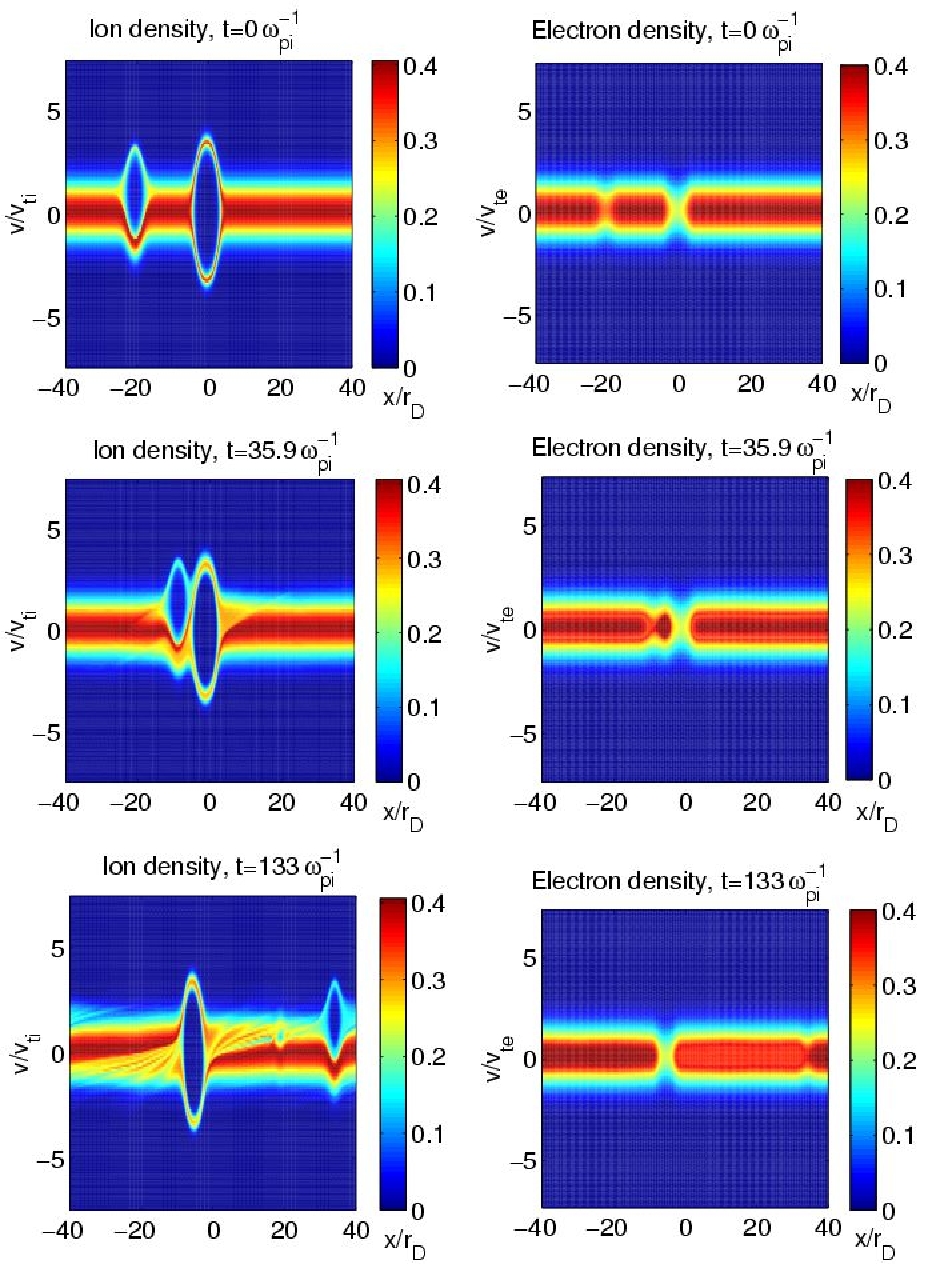}
\caption{
The distribution function for the ions (left panels) and electrons
(right panels) of two colliding ion holes, before the collision
at times $t=0\,\omega_{pi}$ (upper panel) and
$t=35.9\,\omega_{pi}$ (middle panel), and after the collision
at $t=133\,\omega_{pi}$ (lower panel). The color bars go
from dark blue (small values) to dark red (large values).
After \citet{Eliasson_ih_04}.
}
\label{Fig1coll}
\end{figure}

\begin{figure}[htb]
\includegraphics[width=8cm]{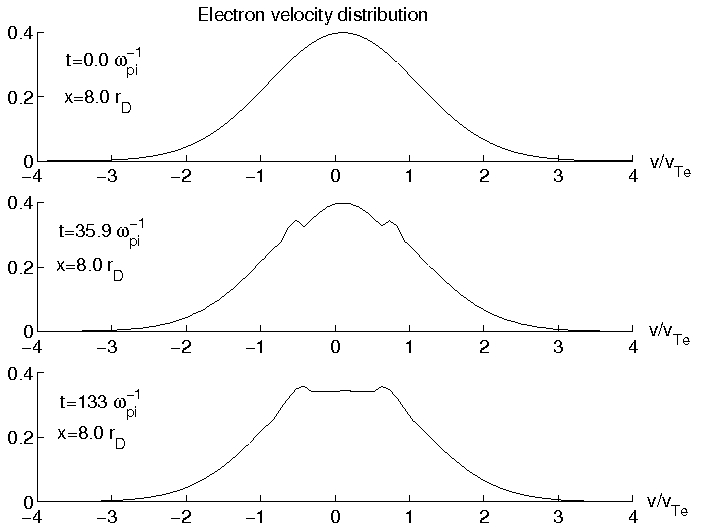}
\caption{
The electron velocity distribution at $x=8\,r_D$, for
$t=0\,\omega_{pi}^{-1}$ (upper panel), $t=35.9\,\omega_{pi}^{-1}$ (middle panel)
and $t=133\,\omega_{pi}^{-1}$ (lower panel).
After \citet{Eliasson_ih_04}.
}
\label{Fig2coll}
\end{figure}

Figure \ref{Fig1coll} displays the features of the ion and electron distribution functions
for two colliding ion holes, where initially (upper panels) the left ion hole
propagates with the speed $u_0=0.9\,v_{Ti}$ where $v_{Ti}=\sqrt{k_B T_i/m_i}$ is the ion thermal
speed, and the right ion hole is standing. The ion and electron distribution functions associated with the ion holes
are shown before collision at times $t=0\,\omega_{pi}^{-1}$ (upper panels)
and $t=35.9\,\omega_{pi}^{-1}$ (middle panels), and after collision at time
$t=133\,\omega_{pi}^{-1}$ (lower panels), where $\omega_{pi}=(n_0 e^2/\varepsilon_0 m_i)^{1/2}$ is the ion plasma frequency.  Figure \ref{Fig1coll} exhibits that the ion holes
undergo collisions without being destroyed; thus they are robust structures.
As can be seen in the right panels of Fig \ref{Fig1coll}, the electrons have a non-Maxwellian,
flat topped distribution in the region between the ion holes after collision has taken place.
The velocity distribution function is plotted as a function of $v/v_{Te}$ in Fig. \ref{Fig2coll} at
$x=8.0\,r_{De}$. We see that the initial Maxwellian distribution (the upper panel) changes
to a distribution with beams at $v\approx \pm 0.6\,v_{Te}$ (the middle panel) slightly
before collision, and to a flat-top distribution with two maxima after collision
(the lower panel).  The reason for this phenomenon is that the two ion holes are associated
with negative electrostatic potentials, and the electrons entering the region between the
ion holes after collision must have a large enough kinetic energy to cross the potential
barriers that are set up by the ion holes. Therefore, the region between the ion holes
becomes excavated of low energy electrons. Also here the features of the electron distribution function survives on the ion time-scale, much longer than the electron plasma period.

\section{The two-dimensional Vlasov equation}
We here discuss the extension of the Fourier transform method to
the Fourier transformed Vlasov equation in two spatial and two
velocity dimensions, including external and self-consistent magnetic
fields. In the design of well-posed absorbing boundary conditions
in the Fourier transformed velocity space, special care has to be
taken with the magnetic field, which enters into the formulation of the
boundary condition.

\subsection{The two-dimensional Vlasov-Poisson system}

The two-dimensional Vlasov-Poisson system for electrons reads
\begin{align}
  \label{vlasov1}
  &\frac{\d f}{\d t}+v_1\frac{\d f}{\d x_1}+v_2\frac{\d f}{\d x_2}
  -\frac{e}{m}\(
  E_1\frac{\d f}{\d v_1}+E_2\frac{\d f}{\d v_2}+Bv_2\frac{\d f}{\d v_1}
  -Bv_1\frac{\d f}{\d v_2}\)=0 \\
  &E_1=-\frac{\d\Phi}{\d x_1},\qquad E_2=-\frac{\d\Phi}{\d x_2}
  \\
  \label{vlasov1_2}
  &-\(\frac{\d^2\Phi}{\d x_1^2}+\frac{\d^2\Phi}{\d x_2^2}\)=
  \frac{e}{\epsilon_0}\left[n_0-\intInfty\intInfty f
  \,\diff v_1\,\diff v_2\right]
\end{align}
where $n_0$ is the neutralizing heavy ion density background, fixed uniformly in space.
The external magnetic
field $\B(x_1,x_2,t)$ is here directed along the $x_3$ axis,
perpendicular to the $(x_1,x_2)$ plane, and
the electrostatic potential is calculated self-consistently from
Poisson's equation.

By using the Fourier transform pair
\begin{align}
  \nonumber
  f(x_1,x_2,v_1,v_2,t)&=\intInfty \intInfty \f(x_1,x_2,\eta_1,\eta_2,t)
    \mathrm{e}^{-\i (\eta_1 v_1+\eta_2 v_2)}\,\diff\eta_2\,\diff\eta_1
  \\
  \f(x_1,x_2,\eta_1,\eta_2,t)&=\frac{1}{(2\pi)^2}\intInfty \intInfty f(x_1,x_2,v_1,v_2,t)
    e^{\i(\eta_1 v_1+\eta_2 v_2)}\,\diff v_2\,\diff v_1
\end{align}
the system (\ref{vlasov1})--(\ref{vlasov1_2}) is transformed into
\begin{align}
  \label{vlasov2}
  &\frac{\d \f}{\d t}-\i\frac{\d^2\f}{\d x_1 \d\eta_1}
  -\i\frac{\d^2\f}{\d x_2 \d\eta_2}+
  \frac{e}{m}\bigg[\i (E_1\eta_1+E_2\eta_2) \f+
  B\eta_1\frac{\d \f}{\d \eta_2}-B\eta_2\frac{\d \f}{\d \eta_1}\bigg]=0,
  \\
  &E_1=-\frac{\d\Phi}{\d x_1},\qquad E_2=-\frac{\d\Phi}{\d x_2},
  \\
  \label{vlasov2_2}
  &-\(\frac{\d^2\Phi}{\d x_1^2}+\frac{\d^2\Phi}{\d x_2^2}\)=
  \frac{e}{\epsilon_0}\left[n_0-(2\pi)^2
  \f_{\eta_1=\eta_2=0}\right].
\end{align}

The systems (\ref{vlasov1})--(\ref{vlasov1_2}) and (\ref{vlasov2})--(\ref{vlasov2_2}) can be
cast into dimensionless form
\begin{align}
  \label{vlasov3}
  &\frac{\d f}{\d t}+v_1\frac{\d f}{\d x_1}+v_2\frac{\d f}{\d x_2}
  -\(
  E_1\frac{\d f}{\d v_1}+E_2\frac{\d f}{\d v_2}+Bv_2\frac{\d f}{\d v_1}
  -Bv_1\frac{\d f}{\d v_2}\)=0,
  \\
  \label{vlasov3_2}
  &E_1=-\frac{\d\Phi}{\d x_1}, \qquad E_2=-\frac{\d\Phi}{\d x_2}
  \\
  \label{vlasov3_3}
  &-\(\frac{\d^2\Phi}{\d x_1^2}+\frac{\d^2\Phi}{\d x_2^2}\)=
  1-\intInfty\intInfty f
  \,\diff v_1\,\diff v_2
  \intertext{and}
  \label{vlasov4}
  &\frac{\d \f}{\d t}
  -\i\frac{\d^2\f}{\d x_1 \d\eta_1}
  -\i\frac{\d^2\f}{\d x_2 \d\eta_2}+
  \i (E_1\eta_1+E_2\eta_2) \f+
  B\eta_1\frac{\d \f}{\d \eta_2}-B\eta_2\frac{\d \f}{\d \eta_1}=0
  \\
  \label{vlasov4_1}
  &E_1=-\frac{\d\Phi}{\d x_1},\qquad E_2=-\frac{\d\Phi}{\d x_2}
  \\
  \label{vlasov4_2}
  &-\(\frac{\d^2\Phi}{\d x_1^2}+\frac{\d^2\Phi}{\d x_2^2}\)=
  1-(2\pi)^2 \f_{\eta_1=\eta_2=0}
\end{align}
respectively, where time $t$ has been normalized by $\w_{\mathrm{pe}}^{-1}$, the velocity variables $v_1$ and $v_2$ by $v_{\mathrm{th,e}}$, the Fourier transformed velocity variables
$\eta_1$ and $\eta_2$ by $v_{\mathrm{th,e}}^{-1}$, the spatial variables
$x_1$ and $x_2$ by $r_{\mathrm{De}}$, the Fourier transformed distribution function $\f$ by $n_0$, the function $f$ by $n_0 v_{\mathrm{th,e}}^{-2}$, the electric field components $E_1$ and $E_2$ by
$v_{\mathrm{th,e}}^2 r_{\mathrm{De}}^{-1}m_e/e$, the electric potential $\Phi$ by
$v_{\mathrm{th,e}}^2 (m/e)$, and the magnetic field $B$ by $\w_{\rm pe} m_{\rm e} /e$.

In order to adapt the system (\ref{vlasov4})--(\ref{vlasov4_2})
for numerical simulations, the computational domain is restricted to
${0\le x_1 < L_1}$, ${0\le x_2 < L_2}$, $0\le\eta_1\le\eta_{{\rm max},1}$ and
$-\eta_{{\rm max},2}\le\eta_2\le\eta_{{\rm max},2}$.
For negative $\eta_1$, the symmetry $\f(x_1,x_2,\eta_1,\eta_2,t)=\f^*(x_1,x_2,-\eta_1,-\eta_2,t)$ is used
to obtain function values, if needed, owing to that the original distribution function $f(\x,\v,t)$ is real-valued. It is therefore not necessary to represent the solution for negative $\eta_1$
on the numerical grid. In the $x_1$ and $x_2$ directions, the periodic boundary conditions
\begin{align}
\f(x_1+L_1,x_2,\eta_1,\eta_2,t)&=\f(x_1,x_2,\eta_1,\eta_2,t)
\intertext{and}
\f(x_1,x_2+L_2,\eta_1,\eta_2,t)&=\f(x_1,x_2,\eta_1,\eta_2,t)
\end{align}
respectively, are used.

\subsection{Outflow boundary conditions in Fourier transformed velocity space}
\label{sec:bound}

Similar to the one-dimensional Vlasov-Poisson system discussed above, we wish to
design absorbing artificial boundary conditions in the Fourier transformed velocity space, so that
the highest oscillations in velocity space can be captured at the boundary and removed from the
calculation. Hence, at the boundaries at $\eta_1=\eta_{{\rm max},1}$ and
$\eta_2=\pm\eta_{{\rm max},2}$, the strategy is to let outgoing waves pass
over the boundaries, and to set incoming waves to zero. We explore the idea by studying
the reduced initial value problem with only a constant magnetic field ${B=B_0}$,
\begin{align}
  \label{vlasov5}
  &\frac{\d \f}{\d t}
  -\i\frac{\d^2\f}{\d x_1 \d\eta_1}
  -\i\frac{\d^2\f}{\d x_2 \d\eta_2}
  +B_0\eta_1\frac{\d \f}{\d \eta_2}-B_0\eta_2\frac{\d \f}{\d \eta_1}=0 \\
  &f(x_1,x_2,\eta_1,\eta_2,0)=f_0(x_1,x_2,\eta_1,\eta_2)
\end{align}
A Fourier transform in space ($\partial/\partial x_1 \rightarrow ik_{x1}$ and $\partial/\partial x_2 \rightarrow ik_{x2}$) gives a new differential equation for the
unknown function $\widetilde{f}(k_{x1},k_{x2},\eta_1,\eta_2,t)$,
\begin{align}
  \label{vlasov6}
  &\frac{\d \widetilde{f}}{\d t}+(k_{x1}-B_0\eta_2)\frac{\d\widetilde{f}}{\d\eta_1}
  +(k_{x2}+B_0\eta_1)\frac{\d\widetilde{f}}{\d\eta_2}=0 \\
  &\widetilde{f}(k_{x1},k_{x2},\eta_1,\eta_2,t)_{t=0}=\widetilde{f}_0(k_{x1},k_{x2},\eta_1,\eta_2)
\end{align}
This is a hyperbolic equation for which the initial values are transported
along the characteristic curves, given by
\begin{align}
  \label{char1}
  \frac{\diff \eta_1(t)}{\diff t}&=k_{x1}-B_0\eta_2(t) \\
  \frac{\diff \eta_2(t)}{\diff t}&=k_{x2}-B_0\eta_1(t)
\end{align}
Along the boundary $\eta_1=\eta_{{\rm max},1}$, Eq.~(\ref{char1})
describes an \emph{outflow} of data when ${k_{x1}-B_0\eta_2\geq 0}$ and
an \emph{inflow} of data when $k_{x1}-B_0\eta_2<0$. A well-posed
boundary condition is to set the inflow to zero at the boundary, i.e.,
\begin{align}
  \widetilde{f}_{\eta_1=\eta_{{\rm max},1}}=0, \hspace{0.5cm} k_{x1}-B_0\eta_2<0
\end{align}
which can be expressed with the help of the Heaviside step function $H$ as
\begin{equation}
  \widetilde{f}=H(k_{x1}-B_0\eta_2)\widetilde{f},
    \hspace{0.5cm} \eta_1=\eta_{{\rm max},1}
    \label{vlasov71}
\end{equation}
where
\begin{equation}
\label{vlasov8}
  H(k_{x1}-B_0\eta_2)=\left\{
  \begin{array}{l}
    1, \hspace{0.5cm} k_{x1}-B_0\eta_2 \ge 0
  \\
    0, \hspace{0.5cm} k_{x1}-B_0\eta_2 < 0
  \end{array}
  \right.
\end{equation}
The boundary condition (\ref{vlasov71}) allows
outgoing waves to pass over the boundary and to be removed, while incoming waves
are set to zero; the removal of the outgoing waves corresponds to the
losing of information about the finest structures in velocity space.

Inverse Fourier transforming Eq.~(\ref{vlasov8}) then
gives the boundary condition for the original problem
(\ref{vlasov5}) as
\begin{equation}
 \label{vlasov9}
  \f=\F_1^{-1}H(k_{x1}-B_0\eta_2)\F_1\f,
    \hspace{0.5cm} \eta_1=\eta_{{\rm max},1}
\end{equation}
The operator $\F_1^{-1}H(k_{x1}-B_0\eta_2)\F_1$ is a projection operator
which removes incoming waves at the boundary ${\eta_1=\eta_{{\rm max},1}}$. Similarly,
the boundary conditions at $\eta_2=\pm\eta_{{\rm max},2}$ becomes
\begin{align}
  \f&=\F_2^{-1}H(k_{x2}+B_0\eta_1)\F_2\f,
    \hspace{0.5cm} \eta_2=\eta_{{\rm max},2}
  \intertext{and}
  \label{vlasov9_2}
  \f&=\F_2^{-1}H(-k_{x2}-B_0\eta_1)\F_2\f,
    \hspace{0.5cm} \eta_2=-\eta_{{\rm max},2}
\end{align}
respectively.

In order to find well-posed boundary conditions in the $\eta_1$ and $\eta_2$ directions
in the case when $B=B(x_1,x_2,t)$ varies in time $t$ and in the $x_1$ and $x_2$ space with the periodicities
$L_1$ and $L_2$, respectively,
Eq.~(\ref{vlasov4}) is rewritten in an equivalent form as
\begin{align}
  \label{vlasov4_4}
  \begin{split}
  \frac{\d\f}{\d t}
  -\i\theta_1\frac{\d}{\d \eta_1}\(\frac{\d}{\d x_1}-\i\eta_2 B_{01}\)(\f
  \theta_1^{-1}) -\i\theta_2\frac{\d}{\d \eta_2}\(\frac{\d}{\d x_2}+\i\eta_1 B_{02}\)(\f
  \theta_2^{-1})
  +\i(E_1\eta_1+E_2\eta_2)\f=0
  \end{split}
\end{align}
where we introduced the spatially averaged magnetic fields
\begin{align}
  \label{B01}
  & B_{01}(x_2,t)=\frac{1}{L_1}\int_0^{L_1}B\,\diff x_1, \qquad B_{02}(x_1,t)=\frac{1}{L_2}\int_0^{L_2}B\,\diff x_2
\end{align}
and the phase factors
\begin{align}
  &\theta_1=\exp\bigg[\i\eta_2\int_0^{x_1}(B-B_{01})\,\diff x_1\bigg],
  \qquad
  \theta_2=\exp\bigg[-\i\eta_1\int_0^{x_2}(B-B_{02})\,\diff x_2\bigg].
\end{align}
We note that if $B$ is periodic and continuous in the
$x_1$ and $x_2$ directions, then the integrals
$\int_0^{x_1}(B-B_{01})\diff x_1$ and $\int_0^{x_2}(B-B_{02})\diff x_2$
are also periodic and continuous functions, and the $x_1$ and $x_2$ derivatives
can be approximated accurately by using the pseudo-spectral method.

By studying the flow of data in the $\eta_1$ direction for the
function $\phi_1=\f\theta_1^{-1}$
and the flow of data in the $\eta_2$ direction for the
function $\phi_2=\f\theta_2^{-1}$,
one can find outflow boundary condition in the $\eta_1$ and
$\eta_2$ directions, similar to the conditions (\ref{vlasov9})--(\ref{vlasov9_2}), as
\begin{align}
  \label{eta_max1}
  \begin{split}
  \f &=\theta_1
  \F_1^{-1}H(k_{x1}-\eta_2 B_{01})\F_1 (\f\theta_1^{-1}),
  \hspace{0.5cm} \eta_1=\eta_{{\rm max},1}
  \end{split}
  \\
  \label{eta_max21}
  \begin{split}
  \f &=\theta_2
  \F_2^{-1}H(k_{x2}+\eta_1 B_{02})\F_2
  (\f\theta_2^{-1}),
  \hspace{0.5cm} \eta_2=\eta_{{\rm max},2}
  \end{split}
  \\
  \label{eta_max22}
  \begin{split}
  \f &=\theta_2
  \F_2^{-1}H(-k_{x2}-\eta_1 B_{02})\F_2 (\f\theta_2^{-1}),
  \hspace{0.5cm} \eta_2=-\eta´_{{\rm max},2}
  \end{split}
\end{align}
respectively. In the case when $B$ is independent of $x_1$ and $x_2$, the
boundary conditions (\ref{eta_max1})--(\ref{eta_max22}) reduce to the
conditions (\ref{vlasov9})--(\ref{vlasov9_2}).
The boundary operators are projection operators, which allow outgoing
waves to pass over the $\Eta$ boundaries of the domain and to be removed
from the domain, while incoming waves are set to zero. The well-posedness
of these boundary conditions was proven by showing that a positive definite
energy integral is non-increasing in time \citep{Eliasson_02}.

\subsection{Electron Bernstein and upper hybrid waves in magnetized plasmas}
\label{sec:Bernstein}
We will here give examples of some effects related to linear electron oscillations in
magnetized plasmas. These include electron Bernstein modes that are exactly undamped according
to Landau theory. Due to this fact, there is a recurrence effect in a
weakly magnetized plasma, namely, there appears that waves can be periodically
Landau damped only to recur in the plasma at a later time. We here compare
simulation results with linear theory, in the form of dispersion corves for Bernstein
mode waves and time-dependent analytic solutions of the Vlasov equation for a magnetized
electron-ion plasma with immobile ions.

\begin{figure}[htb]
  \centering
  \subfigure[Dispersion diagram]{
      \includegraphics[height=6cm]{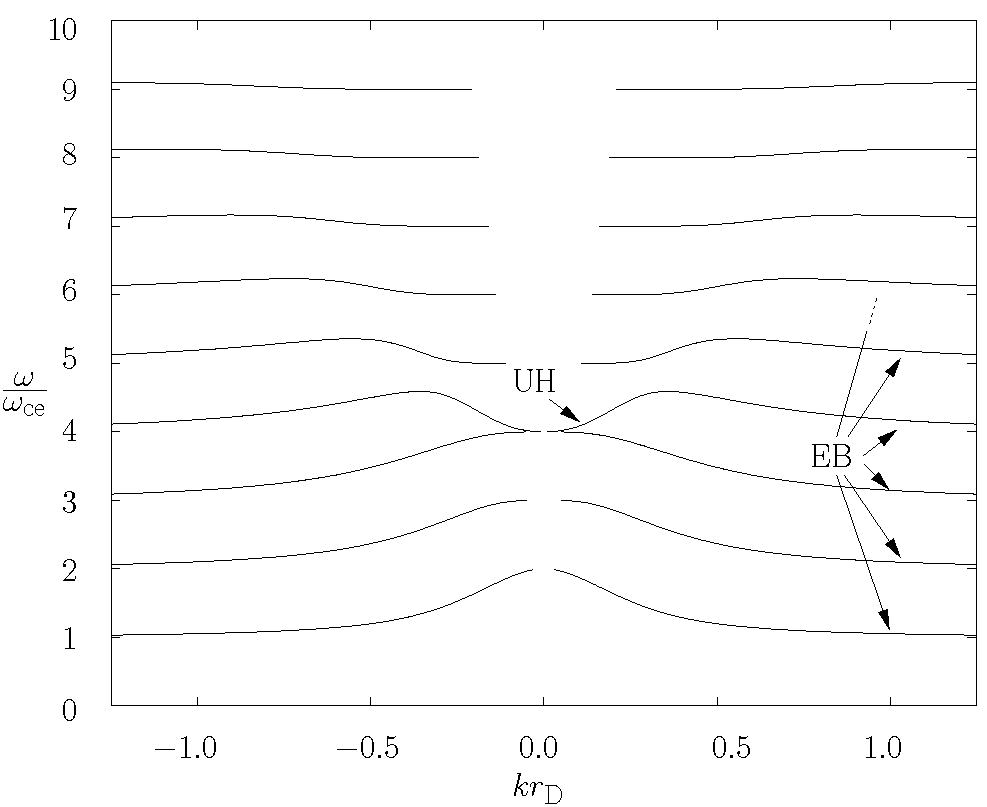}
    \label{fig:Bernstein_disp}
  }
  \subfigure[Power spectrum]{
      \includegraphics[height=6cm]{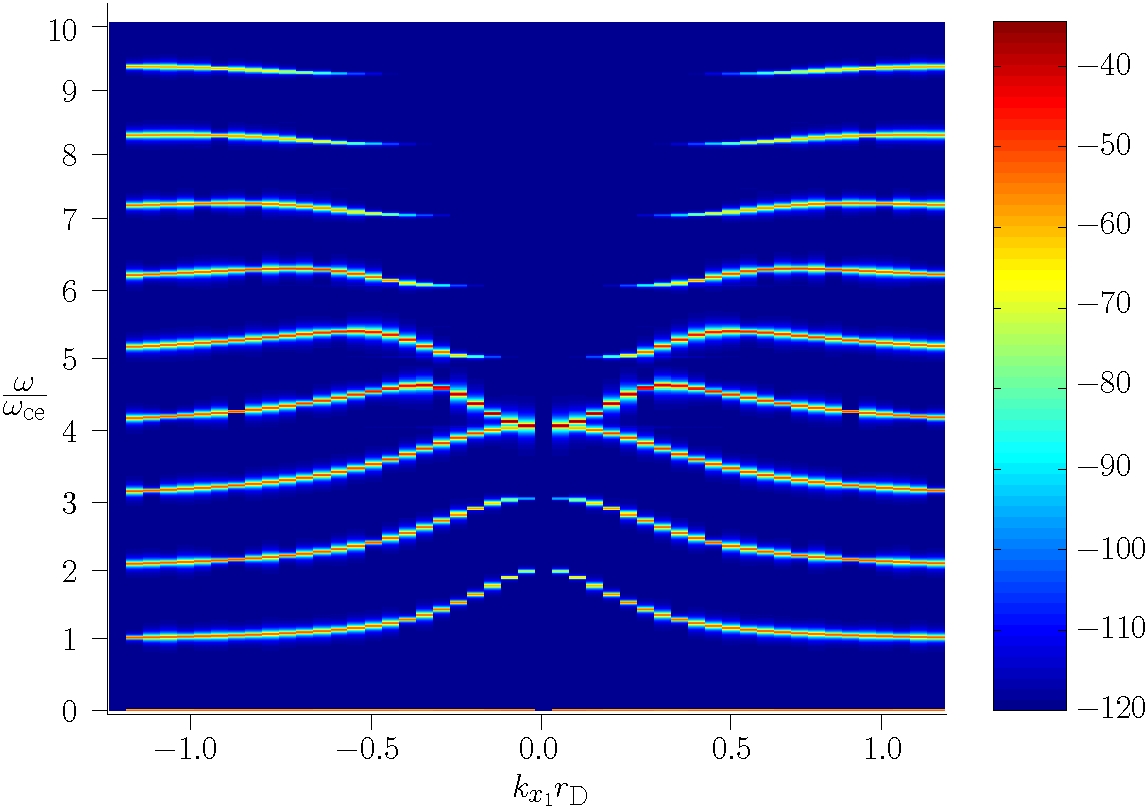}
    \label{fig:Bernstein_ES}
  }
  \caption{Dispersion diagram and power spectrum (decibel)
  for electrostatic electron Bernstein (EB) and upper hybrid (UH) waves;
  $\wuh=4\wce$. After \citet{Eliasson_08}.}
  \label{fig:Bernstein}
\end{figure}
The dispersion relation for the linear upper hybrid and
electron Bernstein modes in a Maxwellian plasma
is given by \citet{Crawford_65} as
\begin{align}
  \label{Bern_disp}
  1+\bigg(\frac{\wpe}{\wce}\bigg)^2\exp(-\lambda)\int_{\psi=0}^{\pi}
  \frac{\sin(\psi\w')\sin(\psi)\exp[-\lambda\cos(\psi)]}{\sin(\pi\w')}\diff\psi &=0
\end{align}
where
\begin{align}
  \w'&=\w/\wce,
  \\
  \lambda&=(k\rD)^2 \bigg(\frac{\wpe}{\wce}\bigg)^2,
  \\
  k^2&=k_{x1}^2+k_{x2}^2 .
\end{align}
Solving for $\w'$ in Eq.~(\ref{Bern_disp}) gives the relation between $\w$ and $k$.
Figure ~\ref{fig:Bernstein_disp} shows the dispersion curves for the case $\wuh=4\wce$,
and Fig.~\ref{fig:Bernstein_ES} shows the power spectrum in space and time from a simulation
of the Vlasov equation with the same parameters. In the simulation, the obtained time
series of the electric field component $E_1$ was Fourier transformed in $x_1$ space and
in time (using a Hamming window), to produce a power spectrum in a logarithmic scale.

As can be seen in Fig.~\ref{fig:Bernstein_ES}, the electrostatic energy is concentrated at the
linear Bernstein eigen-modes, in good agreement with theory.
In the long wavelength limit $k r_D\ll 1$, Eq.~(\ref{Bern_disp}) reduces to the dispersion relation
\begin{align}
  \label{wuh}
  \w^2=\wuh^2 + \frac{3 \vthe^2 k^2}{\omega^2-3\omega_{ce}^2}
\end{align}
where $\wuh=(\wpe^2+\wce^2)^{1/2}$ is the upper hybrid frequency. However, taking electromagnetic effects
into account, there are corrections in the long wavelength limit where the upper hybrid
waves go over to the electromagnetic $Z$ mode waves which we will discuss below.
A zero-frequency ($\w=0$) mode, which is not a solution of the dispersion relation (\ref{Bern_disp}),
can be seen in the power spectrum in Fig.~\ref{fig:Bernstein_ES};
this ``convective mode'' has earlier been observed in numerical PIC simulations by
\citet{Kamimura_78}, and theoretically by \citet{Sukhorukov_97}.
In terms of Landau theory, this mode is related
to a pole in the initial condition, and not to
a solution of the dispersion relation (\ref{Bern_disp}).

The simulation was restricted to one spatial dimension, along the $x_1$ axis, where
the simulation domain was set to ${0\le x_1/\rD \le 40\pi}$,
${0\le\eta_1\vthe\le 15}$ and ${0\le\eta_2\vthe\le 15}$ and the number of intervals
${N_{x1}=50}$, ${N_{\eta_1}=60}$ and ${2N_{\eta_2}=120}$, respectively.
The initial condition was set to
\begin{align}
  \f(x_1,x_2,\eta_1,\eta_2,0)=n(x_1)\f_0(\eta_1,\eta_2)
\end{align}
where the perturbed relative density was set to a sum of
waves with all possible wavenumbers,
\begin{align}
  n(x_1)=\bigg[1+A\sum_{i_1=1}^{24}i_1\sin(0.05 i_1 x_1/\rD)\bigg]
\end{align}
with the amplitude set to $A=0.0001$, giving an almost linear problem.
The Fourier transformed velocity distribution was set to a Maxwellian,
\begin{align}
  \f_0(\eta_1,\eta_2)&=\frac{n_0}{(2\pi)^2}\exp\bigg[-\frac{1}{2}(\eta_1^2+\eta_2^2)\vthe^2\bigg]
\end{align}
In velocity $\v$ space, the Maxwellian would be
\begin{align}
  f_0(v_1,v_2)&=\frac{n_0}{\vthe^2}\frac{1}{2\pi}\exp\bigg[-\frac{1}{2\vthe^2}(v_1^2+v_2^2)\bigg]
\end{align}
The external magnetic field was kept constant in the simulation, with the ratio $\wpe/\wce=\sqrt{15}$
(giving $\wuh=4\wce$). The number of time steps taken was $N_t=42\,530$ and the end time $t_{\mathrm{end}}=1737\wpe^{-1}$ with a fixed timestep.

\begin{figure}[htb]
  \centering
  \subfigure[Simulation results]{
      \includegraphics[width=8cm,height=9cm]{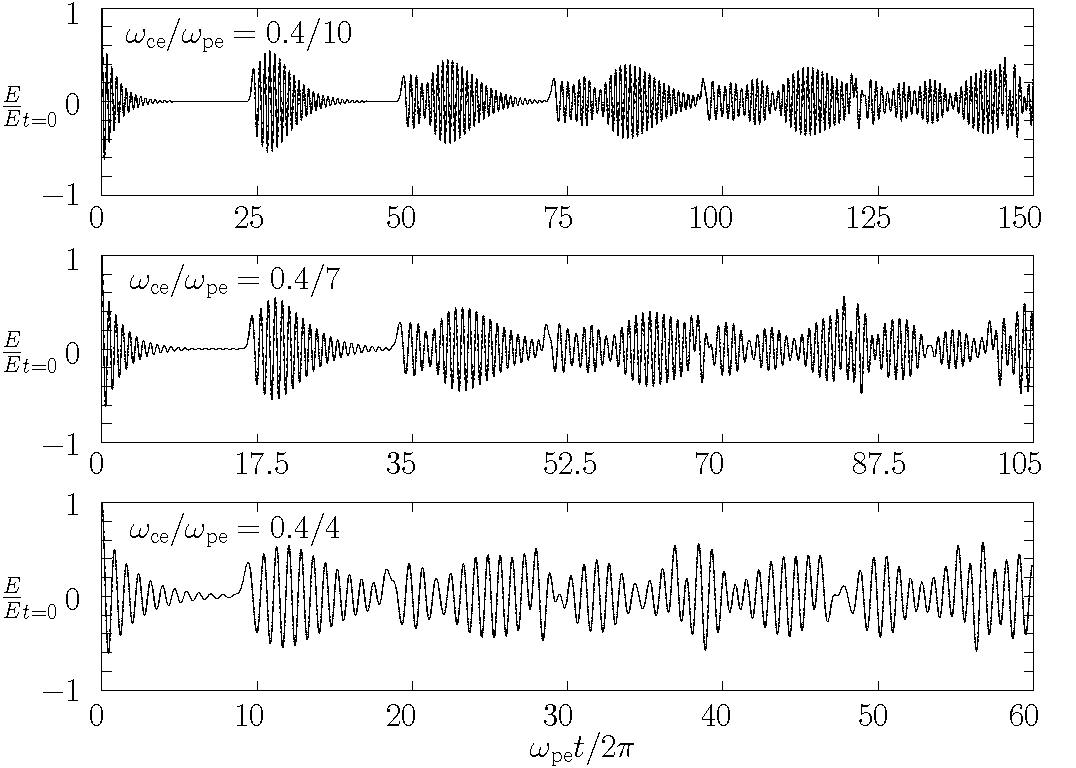}
    \label{fig:BL_simulation}
  }
  \subfigure[Analytic solutions]{
      \includegraphics[width=8cm,height=9.1cm]{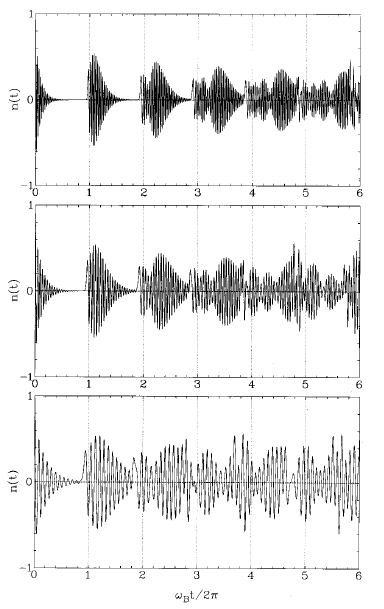}
    \label{fig:BL_analytic}
  }
  \caption{Pseudo-periodically damping and recurrence of
  electrostatic waves propagating perpendicularly to the magnetic field in a weakly magnetized plasma; the so-called Bernstein-Landau paradox. (a): Simulation results for different magnetic fields. (b): Analytic solutions by \citet{Sukhorukov_97}.}
  \label{fig:BL_paradox}
\end{figure}
According to linear theory, wave solutions of the unmagnetized Vlasov
equation exhibit collision-less damping, while in magnetized plasma
waves propagating perpendicularly to the magnetic field $\Bext$ are
exactly undamped, no matter how weak the magnetic field is.
This is the so-called Bernstein-Landau paradox; it seems
that the theory for magnetized plasma does not converge smoothly
to the theory for unmagnetized plasma when the magnetic field strength decreases.

This problem was investigated theoretically by \citet{Sukhorukov_97}
who derived analytical solutions of the problem.
In Fig. \ref{fig:BL_paradox}, we have compared a numerical solution of
the Vlasov equation with the analytic solution of \citet{Sukhorukov_97} for
a wave with the wavenumber $k_x=0.4\,\rD^{-1}$ and with different values on the magnetic field,
such that $\wce/\wpe=0.4/10$, $0.4/7$ and $0.4/4$, where $\wce=eB_{\mathrm{ext}}/\me$.
The numerical result obtained from the Vlasov simulation
can be seen in Fig. \ref{fig:BL_simulation} show excellent agreement with
the analytic results of \citet{Sukhorukov_97} in Fig. \ref{fig:BL_analytic}. For the numerical results,
where the electric field (normalized by its initial value) at $x=0$ is plotted as a
function of time. The upper panels show the case
with the weakest magnetic field $B_{\mathrm{ext}}$ and the lower panel
the strongest magnetic field. The horizontal time axis
is scaled so that the tick marks are placed at each gyro period;
in the upper panel the gyro period is $t_{\mathrm{gyro}}=(25\X 2\pi)\wpe^{-1}$,
in the middle panel $t_{\mathrm{gyro}}=(17.5\X 2\pi)\wpe^{-1}$ and in
the lower panel $t_{\mathrm{gyro}}=(10\X 2\pi)\wpe^{-1}$.
As can be seen in Fig.~\ref{fig:BL_paradox}, the waves exhibit
damping within the first gyro period, followed by a recurrence
of the wave, which is again damped, etc, in an increasingly
irregular pattern.
In the upper panels of Fig. \ref{fig:BL_paradox}, with the weakest
magnetic field, the field has the time to perform the largest number
of oscillations within each gyro period and the electric field
is also strongest damped before recurring at each gyro period.
The ``paradox'' is resolved in the following manner: The
waves exhibit damping within the first gyro period, given
by $t_{\mathrm{gyro}}=2\pi/\wce$, where $\wce=e B_{\mathrm{ext}}/\me$, and then the wave
recurs the first time, followed by a new damping, et.c.
In the limit of a vanishing magnetic field, $B_{\mathrm{ext}}\rightarrow 0$, it
follows that $\wce \rightarrow 0$ and the gyro-period goes to infinity,
$t_{\mathrm{gyro}}\rightarrow \infty$, and hence the wave will be damped, since
it will take an infinite time for the first recurrence to occur.

In the present numerical simulation, the initial
condition was set to
\begin{align}
  \f(x_1,x_2,\eta_1,\eta_2,0)&=n(x_1)\f_0(\eta_1,\eta_2)
\end{align}
with the relative perturbation
\begin{align}
  n(x_1)=[1+A\sin(k_{x1}x_1)]
\end{align}
and with the $x$-component of the wave vector set to
$k_{x1}=0.4\,r_{\mathrm{D}}^{-1}$. The amplitude of the wave was
set to $A=0.0001$, making the problem close to linear.
The Fourier transformed velocity distribution was set to be a Maxwellian
function,
\begin{align}
  \f(\eta_1,\eta_2)&=
  \frac{n_0}{(2\pi)^2}\exp\bigg[-\frac{1}{2}(\eta_1^2+\eta_2^2)\vthe^2\bigg]
\end{align}
The domain
was set to $0\leq x_1/\rD \leq 2\pi/0.4$,
$0\leq \eta_1\vthe\leq 30$ and $-30\leq \eta_2\vthe \leq 30$,
and the number of intervals
$N_{x1}=4$, $N_{\eta_1}=N_{\eta_2}=150$.
\subsection{Electromagnetic waves perpendicular to the magnetic field lines}
\label{sec:vlasovEM}

We restrict the Vlasov-Maxwell system to two spatial and two velocity dimensions,
where the particles move in the $(x_1,x_2)$ plane, the electric field $\E(x_1,x_2,t)$ is
directed in the $(x_1,x_2)$ plane and the magnetic fields $\Bext$ and $\B(x_1,x_2,t)$
are directed in the $x_3$ direction, perpendicular to the $(x_1,x_2)$ plane.
We use the same normalization of variables as \citep{Eliasson_03}, i.e.,
$t$ is normalized by $\wpe^{-1}$, $v_1$ and $v_2$ by $\vthe$,
$x_1$ and $x_2$ by $\rD$,
$\eta_1$ and $\eta_2$ by $\vthe^{-1}$,
$\Fe$ and $\Fi$ by $n_0$,
$f_{\mathrm{e}}$ and $f_{\mathrm{i}}$ by $n_0 \vthe^{-2}$,
$E_1$ and $E_2$  by $\vthe^2 \rD^{-1}(\me/e)$,
$\Phi$ by $\vthe^2 (\me/e)$,
$B_{\mathrm{ext}}$ and $B$ by $\wpe (\me/e)$,
$A_1$ and $A_2$ by $\vthe {\me}/{e}$,
$\Gamma_1$ and $\Gamma_2$ by $\wpe \vthe {\me}/{e}$,
$\rho$ by $n_0 e$,
$j_1$ and $j_2$ by $\vthe n_0 e $,
in which the Fourier transformed Vlasov equation
for the ions and electrons attain the dimensionless form
\begin{align}
  \label{Vlasov_ion2}
  &\!\!\!\!\!\!\!\!\!\!\frac{\d \Fi}{\d t}-\i\frac{\d^2\Fi}{\d x_1 \d\eta_1}
  -\i\frac{\d^2\Fi}{\d x_2 \d\eta_2}-
  \frac{\me}{\mi}\bigg[\i (E_1\eta_1+E_2\eta_2) \Fi+
  (B_{\mathrm{ext}}+B)
  \bigg(\eta_1\frac{\d \Fi}{\d \eta_2}-\eta_2\frac{\d \Fi}{\d \eta_1}\bigg)\bigg]=0
  \\
  &\!\!\!\!\!\!\!\!\!\!\frac{\d \Fe}{\d t}-\i\frac{\d^2\Fe}{\d x_1 \d\eta_1}
  -\i\frac{\d^2\Fe}{\d x_2 \d\eta_2}+
  \bigg[\i (E_1\eta_1+E_2\eta_2) \Fe+
  (B_{\mathrm{ext}}+B)
  \bigg(\eta_1\frac{\d \Fe}{\d \eta_2}-\eta_2\frac{\d \Fe}{\d \eta_1}\bigg)\bigg]=0
\end{align}
respectively. The
electromagnetic wave equations (\ref{Lorentz1})--(\ref{Lorentz2}) take the form
\begin{align}
  \Dtt{\Phi}-\bigg(\frac{c}{\vthe}\bigg)^2\bigg(\frac{\d^2 \Phi}{\d x_1^2}+\frac{\d^2 \Phi}{\d x_2^2}\bigg)&=
    \bigg(\frac{c}{\vthe}\bigg)^2\rho\\
  \Dtt{\A}-\bigg(\frac{c}{\vthe}\bigg)^2\bigg(\frac{\d^2 \A}{\d x_1^2}+\frac{\d^2 \A}{\d x_2^2}\bigg)&=\vec{j}
\end{align}
and the first-order system (\ref{Lorentz3})--(\ref{Lorentz3_2}) takes the form
\begin{align}
  \label{Lorentz4}
  \Dt{A_1}&=\Gamma_1 & \Dt{\Gamma_1}&=\bigg(\frac{c}{\vthe}\bigg)^2
  \bigg(\frac{\d^2 A_1}{\d x_1^2}+\frac{\d^2 A_1}{\d x_2^2}\bigg)+j_1\\
  \label{Lorentz4_1}
  \Dt{A_2}&=\Gamma_2 & \Dt{\Gamma_2}&=\bigg(\frac{c}{\vthe}\bigg)^2
  \bigg(\frac{\d^2 A_2}{\d x_1^2}+\frac{\d^2 A_2}{\d x_2^2}\bigg)+j_2 \\
  \label{Lorentz4_2}
  -\bigg(\frac{\d^2\Phi}{\d x_1^2}+\frac{\d^2\Phi}{\d x_2^2}\bigg)
  &=\rho+\frac{\d\Gamma_1}{\d x_1}+\frac{\d\Gamma_2}{\d x_2}.
\end{align}
The electric and magnetic fields are calculated as
\begin{align}
  E_1&=-\frac{\d \Phi}{\d x_1}-\Gamma_1 & E_2&=-\frac{\d \Phi}{\d x_2}-\Gamma_2
  \intertext{and}
  \label{rot_A}
  B&=B_0 + \frac{\d A_2}{\d x_1} - \frac{\d A_1}{\d x_2},
\end{align}
respectively. The charge and current densities are calculated
from the ion and electron distribution functions as
\begin{align}
  \label{rho}
  \rho&= (2\pi)^2(\Fi-\Fe)_{\eta_1=\eta_2=0}=(2\pi)^2(\Re(\Fi)-\Re(\Fe))_{\eta_1=\eta_2=0}\\
  \label{current1}
  j_1 &=-\i (2\pi)^2 \bigg(\frac{\d\Fi}{\d \eta_1}-\frac{\d\Fe}{\d \eta_1}\bigg)_{\eta_1=\eta_2=0}
    =(2\pi)^2 \bigg[\frac{\d\Im(\Fi)}{\d \eta_1}-\frac{\d\Im(\Fe)}{\d \eta_1}\bigg]_{\eta_1=\eta_2=0} \\
  \label{current2}
  j_2 &=-\i (2\pi)^2 \bigg(\frac{\d\Fi}{\d \eta_2}-\frac{\d\Fe}{\d \eta_2}\bigg)_{\eta_1=\eta_2=0}
    =(2\pi)^2 \bigg[\frac{\d\Im(\Fi)}{\d \eta_2}-\frac{\d\Im(\Fe)}{\d \eta_2}\bigg]_{\eta_1=\eta_2=0}
\end{align}
respectively, and where $\Im$ and $\Re$ denote the imaginary and real parts
parts, respectively; the last equalities in Eqs.~(\ref{rho})--(\ref{current2})
follow from the symmetry condition where the real parts of the distribution functions are
even and the imaginary parts are odd with respect to $\Eta$.

We here consider electromagnetic waves that are propagating perpendicularly
to the magnetic field lines. The waves have the electric field component in the plane
perpendicular to the magnetic field, while the wave magnetic field is parallel to the
external magnetic field. This configuration support the high-frequency X mode waves
(X stands for ``extraordinary'') that can also propagate in vacuum as light, and the Z mode waves
that connect to the upper hybrid resonance at short wavelengths shorter than the electron
inertia length $\lambda_e=c/\omega_{pe}$. In the low-frequency regime
we have magnetosonic waves that connect to the lower hybrid resonance at wavelengths shorter
than the electron inertial length. The O mode wave (O stands for "ordinary") has a component of
the electric field along the background magnetic field direction, and cannot be simulated
in the two-dimensional model discussed here, since it needs electron dynamics also along
the magnetic field lines. The theoretical predictions for high- and low-frequency waves
are compared with a Vlasov simulation in one spatial dimension (along the $x_1$ direction)
and two velocity dimensions, of waves propagating perpendicularly to the external magnetic field.

\begin{figure}[htb]
  \centering
  \subfigure[Dispersion diagram]{
      \includegraphics[height=6cm]{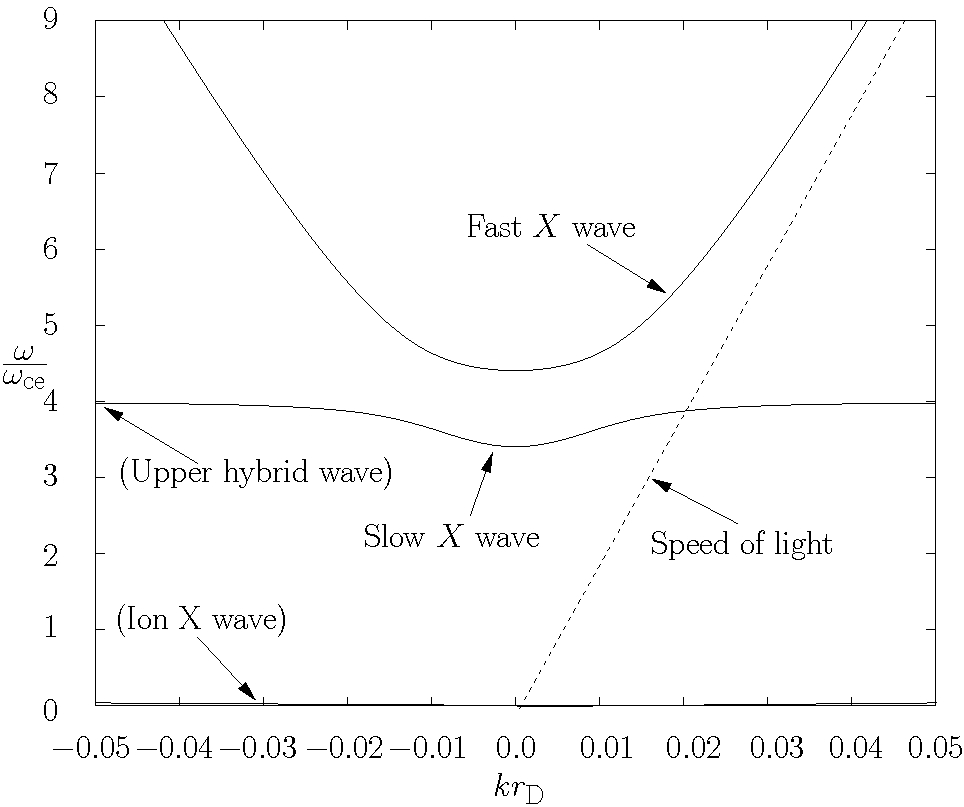}
    \label{fig:Xdisp}
  }
  \subfigure[Power spectrum]{
      \includegraphics[height=6cm]{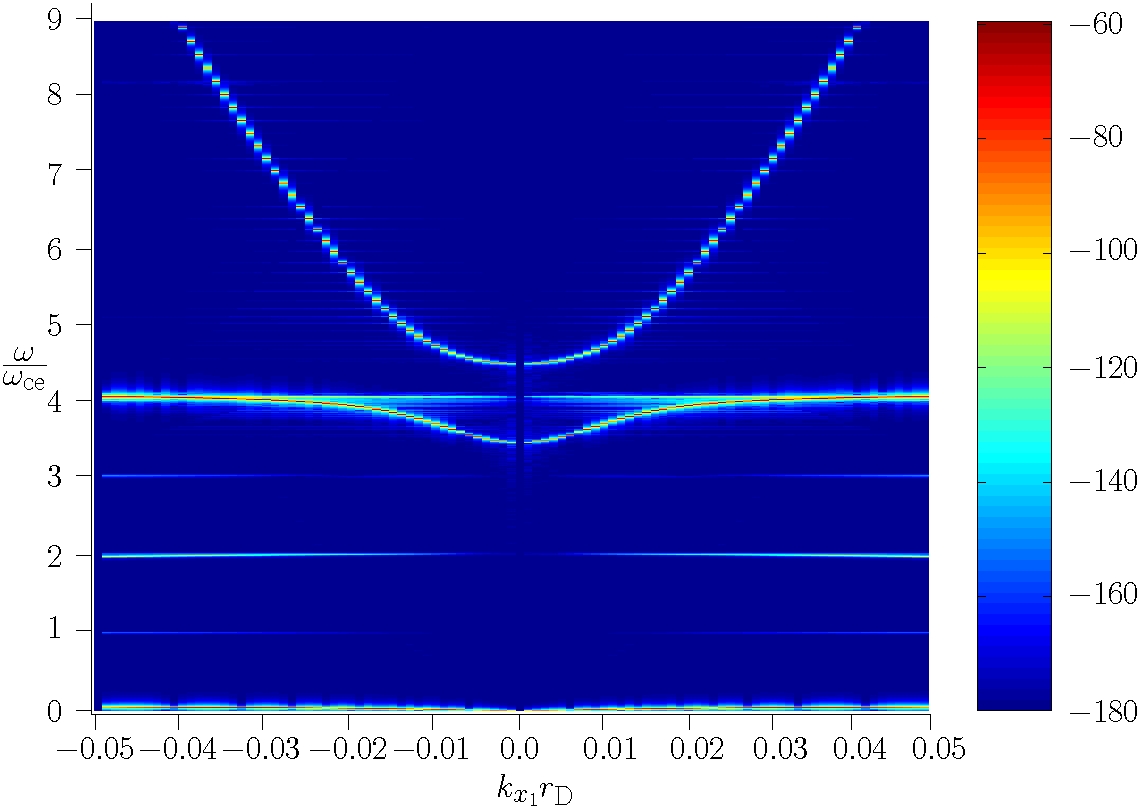}
    \label{fig:Bernstein_Z}
  }
  \caption{Dispersion diagram for the high frequency electromagnetic
  extraordinary mode, obtained from cold plasma fluid theory,
  and power spectrum (decibel) of the transverse part $E_2$ of
  the electric field obtained from Vlasov simulation;
  $\wuh=4\wce$. After \citet{Eliasson_03}.}
  \label{fig:Bernstein_Xmode}
\end{figure}

In Fig.~\ref{fig:Xdisp} we show a comparison between dispersion curves
obtained from the dispersion relation of cold fluid theory \citep{Goldston_97}
\begin{align}
  \frac{c^2 k^2}{\w^2} &= 1 - \frac{\wpe^2(\w^2-\wpe^2)}{\w^2(\w^2-\wuh^2)}
  \label{X_mode}
\end{align}
and a Vlasov simulation, for the case $\wuh/\wce=4$ from which it follows
that ${\wpe/\wce=\sqrt{15}}$. In the simulation, we used the ratio $c/\vthe=50$ between
the speed of light and the electron thermal speed.
For large $k$, the we see that the fast X mode approaches the speed of light, while the
Z mode wave approaches the upper hybrid oscillation, with
frequency ${\w^2=\wuh^2=\wpe^2+\wce^2}$ for large $k$. In the
short wave length limit (very large $k$), thermal and kinetic effects
are important, and the Z mode wave goes smoothly over to
one of the upper hybrid and one of the electron Bernstein waves.
The energy for the high frequency waves in Fig.~\ref{fig:Bernstein_Z} is concentrated at
the linear dispersion curves for the fast and slow $X$ modes, displayed in Fig.~\ref{fig:Xdisp},
in good agreement with theory.
In Fig.~\ref{fig:Bernstein_Z}, one can also see some weakly excited waves at the
gyro harmonics $\w/\wce=1,\,2,\,3,\,4$, which are waves not covered by the dispersion curves
in Fig.~\ref{fig:Xdisp}, obtained from the cold plasma fluid model.
The weakly excited $\w/\wce=1$ mode is an electromagnetic effect \citep{Puri_73} which can
not be seen in the electrostatic case shown in Fig.~\ref{fig:Bernstein}
on page~\pageref{fig:Bernstein}.

\begin{figure}[htb]
  \centering
  \subfigure[Dispersion diagram]{
      \includegraphics[height=6cm]{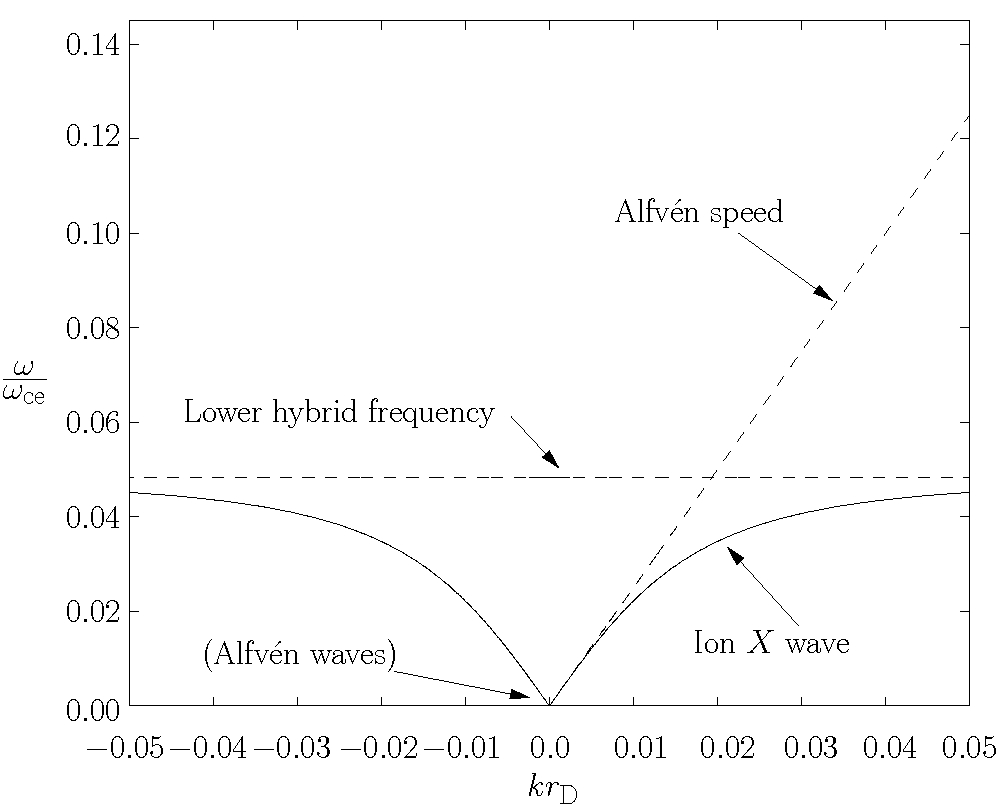}
    \label{fig:ion_Xdisp}
  }
  \subfigure[Power spectrum]{
      \includegraphics[height=6cm]{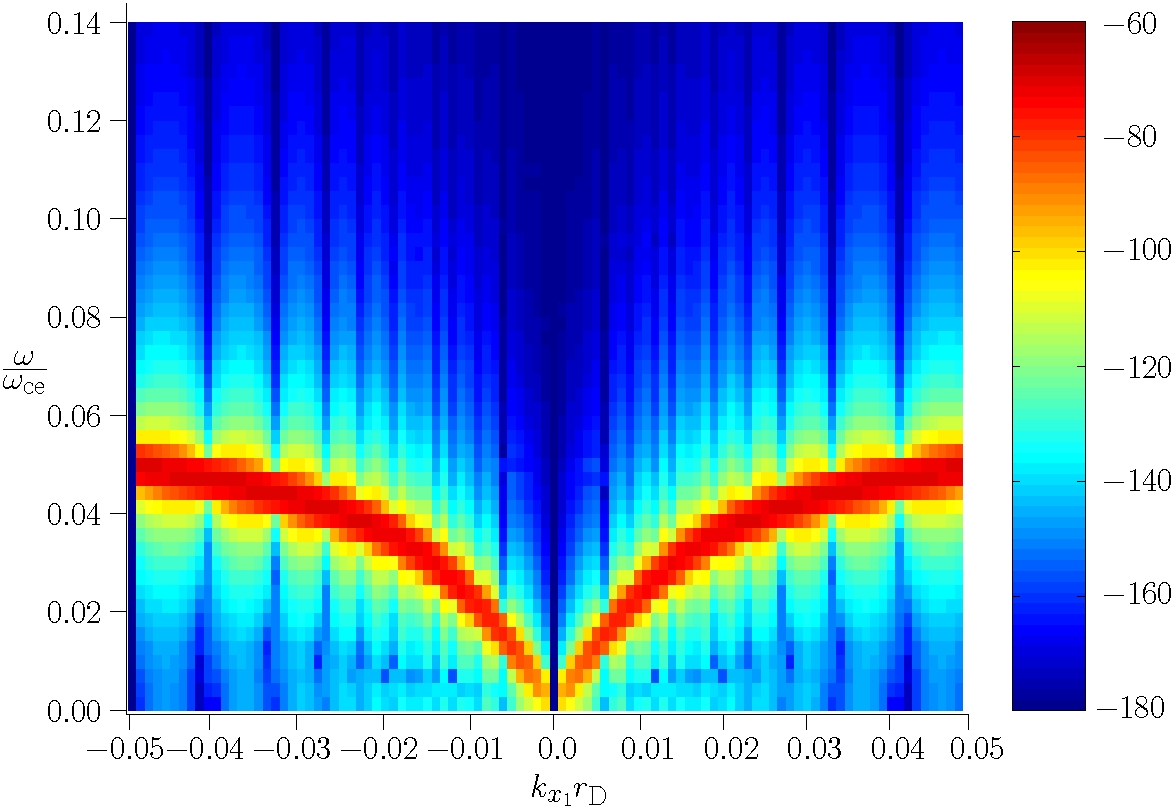}
    \label{fig:Bernstein_LH}
  }
  \caption{Dispersion diagram for the ion
  extraordinary wave, obtained from cold plasma fluid theory,
  and power spectrum (decibel) of the transverse part $E_2$ of
  the electric field obtained from the Vlasov simulation. The
  waves go over to Alfv{\'e}n waves at small $k_{x1}$
  and to lower hybrid hybrid oscillations at large $k_{x1}$;
  $\wuh=4\wce$, $\mi/\me=400$. After \citet{Eliasson_03}.}
  \label{fig:ion_Xmode}
\end{figure}

In Fig. \ref{fig:ion_Xmode}, we compare theory with a closeup of the
low-frequency part of the energy spectrum obtained in the Vlasov simulation.
For the low frequency electromagnetic waves perpendicular to the
magnetic field, An approximate dispersion relation obtained from
a cold fluid description of ions and electrons is given by \citet{Goldston_97} as
\begin{align}
  \label{ion_X}
  \frac{c^2 k^2}{\w^2}\bigg(
  \frac{\wci^2-\w^2+\wpi^2}{\wci^2-\w^2}+\frac{\wpe^2}{\wce^2}
  \bigg) &= \bigg(
  \frac{\wci^2-\w^2+\wpi^2}{\wci^2-\w^2}+\frac{\wpe^2}{\wce^2}
  \bigg)^2-\bigg(\frac{\wpi^2 \w}{\wci(\wci^2-\w^2)}\bigg)^2
\end{align}
with $c/\vthe=50$, $\wuh/\wce=4$ and $\wpe/\wce=\sqrt{15}$.
For numerical efficiency we use the mass ratio $\mi/\me=400$ between the ion
and electron masses, which gives the ratios $\wpi/\wce=\sqrt{15}/20$ and $\wci/\wce=1/400$.
Eq.~(\ref{ion_X}) is solved for $\w/\wce$ and displayed in Fig.~\ref{fig:ion_Xdisp}.
For large $k_{x1}$, the dispersion curve approaches the
lower hybrid frequency $\wlh$, approximately
given by
\begin{align}
  \wlh^{-2}=\wpi^{-2}+(\wci\wce)^{-1}
\end{align}
and which is indicated in Fig.~\ref{fig:ion_Xdisp}.
For very small $k_{x1}\lambda_e < 1$, we see in Fig.~\ref{fig:ion_Xdisp} that
the dispersion curve approaches the one for Alfv{\'e}n waves,
governed by the dispersion relation
\begin{align}
  \w^2=v_A^2k^2,
\end{align}
where $v_A=c\omega_{ci}/\omega_{pi}$ is the Alfv\'en speed.
The energy spectrum for the low frequency waves in Fig.~\ref{fig:Bernstein_LH}
shows good similarity with the dispersion curve for the low frequency wave in
Fig.~\ref{fig:ion_Xdisp}. The width of the energy bands in the power spectrum
is the frequency resolution obtained in the simulation; a longer simulation would
resolve the waves more. The frequencies of the waves in Fig.~\ref{fig:Bernstein_LH}
for large $k$ is slightly higher than the corresponding frequencies in
the dispersion diagram in Fig.~\ref{fig:ion_Xdisp}, which probably is a thermal
effect, not included in the cold plasma fluid model.

In the Vlasov simulation presented here, the simulation domain was
restricted to one spatial dimension with ${0\le x_1 \le 2000\pi}$
and two velocity dimensions, plus time.
The Fourier transformed velocity domain was ${0\le\eta_1 \le 10}$ and ${-10\le\eta_2 \le 10}$ for the electrons, and ${0\le\eta_1\le 200}$ and ${-200\le\eta_2\le 200}$ for the ions, with the number of intervals
${N_{x1}=100}$ in space and ${N_{\eta_1}=30}$ and ${2N_{\eta_2}=60}$ in the Fourier transformed velocity space.
The initial conditions for the electrons and ions were set to $\f_{\mathrm{e}}(x_1,x_2,\eta_1,\eta_2,0)=n(x)\f_{\mathrm{e},0}(\eta_1,\eta_2)$
and
$\f_{\mathrm{i}}(x_1,x_2,\eta_1,\eta_2,0)=n(x)\f_{\mathrm{i},0}(\eta_1,\eta_2)$,
respectively, with the
density perturbation
\begin{align}
  n(x)=1+10^{-7}\sum_{i_1=1}^{49}i_1\sin(0.05 i_1 x_1)
\end{align}
so that all wavemodes were excited with low amplitudes, making the problem close to linear.
The electron and ion distribution functions were set to
\begin{align}
  \f_{\mathrm{e},0}(\eta_1,\eta_2)&=\frac{1}{(2\pi)^2}
    \exp\bigg[-\frac{1}{2}(\eta_1^2+\eta_2^2)\bigg]
  \intertext{and}
  \label{ion_Maxwell}
  \f_{\mathrm{i},0}(\eta_1,\eta_2)&=\frac{1}{(2\pi)^2}
    \exp\bigg[-\frac{1}{2}(\eta_1^2+\eta_2^2)\bigg(\frac{\vthe}{\vthi}\bigg)^{-2}\bigg]
\end{align}
respectively.
The number of time steps taken was $N_t=97\,280$; the end time was $t_{\mathrm{end}}=8340$.
No numerical dissipation was used. The ion-electron mass ratio was $\mi/\me=400$ and the
ion and electron temperatures were set equal, $\Ti=\Te$, giving the
factor $(\vthe/\vthi)^{-2}=1/400$ in Eq.~(\ref{ion_Maxwell}).

\section{The three-dimensional Vlasov equation\label{sec:vlasov3}}
We here discuss the extension of the Fourier transform method to
the Fourier transformed Vlasov equation in full three spatial and three
velocity dimensions, including external and self-consistent magnetic
fields. Again, in the design of well-posed absorbing boundary conditions
in the Fourier transformed velocity space, special care has to be
taken with the magnetic field, which enters into the formulation of the
boundary condition.

The Fourier transformed Vlasov equation (\ref{3D_eta1}) can be cast into the
normalized form
\begin{equation}
  \frac{\d \f_{\a}}{\d t}-\i\D_{\x}\cdot\D_{\Eta}\f_{\a}
 -\frac{Q_{\a}}{M_{\a}}\left\{\i\E\cdot\Eta\f_{\a}+
 \D_{\Eta}\cdot\left[(\B\times\Eta)\f_{\a}\right]\right\} =0,
  \label{vlasov_normalized}
\end{equation}
where $t$ is normalized by $\w_{\mathrm{pe}}^{-1}$,
$\v$ by $v_{\rm Te}$,
$\x$ by $r_{\rm De}$,
$\Eta$ by $v_{\rm Te}^{-1}$,
$\f_\a$ by $n_0$,
$\E$ by $v_{\rm Te}^2 r_{\rm De}^{-1}(m_{\rm e}/e)$,
and $\B$ by $\w_{\rm pe} (m_{\rm e}/e)$.
Here we defined $Q_\alpha=q_\alpha/e$ and $M_\alpha=m_\alpha/m_{\rm e}$, where we assume electrons and singly charged ions, so that $Q_{\rm i}=1$, $Q_{\rm e}=-1$, $M_{\rm i}=m_{\rm i}/m_{\rm e}$, and
$M_{\rm e}=1$. The equations for the potentials,
(\ref{Lorentz3}) and (\ref{Lorentz3_2}), take normalized form
\begin{align}
\frac{\partial {\bf A}}{\partial t}&={\bf \Gamma},
&\frac{\partial{\bf \Gamma}}{\partial t}&=\frac{c^2}{v_{\rm
Te}^2}\nabla_{\x}^2{\bf A}+{\bf j}, \label{normalized_potentials}
\\
-\nabla_{\x}^2\Phi&=\rho+\D_{\x}\cdot {\bf \Gamma}, \label{PPPhi}
\intertext{and the electric and magnetic fields (\ref{E_grad}) and (\ref{B_rot}) are obtained as}
&\E=-\D_{\x} \Phi-{\bf \Gamma}, \label{EEE}
\\
&{\bf B}={\bf B}_0+\D_{\x}\times{\bf A}, \label{BBB}
\end{align}
where $\Phi$ is normalized by $v_{\rm Te}^2 (m_{\rm e}/e)$,
${\bf A}$ by $v_{\rm Te}(m_{\rm e}/e)$,
${\bf \Gamma}$ by $\omega_{\rm pe}v_{\rm Te}(m_{\rm e}/e)$.
$\rho$ by $en_0$, and ${\bf j}$ by $e n_0 v_{\rm Te}$.
Using Eqs. (\ref{3D_density}) and (\ref{3D_velocity}) in Eqs.
(\ref{3D_rho}) and (\ref{3D_j}), the normalized charge and current densities
are obtained as
\begin{align}
  \label{charge}
  \rho    &=(2\pi)^3[\Re(\f_{\rm i})_{\Eta=\mathbf{0}}-\Re(\f_{\rm e})_{\Eta=\mathbf{0}} ],
  \intertext{and}
  {\bf j} &=(2\pi)^3 [\D_{\Eta}\Im(\f_{\rm i})_{\Eta=\mathbf{0}}-\D_{\Eta}\Im(\f_{\rm e})_{\Eta=\mathbf{0}}] ,
  \label{current}
\end{align}
respectively.
%
\subsection{Restriction to a bounded domain}
 In order to adapt the Fourier
transformed Vlasov Maxwell system for numerical simulations, it must
be restricted to a bounded domain. The computational domain is
restricted to ${0\le x_1 < L_1}$, ${0\le x_2 < L_2}$, ${0\le x_3 <
L_3}$, $0\le\eta_1\le\eta_{{\rm max},{1\a}}$,
$-\eta_{{\rm max},2\a}\le\eta_2\le\eta_{{\rm max},2\a}$, and
$-\eta_{{\rm max}, 3\a}\le\eta_3\le\eta_{{\rm max},3\a}$. Here $\a$ (equal to $e$
for electrons and $i$ for ions) is introduced so that different
domain sizes can be used in $\Eta$ space for the ion and electron
distribution functions. For negative $\eta_1$, the symmetry
$\f(x_1,x_2,x_3,\eta_1,\eta_2,\eta_3,t)=\f^*(x_1,x_2,x_3,-\eta_1,-\eta_2,-\eta_3,t)$
is used to obtain function values; it is therefore
not necessary to numerically represent the solution for negative
$\eta_1$ if the solution is represented for negative $\eta_2$ and
$\eta_3$.

\subsection{Outflow boundary conditions in Fourier transformed velocity space}
\label{sec:wellposed}
In this section, we will derive well-posed boundary conditions for
the Vlasov equation in $\Eta$ space. Writing out the terms of the
Fourier transformed Vlasov equation (\ref{vlasov_normalized}), we
have
\begin{equation}
  \begin{split}
  &\frac{\d \f_\a}{\d t}
  -\i\frac{\d^2\f_\a}{\d x_1 \d\eta_1}
  -\i\frac{\d^2\f_\a}{\d x_2 \d\eta_2}
  -\i\frac{\d^2\f_\a}{\d x_3 \d\eta_3}
   -\frac{Q_\a}{M_\a}\bigg[\i(E_1\eta_1+E_2\eta_2+E_3\eta_3)\f_\a
   \\
  &+(B_{2}\eta_3-B_{3}\eta_2)\frac{\d \f_\a}{\d \eta_1}
   +(B_{3}\eta_1-B_{1}\eta_3)\frac{\d \f_\a}{\d \eta_2}
   +(B_{1}\eta_2-B_{2}\eta_1)\frac{\d \f_\a}{\d \eta_3}\bigg]=0. \\
  \end{split}
  \label{vlasov_comp}
\end{equation}
In position space, periodic boundary conditions
\begin{align}
\f_\a(x_1+L_1,x_2,x_3,\eta_1,\eta_2,\eta_3,t)&=\f_\a(x_1,x_2,x_3,\eta_1,\eta_2,\eta_3,t),\\
\f_\a(x_1,x_2+L_2,x_3,\eta_1,\eta_2,\eta_3,t)&=\f_\a(x_1,x_2,x_3,\eta_1,\eta_2,\eta_3,t),\\
\f_\a(x_1,x_2,x_3+L_3,\eta_1,\eta_2,\eta_3,t)&=\f_\a(x_1,x_2,x_3,\eta_1,\eta_2,\eta_3,t).
\end{align}
are used for both the distribution functions and the electromagnetic
fields. The artificial boundaries at $\eta_1=\eta_{{\rm max}, 1\a}$,
$\eta_2=\pm\eta_{{\rm max}, 2\a}$ and $\eta_3=\pm\eta_{{\rm max},3\a}$ must be
treated with care so that they do not give rise to reflections of
waves or to instabilities. The strategy is to let outgoing waves
pass over the boundaries, and to set incoming waves to zero. The
problem of separating outgoing waves from incoming waves is solved
by employing the spatial Fourier series expansions (transforms). In
order to explore the idea, one can study the reduced initial value
problem with a constant external magnetic field ${\B=\B_0}$ and a
zero electric field $\E={\bf 0}$,
\begin{align}
  \label{vlasov5_3D}
  \begin{split}
  &\frac{\d \f_\a}{\d t}
  -\i\frac{\d^2\f_\a}{\d x_1 \d\eta_1}
  -\i\frac{\d^2\f_\a}{\d x_2 \d\eta_2}
  -\i\frac{\d^2\f_\a}{\d x_3 \d\eta_3}
   -\frac{Q_\a}{M_\a}\bigg[
    (B_{2,0}\eta_3-B_{3,0}\eta_2)\frac{\d \f_\a}{\d \eta_1}
    \\
   &+(B_{3,0}\eta_1-B_{1,0}\eta_3)\frac{\d \f_\a}{\d \eta_2}
   +(B_{1,0}\eta_2-B_{2,0}\eta_1)\frac{\d \f_\a}{\d \eta_3}\bigg]=0,\\
  \end{split}
\\
  &\f_\a(x_1,x_2,x_3,\eta_1,\eta_2,\eta_3,t)_{t=0}=\f_{\a 0}(x_1,x_2,x_3,\eta_1,\eta_2,\eta_3).
  \end{align}
By introducing the spatial Fourier series pairs in ($x_1$, $x_2$,
$x_3$) space,
\begin{align}
  \label{Fourier1}
  \widetilde{\phi}_{1,i_1}&={\rm F}_1 \phi_1 = \frac{1}{L_1}\int_{0}^{L_1} \phi_1(x_1)
  e^{-\i k_{x_1} x_1} \diff x_1 & &
  \\
  \label{invFourier1}
  \phi_1&={\rm F}_1^{-1} \widetilde{\phi}_1 =
  \sum_{i_1=-\infty}^{\infty} \widetilde{\phi}_{1,i_1} e^{\i k_{x_1} x_1}\\
  k_{x_1}&=\frac{2\pi i_1}{L_1},\quad i_1=0,\,\pm 1,\,\pm
  2,\,\ldots,
  \\
  \widetilde{\phi}_{2,i_2}&={\rm F}_2 \phi_2 = \frac{1}{L_2}\int_{0}^{L_2} \phi_2(x_2)
  e^{-\i k_{x_2} x_2} \diff x_2
  \\
  \phi_2&={\rm F}_2^{-1} \widetilde{\phi}_{2} =
  \sum_{i_2=-\infty}^{\infty} \widetilde{\phi}_{2,i_2} e^{\i k_{x_2} x_2} \\
  \label{Fourier2_2}
  k_{x_2}&=\frac{2\pi i_2}{L_2},\quad i_2=0,\,\pm 1,\,\pm 2,\,\ldots
  \intertext{and}
  \widetilde{\phi}_{3,i_3}&={\rm F}_3 \phi_3 = \frac{1}{L_3}\int_{0}^{L_3} \phi_3(x_3)
  e^{-\i k_{x_3} x_3} \diff x_3,
  \\
  \phi_3&={\rm F}_3^{-1} \widetilde{\phi}_{3} =
  \sum_{i_3=-\infty}^{\infty} \widetilde{\phi}_{3,i_3} e^{\i k_{x_3} x_3}, \\
  \label{Fourier2_3}
  k_{x_3}&=\frac{2\pi i_3}{L_3},\quad i_3=0,\,\pm 1,\,\pm 2,\,\ldots
\end{align}
respectively, and Fourier-transforming Eq.~(\ref{vlasov5_3D}) in all
spatial directions, one obtains a new differential equation for the
unknown function
$\widetilde{f}(k_{x_1},k_{x_2},k_{x_3},\eta_1,\eta_2,\eta_3,t)$,
\begin{align}
  \label{vlasov6_3D}
  \begin{split}
  \frac{\d \widetilde{f_\a}}{\d t}&+
   \bigg[k_{x_1}-\frac{Q_\a}{M_\a}(B_{2,0}\eta_3-B_{3,0}\eta_2)\bigg]\frac{\d\widetilde{f}_\a}{\d\eta_1}
  \\
  &+\bigg[k_{x_2}-\frac{Q_\a}{M_\a}(B_{3,0}\eta_1-B_{1,0}\eta_3)\bigg]\frac{\d\widetilde{f}_\a}{\d\eta_2}
  \\
  &+\bigg[k_{x_3}-\frac{Q_\a}{M_\a}(B_{1,0}\eta_2-B_{2,0}\eta_1)\bigg]\frac{\d\widetilde{f}_\a}{\d\eta_3}
  =0,
  \end{split}
  \intertext{with the initial condition}
  &\widetilde{f}_\a(k_{x_1},k_{x_2},k_{x_3},\eta_1,\eta_2,\eta_3,t)_{t=0}
  =\widetilde{f}_{\a
  0}(k_{x_1},k_{x_2},k_{x_3},\eta_1,\eta_2,\eta_3).
\end{align}
Equation (\ref{vlasov6_3D}) is a hyperbolic equation for which the
initial values are transported along the characteristic curves,
given by
\begin{align}
  \label{char1_3D}
  \frac{\diff \eta_1(t)}{\diff t}&=k_{x_1}-\frac{Q_\a}{M_\a}(B_{2,0}\eta_3-B_{3,0}\eta_2), \\
  \frac{\diff \eta_2(t)}{\diff t}&=k_{x_2}-\frac{Q_\a}{M_\a}(B_{3,0}\eta_1-B_{1,0}\eta_3), \\
  \frac{\diff \eta_3(t)}{\diff t}&=k_{x_3}-\frac{Q_\a}{M_\a}(B_{1,0}\eta_2-B_{2,0}\eta_1).
\end{align}
Along the boundary $\eta_1=\eta_{{\rm max},\a1}$, Eq.~(\ref{char1_3D})
describes an \emph{outflow} of data when
$k_{x_1}-({Q_\a}/{M_\a})(B_{2,0}\eta_3-B_{3,0}\eta_2) > 0$ and an
\emph{inflow} of data when
$k_{x_1}-({Q_\a}/{M_\a})(B_{2,0}\eta_3-B_{3,0}\eta_2)<0$. A
well-posed boundary condition is to set the inflow to zero at the
boundary, i.e.,
\begin{align}
  (\widetilde{f}_\a)_{\eta_1=\eta_{{\rm max},\a1}}=0, \hspace{0.5cm}
  k_{x_1}-\frac{Q_\a}{M_\a}(B_{2,0}\eta_3-B_{3,0}\eta_2)<0,
\end{align}
which can be expressed with the help of the Heaviside step function
$H$ as
\begin{equation}
\label{vlasov71_3D}
  \widetilde{f}_\a=H\bigg[k_{x_1}-\frac{Q_\a}{M_\a}(B_{2,0}\eta_3-B_{3,0}\eta_2)\bigg]\widetilde{f}_\a,
    \hspace{0.5cm} \eta_1=\eta_{{\rm max}, \a1},
\end{equation}
where \begin{equation}
 \label{vlasov8_3D}
  H(\xi)=\left\{
  \begin{array}{l}
    1, \hspace{0.5cm} \xi \ge 0
  \\
    0, \hspace{0.5cm} \xi < 0,
  \end{array}
  \right.
\end{equation} The boundary condition (\ref{vlasov71_3D}) allows outgoing waves to
pass over the boundary and to be removed, while incoming waves are
set to zero; the removal of the outgoing waves corresponds to the
losing of information about the finest structures in velocity space.
Inverse Fourier transforming Eq.~(\ref{vlasov71_3D}) then gives the
boundary condition for the original problem (\ref{vlasov5_3D}) as
\begin{equation}
\label{vlasov9_3D}
  \f_\a={\rm F}_1^{-1}H\bigg[k_{x_1}-\frac{Q_\a}{M_\a}(B_{2,0}\eta_3-B_{3,0}\eta_2)\bigg]{\rm F}_1\f_\a,
    \hspace{0.5cm} \eta_1=\eta_{{\rm max},\a1}.
\end{equation}
The operator ${\rm
F}_1^{-1}H[k_{x_1}-({Q_\a}/{M_\a})(B_{2,0}\eta_3-B_{3,0}\eta_2)]{\rm
F}_1$ is a projection operator which removes incoming waves at the
boundary $\eta_1=\eta_{{\rm max}, \a1}$. Similarly, the boundary conditions
at $\eta_2=\pm\eta_{{\rm max},\a2}$ and $\eta_3=\pm\eta_{{\rm max},\a3}$ become
\begin{align}
  \f_\a&={\rm F}_2^{-1}H\bigg[k_{x_2}-\frac{Q_\a}{M_\a}(B_{3,0}\eta_1-B_{1,0}\eta_3)\bigg]{\rm F}_2\f_\a,
    \hspace{0.5cm} \eta_2=\eta_{{\rm max}, \a2},
  \\
  \f_\a&={\rm F}_2^{-1}H\bigg[-k_{x_2}+\frac{Q_\a}{M_\a}(B_{3,0}\eta_1-B_{1,0}\eta_3)\bigg]{\rm F}_2\f_\a,
    \hspace{0.5cm} \eta_2=-\eta_{{\rm max},\a2},
  \intertext{and}
  \f_\a&={\rm F}_3^{-1}H\bigg[k_{x_3}-\frac{Q_\a}{M_\a}(B_{1,0}\eta_2-B_{2,0}\eta_1)\bigg]{\rm F}_3\f_\a,
    \hspace{0.5cm} \eta_3=\eta_{{\rm max},\a3},
  \\
  \label{vlasov9_2_3D}
  \f_\a&={\rm F}_3^{-1}H\bigg[-k_{x_3}+\frac{Q_\a}{M_\a}(B_{1,0}\eta_2-B_{2,0}\eta_1)\bigg]{\rm F}_3\f_\a,
    \hspace{0.5cm} \eta_3=-\eta_{{\rm max},\a3},
\end{align}
respectively.

In order to find well-posed boundary conditions in $\Eta$ space in
the case when $\B$ varies both in space and time,
Eq.~(\ref{vlasov_comp}) is rewritten in the form
\begin{align}
  \label{vlasov4_4_3D}
  \begin{split}
  \frac{\d\f_\a}{\d t}
  +&\theta_{\a 1}
  \frac{\d}{\d \eta_1}\bigg[-\i\frac{\d}{\d x_1}-\beta_{01}\bigg](\f_\a \theta_{\a 1}^{-1})\\
  +&\theta_{\a 2}
  \frac{\d}{\d \eta_2}\bigg[-\i\frac{\d}{\d x_2}-\beta_{02}\bigg](\f_{\a}\theta_{\a 2}^{-1})\\
  +&\theta_{\a 3}
  \frac{\d}{\d \eta_3}\bigg[-\i\frac{\d}{\d x_3}-\beta_{03}\bigg](\f_{\a}\theta_{\a 3}^{-1})
  +\i(E_1\eta_1+E_2\eta_2+E_3\eta_3)\f_\a=0,
  \end{split}
\end{align}
where the phase factors $\theta_{\a 1}$, $\theta_{\a 2}$ and $\theta_{\a 3}$ are
\begin{align}
  \label{G1}
  \theta_{\a 1}&=\exp\left[\i\int_0^{x_1}(\beta_1-\beta_{01})\,\diff
  x_1\right],
  \\
  \label{G2}
  \theta_{\a 2}&=\exp\left[\i\int_0^{x_2}(\beta_2-\beta_{02})\,\diff
  x_2\right],
  \intertext{and}
  \label{G3}
  \theta_{\a 3}&=\exp\left[\i\int_0^{x_3}(\beta_3-\beta_{03})\,\diff
  x_3\right],
\end{align}
respectively, and where
\begin{align}
  \label{B01_3D}
  & \beta_1=\frac{Q_\alpha}{M_\alpha}(B_2\eta_3-B_3\eta_2),\qquad\beta_{01}=\frac{1}{L_1}\int_0^{L_1}\beta_1\,\diff x_1,
  \\
  \label{B02}
  & \beta_2=\frac{Q_\alpha}{M_\alpha}(B_3\eta_1-B_1\eta_3),\qquad\beta_{02}=\frac{1}{L_2}\int_0^{L_2}\beta_2\,\diff x_2,
  \intertext{and}
  \label{B03}
  & \beta_3=\frac{Q_\alpha}{M_\alpha}(B_1\eta_2-B_2\eta_1),\qquad\beta_{03}=\frac{1}{L_3}\int_0^{L_3}\beta_3\,\diff
  x_3.
\end{align}
The form (\ref{vlasov4_4_3D}) of the Vlasov equation makes it possible
to introduce stable numerical boundary conditions in $\Eta$ space in
a systematic manner. Furthermore, $\int_0^{x_1}(\beta_1-\beta_{01})\diff x_1$,
$\int_0^{x_2}(\beta_2-\beta_{02})\diff x_2$ and
$\int_0^{x_3}(\beta_3-\beta_{03})\diff x_3$ are continuous and
periodic in $\x$ space if ${\bf B}$ is continuous and periodic in
$\x$; this is the reason for the subtraction of the mean values
$\beta_{01}$, $\beta_{02}$ and $\beta_{03}$ in the integrals.

By studying the flow of data in the $\eta_1$, $\eta_2$ and $\eta_3$
directions for $\f_{\a} \theta_{\a 1}^{-1}$, $\f_{\a} \theta_{\a 2}^{-1}$ and $\f_{\a} \theta_{\a 3}^{-1}$,
respectively, one finds the outflow boundary conditions to be
\begin{align}
  \label{eta_max31}
  \f_{\a} &= \theta_{\a 1} {\rm F}_1^{-1}H(k_{x_1}-\beta_{01}){\rm F}_1 (\f_{\a} \theta_{\a 1}^{-1}),
  \hspace{0.5cm} \eta_1=\eta_{{\rm max},\a1},
  \\
  \label{eta_max32}
  \f_{\a} &=\theta_{\a 2} {\rm F}_2^{-1}H(k_{x_2}-\beta_{02}){\rm F}_2 (\f_{\a} \theta_{\a 2}^{-1}),
  \hspace{0.5cm} \eta_2=\eta_{{\rm max},\a2},
  \\
  \label{eta_max33}
  \f_{\a} &=\theta_{\a 2} {\rm F}_2^{-1}H(-k_{x_2}+\beta_{02}){\rm F}_2 (\f_{\a} \theta_{\a 2}^{-1}),
  \hspace{0.5cm}
  \eta_2=-\eta_{{\rm max},\a2},
  \\
  \label{eta_max34}
  \f_{\a} &=\theta_{\a 3} {\rm F}_3^{-1}H(k_{x_3}-\beta_{03}){\rm F}_3 (\f_{\a} \theta_{\a 3}^{-1}),
  \hspace{0.5cm} \eta_2=\eta_{{\rm max},\a2},
  \\
  \label{eta_max35}
  \f_{\a} &=\theta_{\a 3} {\rm F}_3^{-1}H(-k_{x_3}+\beta_{03}){\rm F}_3 (\f_{\a} \theta_{\a 3}^{-1}), \hspace{0.5cm}
  \eta_2=-\eta_{{\rm max},\a2}.
\end{align}
In the case when $\B$ is independent of $\x$ and $t$, the boundary
conditions (\ref{eta_max31})--(\ref{eta_max32}) reduce to the
conditions (\ref{vlasov9_3D})--(\ref{vlasov9_2_3D}). For the case where the domain is extended to
negative $\eta_1$, we also have the boundary condition
\begin{equation}
    \f_{\a} = \theta_{\a 1} {\rm F}_1^{-1}H(-k_{x_1}+\beta_{01}){\rm F}_1 (\f_{\a} \theta_{\a 1}^{-1}),
  \hspace{0.5cm} \eta_1=-\eta_{{\rm max},\a 1},
\end{equation}
which was used at a stage in the numerical algorithm. The well-posedness of these
boundary conditions were proven by using that a positively definite energy
integral was non-increasing \citep{Eliasson_07}.

\subsection{Electromagnetic electron waves}
\label{sec:EM}
\begin{figure}[htb]
 \centering
 \includegraphics[width=15cm]{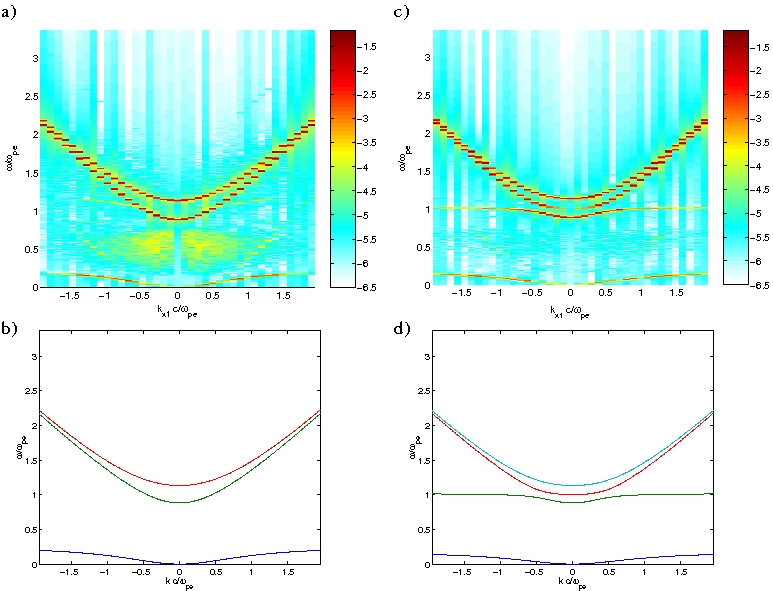}
 \caption{a) Amplitude spectrum (10-log scale) of $E_2$ for waves propagating
 parallel to the external magnetic field ($\theta=0$).
 b) Dispersion relations for the high-frequency R and L waves and
 the low-frequency whistler wave, propagating along the external magnetic field ($\theta=0$).
 c) Amplitude spectrum of $E_2$ for waves propagating with an angle $\theta=\pi/4$ to the external magnetic field,
 and d) dispersion relations for electron waves propagating with an angle of $\theta=\pi/4$ to the
 external magnetic field. Here $\w_{\rm ce}/\w_{\rm pe}=1/4$.  After \citet{Eliasson_07}.}
 \label{fig:helicon}
\end{figure}

\begin{figure}[htb]
 \centering
 \includegraphics[width=15cm]{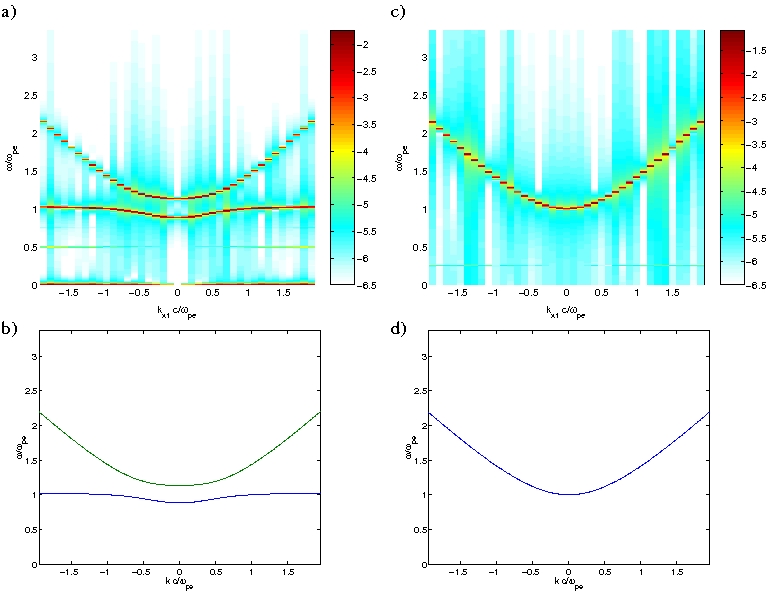}
 \caption{a) Amplitude spectrum (10-log scale) of $E_3$ for waves propagating
 perpendicularly to the external magnetic field ($\theta=\pi/2$),
 with the electric field perpendicular to the external magnetic field, and
 b) dispersion relations for the high-frequency slow and fast extraordinary (X-) modes,
 propagating perpendicularly to the external magnetic field ($\theta=\pi/2$).
 c) Amplitude spectrum of $E_2$ for waves propagating perpendicularly to the external magnetic field ($\theta=\pi/2$),
 with the electric field parallel to the magnetic field, and d) dispersion relations for
 the ordinary (O-) mode propagating perpendicularly ($\theta=\pi/2$) to the magnetic field.
 Here $\w_{\rm ce}/\w_{\rm pe}=1/4$.  After \citet{Eliasson_07}.}
 \label{fig:xomode}
\end{figure}

The general dispersion relation for electron waves in a cold
collisionless plasma with an external magnetic field is given by the
Appleton-Hartree dispersion relation \citep{Stix_92}
\begin{align}
 \frac{c^2k^2}{\w^2}&=1-\frac{2\w_{\rm pe}^2(\w^2-\w_{\rm pe}^2)/\w^2}
 {2(\w^2-\w_{\rm pe}^2)-\w_{\rm ce}^2\sin^2\theta\pm \w_{\rm ce}\Delta},
 \label{Appleton}
\end{align}
where $\Delta=[\w_{\rm ce}^2\sin^4\theta+4\w^{-2}(\w^2-\w_{\rm
pe}^2)^2\cos^2\theta]^{1/2}$ and $\theta$ is the angle between the
external magnetic field and the wave vector. In order to assess that
the Vlasov code reproduces these well-known wave modes in the
plasma, we have simulated electromagnetic waves propagating at
different angles to the magnetic field; see Figs.~\ref{fig:helicon}
and \ref{fig:xomode}. In the simulations, we restricted the problem
to one spatial dimension, along the $x_1$ axis, and used three
velocity dimensions. The initial condition for the electron
distribution function was a Maxwellian distribution (in normalized
units)
\begin{equation}
   \f_{\rm e}=(2\pi)^{-3}\exp\left[-\frac{(\eta_1^2+\eta_2^2+\eta_3^2)}{2}\right]
\end{equation}
while for the ions we used
\begin{equation}
   \f_{\rm i}=(2\pi)^{-3}\exp\left[-\frac{(\eta_1^2+\eta_2^2+\eta_3^2)}{2}
   \left(\frac{T_{\rm i} m_{\rm e}}{T_{\rm e} m_{\rm i}}\right)^{1/2}\right],
\end{equation}
with $T_{\rm i}/T_{\rm e}=100$ and $m_{\rm i}/m_{\rm e}=1836$, and
we chose $c/v_{\rm Te}=50$ for the electromagnetic waves. The
magnetic field strength was chosen such that $\omega_{\rm
ce}/\omega_{\rm pe}=1/4$, i.e. in our scaled unit we have $|{\bf
B}_0|=1/4$. A low-amplitude noise (random numbers) were added to the
vector potential ${\bf A}$ and to ${\bf \Gamma}$ so that all wave
modes in the system were excited. The numerical parameters were
chosen as $N_{x_1}=40$, $N_{x_2}=N_{x_3}=1$, $L_1=\pi\times 10^3$
(corresponding to $20\pi\,c/\w_{\rm pe}$ in dimensional units),
$N_{\eta_1}=N_{\eta_2}=N_{\eta_3}=20$,
$\eta_{{\rm max},{\rm e}1}=\eta_{{\rm max},{\rm e} 2}=\eta_{{\rm max},{\rm e} 3}=6$,
and
$\eta_{{\rm max},{\rm i}1}=\eta_{{\rm max},{\rm i} 2}=\eta_{{\rm max},{\rm i} 3}=200$. The simulations were
run with $8000$ time-steps with the fixed time interval $\Delta
t=0.14$. In Figs.~\ref{fig:helicon} and \ref{fig:xomode}, we have
Fourier transformed the electric field in space and time (with a
Gaussian time window) to obtain the spatio-temporal wave spectrum.
In panel a) of Fig. \ref{fig:helicon}, we show the power spectrum
for the transversal electric field component $E_2$, for waves
propagating along the magnetic field lines. It is clearly seen that
the wave energy is concentrated along the dispersion curves of the
electromagnetic right-hand (R) and left-hand (L) circularly
polarized waves, shown in in panel b). They are given by the
dispersion relation
\begin{align}
 \frac{c^2k^2}{\w^2}&=1-\frac{\w_{\rm pe}^2/\w^2}{1-\w_{\rm ce}/\w}
 \label{helicon1}
 \intertext{and}
 \frac{c^2k^2}{\w^2}&=1-\frac{\w_{\rm pe}^2/\w^2}{1+\w_{\rm ce}/\w}
 \label{helicon2}
\end{align}
respectively, obtained by setting $\theta=0$ in Eq.
(\ref{Appleton}). The R-wave is divided into a high-frequency branch
(having the highest frequency) and the low-frequency electron
whistler branch. We next made a simulation of waves propagating
obliquely to the external magnetic field, which was chosen as
$(B_{01},\,B_{02},\,B_{03})=(0.25,\,0.25,\,0)/\sqrt{2}$. The
resulting amplitude spectrum of $E_2$ is presented in panel c) of
Fig. \ref{fig:helicon}, while the solutions of the dispersion
relation (\ref{Appleton}) is plotted in panel d). Here, we can see
the emergence of the slow $X$-mode which has a resonance somewhat
higher than the plasma frequency $\omega=\omega_{\rm pe}$. Comparing
panel c) and d), we see that the wave energy is concentrated at the
dispersion curves. In Fig.~(\ref{fig:xomode}), we are considering
waves propagating perpendicularly to the magnetic field. Here, the
external magnetic field is given by
$(B_{01},\,B_{02},\,B_{03})=(0,\,0.25,\,0)$, and the energy spectrum
in panel a) and c) are for the perpendicular (to the magnetic field
direction) and parallel electric field components $E_3$ and $E_2$,
respectively. The wave energy is concentrated at the dispersion
curves for the cold plasma fast and slow X-modes displayed in panel
b) and the O-mode plotted in panel d). The cold plasma dispersion
relation for the ordinary (O-) mode and extraordinary (X-) mode,
perpendicular to the magnetic field, is given by
\begin{align}
 \frac{c^2k^2}{\w^2}&=1-\frac{\w_{\rm pe}^2}{\w^2}
 \label{omode}
 \intertext{and}
 \frac{c^2k^2}{\w^2}&=1-\frac{\w_{\rm pe}^2}{\w^2}
 \frac{(\w^2-\w_{\rm pe}^2)}{(\w^2-\w_{\rm pe}^2-\w_{\rm ce}^2)},
 \label{xmode}
\end{align}
respectively, obtained by setting $\theta=\pi/2$ in Eq.
(\ref{Appleton}). Also seen in panel a) at $\omega/\omega_{\rm
pe}=0.5$ is an excitation of an electron Bernstein mode which starts
at $\omega=2\omega_{ce}$ at small wavenumbers and has a resonance at
$\omega=\omega_{\rm ce}$ for large wavenumbers.

\subsection{Temperature anisotropy driven whistler instability}
\label{whistler_instab}
\begin{figure}[htb]
 \centering
 \includegraphics[width=10cm]{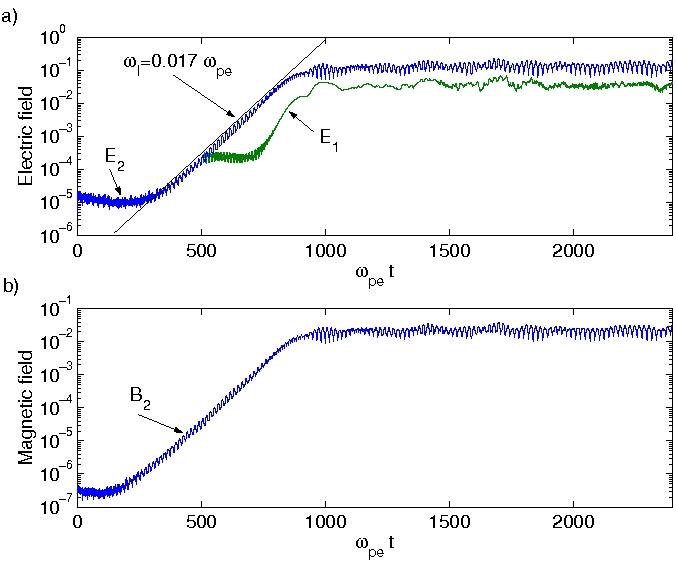}
 \caption{
 a) The maximum amplitude of the parallel and perpendicular electric field
 components $E_1$ and $E_2$, respectively, and b) the perpendicular magnetic
 field component $B_2$ (The parallel electric field $E_1$ is shown only for
 times $t>500\,\omega_{\rm pe}^{-1}$). After \citet{Eliasson_07}.
 }
 \label{fig:whistler_instab1}
\end{figure}

\begin{figure}[htb]
 \centering
 \includegraphics[width=15cm]{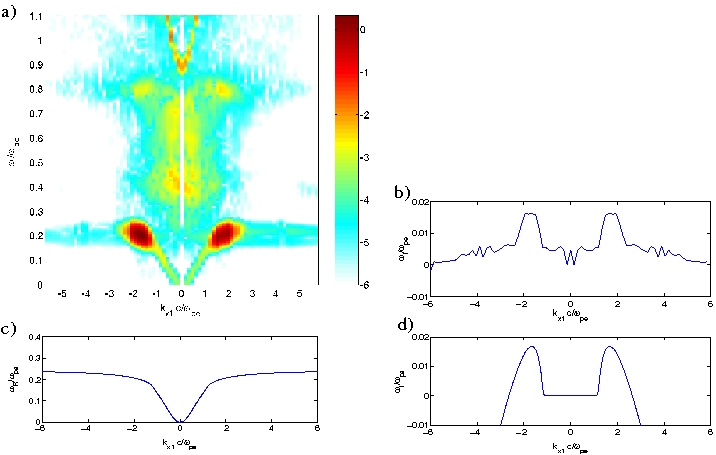}
 \caption{ a) The spatio-temporal amplitude spectrum of the electric field component $E_2$,
 and b) the spatial amplitude spectrum at time $t=700\,\omega_{\rm pe}$.
 c) The real and d) imaginary parts of the frequency for whistler waves, obtained from the
 dispersion relation (\ref{eq:whistler_instab}), in an anisotropic plasma
 with $T_\perp/T_\parallel=8$, $c/v_{{\rm Te}\parallel}=50$ and $\w_{\rm ce}/\w_{\rm pe}=0.25$. After \citet{Eliasson_07}.
 }
 \label{fig:whistler_instab2}
\end{figure}
In magnetized plasmas, there are often different temperatures
parallel and perpendicular to the magnetic field direction. In this
case, we may have a firehose instability if $T_{{\rm
e}\parallel}>T_{{\rm e}\perp}$, or a whistler instability if
$T_{{\rm e}\perp}>T_{{\rm e}\parallel}$. The latter case can have
relevance both for the sun and for the Earth's magnetosheath
\citep{Zhao96,Gosling89}. In order to study the growth and saturation
of the whistler instability, we carried out a simulation where the
initial condition for the electrons was taken to be a bi-Maxwellian
distribution function, where we used the temperature ratio $T_{{\rm
e}\perp}/T_{{\rm e}\parallel}=8$. In the Fourier transformed
velocity variables used in the simulation, the electron and ion
distribution function takes the form
\begin{equation}
   \f_{\rm e}=(2\pi)^{-3}\exp\left[-\frac{(\eta_1^2+8\eta_2^2+8\eta_3^2)}{2}\right],
\end{equation}
while for the ions we used
\begin{equation}
   \f_{\rm i}=(2\pi)^{-3}\exp\left[-\frac{(\eta_1^2+\eta_2^2+\eta_3^2)}{2}
   \left(\frac{T_{\rm i} m_{\rm e}}{T_{\rm e} m_{\rm i}}\right)^{1/2}\right],
\end{equation}
with $T_{\rm i}/T_{{\rm e}\parallel}=100$ and $m_{\rm i}/m_{\rm
e}=1836$. We use the same numerical parameters as in section
\ref{sec:EM}, except that
 $\eta_{{\rm max},{\rm e}2}=\eta_{{\rm max},{\rm e}3}=3$ and we use a higher resolution in space such that
$N_{x1}=80$. An adaptive time step was used in the simulation to maintain numerical stability \citep{Eliasson_07}. The numerical results are
displayed in Figs.~\ref{fig:whistler_instab1} and \ref{fig:whistler_instab2}.
The time-dependence of the maximum amplitude (over the spatial
domain) of the perpendicular electric and magnetic field components
$E_2$ and $B_2$ are shown in Fig.~\ref{fig:whistler_instab1}, and we
see an exponential growth of the perpendicular electric field
component $E_2$, with a growth rate of $\omega_{\rm I}\approx
0.017\omega_{\rm pe}$ [indicated by the solid line in panel a)] of
both the electric and magnetic field. The initially almost purely
electromagnetic waves saturate nonlinearly by exciting the
electrostatic field component $E_1$ (see the upper panel of
Fig.~\ref{fig:whistler_instab1}), and the amplitude of the
perpendicular magnetic field fluctuations are at this point about 10
\% of the external magnetic field. In order to compare the
simulation result with theory, we have plotted the spatio-temporal
amplitude spectrum of the perpendicular electric field component
$E_2$ in panel a) of Fig.~\ref{fig:whistler_instab2} (in a 10-log
scale) where the Fourier transform in time was taken for waves
between $\omega_{\rm pe}t=0$ and $\omega_{\rm pe}t=600$, with a
Gaussian time window. In panel b), we have plotted the quantity
$\omega_{\rm I}=(1/t_1){\rm ln}[|\widehat{E}_2(t_1)|]+$constant
where $\widehat{E}_2$ is the spatial Fourier transform of the
electric field component $E_2$ at time $t=t_1$ and
$t_1=700\,\omega_{pe}^{-1}$. Panel b) gives a rough estimate of the
growth rate for different wavenumbers in the simulation. We see that
there is a significant growth rate of waves with wavenumbers between
$kc/\w_{\rm pe}\approx 1.2$ and $kc/\w_{\rm pe}\approx 2.6$. We have
solved the dispersion relation for whistler waves in a plasma with a
bi-Maxwellian electron distribution. In the one-dimensional case,
and for immobile ions, it is \citep{Stix_92}
\begin{equation}
 \frac{k_{\parallel}^2c^2}{\omega_{\rm pe}^2}=\frac{T_{{\rm e}\perp}}{T_{{\rm e}\parallel}}-1+
 \frac{(\omega-\omega_{\rm ce})T_{{\rm e}\perp}/T_{{\rm e}\parallel}+
 \omega_{\rm ce}}{\sqrt{2}k_{\parallel}v_{{\rm Te}\parallel}}
 Z\left(\frac{\omega-\omega_{\rm ce}}{\sqrt{2}k_{\parallel}v_{{\rm Te}\parallel}} \right)
 \label{eq:whistler_instab}
\end{equation}
where $v_{{\rm Te}\parallel}=(k_{\rm B} T_{{\rm e}\parallel}/m_{\rm
e})^{1/2}$ is the parallel electron thermal speed and
\begin{equation}
  Z(\xi)=\i \sqrt{\pi}\exp(-\xi^2)[1+{\rm erf}(\i \xi)]
\end{equation}
is the plasma dispersion function. In the dispersion relation
(\ref{eq:whistler_instab}), we have neglected the electromagnetic
displacement current term [corresponding to the first term in the
right-hand side of Eq. (\ref{helicon1})]. In panels c) and d) of
Fig. \ref{fig:whistler_instab2}, we have plotted the real and
imaginary parts of the frequency, obtained from the dispersion
relation (\ref{eq:whistler_instab}), where we have used the
simulation parameters $\omega_{\rm ce}/\omega_{\rm pe}=0.25$,
$T_{{\rm e}\perp}/T_{{\rm e}\parallel}=8$ and $c/v_{{\rm
Te}\parallel}=50$. Comparing panels a) and c), we see that for the
undamped waves at small wavenumbers, the wave energy of the waves in
the simulation is located along the dispersion curve of the whistler
wave. Panels b) and d) show that the spectrum of the growing waves
matches approximately the waves with positive growth rate obtained
from the dispersion relation (\ref{eq:whistler_instab}). We note
that the maximum growth rate $\omega_{\rm I}\approx
0.017\,\omega_{\rm pe}$ in panel d) of
Fig.~\ref{fig:whistler_instab2} agrees well with the measured growth
rate in panel a) of Fig.~\ref{fig:whistler_instab1}.

\section{Extensions to incorporate relativistic and quantum effects}
We here very briefly discuss the extensions of the Fourier technique for relativistic and quantum
Vlasov equations.
\subsection{The relativistic Vlasov equation}
\begin{figure}[htb]
\centering
\includegraphics[width=8cm]{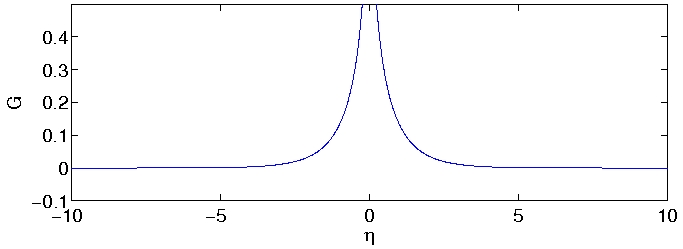}
\caption{The function $G(\eta)$.} \label{Fig1}
\end{figure}
In the relativistic Vlasov equation, the relativistic gamma factor comes into play for
particles moving close to the speed of light. This complicates the use of the Fourier transform
technique to solve the Vlasov equation. As a model example we consider the one-dimensional
Vlasov-Poisson system
\begin{align}
  &\frac{\partial f}{\partial t}+\frac{p}{\gamma}
  \frac{\partial f}{\partial x}
  -E\frac{\partial f}{\partial p}=0,
  \\
  &\frac{\partial E}{\partial x}=1-\int_{-\infty}^{\infty} f\,dp
\end{align}
where the relativistic gamma factor is $\gamma=\sqrt{1+p^2}$, and
$f$ describes the distribution of electrons in $(x,p)$ space
where $p$ is the momentum. Here the distribution function $f$ has been normalized by $n_0/c$, time $t$ by $\omega_{pe}^{-1}$, space $x$ by $\lambda_e=c/\omega_{pe}$, momentum $p$ by $m_e c$, and the electric field $E$ by $m_e c^2/e \lambda_e$.
  Using the Fourier transform pair
\begin{align}
f(x,p,t)&=\int_{-\infty}^{\infty}\widehat{f}(x,\eta,t)e^{-i\eta
p}\,d\eta,
\\
\widehat{f}(x,\eta,t)&=\frac{1}{2\pi}\int_{-\infty}^\infty
f(x,v,t)e^{i\eta p}\,dp,
\end{align}
we have the Fourier transformed Vlasov-Poisson system
\begin{align}
&\frac{\partial \widehat{f}}{\partial t}-
i\frac{\partial^2}{\partial x\partial\eta}(G*\widehat{f}) +iE\eta
\widehat{f}=0,
\label{relvlasov}
\\
&\frac{\partial E}{\partial x}=1-2\pi\widehat{f}(x,\eta,t)_{\eta=0},
\end{align}
where $*$ denotes convolution over $\eta$ space. Here
\begin{equation}
  \begin{split}
  G&=\frac{1}{2\pi}\int_{-\infty}^\infty\gamma^{-1}e^{i\eta p}\,dp=\frac{1}{\pi}{\rm K}_0(|\eta|),
  \end{split}
\end{equation}
where $K_0$ is the Bessel function of second kind of order $0$. The function
$G(\eta)$, plotted in Fig.~\ref{Fig1}, grows like
$[-\gamma_E-{\rm ln}(|\eta|/2)]/\pi$ for $|\eta|<1$, where $\gamma_E\approx 0.5772156649$ is the
Euler-Mascheroni constant, and falls off like $\exp(-|\eta|)(2\pi|\eta|)^{-1/2}$ for $|\eta|>1$, and has the property that $\int_{-\infty}^{\infty} G\,d\eta=1$. The relativistic corrections
are contained in $G$. For a weakly relativistic plasma, the
distribution function $\widehat{f}$ is much wider and smoother in
$\eta$ space than $G$, and hence we have $G*\widehat{f}\approx
\widehat{f}$, i.e., $G$ then has the property of Dirac's delta
function and we retain the non-relativistic Fourier transformed Vlasov equation treated by
\citet{Eliasson_01}. The numerical implementation of the convolution by $G$ and the well-posed
absorbing boundary conditions in $\eta$ space are unsolved problems,
and so are the extensions to higher dimensions. Since the function
$G$ in Fig. \ref{Fig1} falls of exponentially for large $\eta$, the convolution integral
in (\ref{relvlasov}) can possibly be approximated by a truncated operator
with compact support, and the problem with absorbing boundary conditions could
potentially be solved using normal mode analysis similar to
that of \citet{Engquist_77,Engquist_79}.

\subsection{The Quantum Vlasov/Wigner equation}

The quantum analogue to the Vlasov-Poisson system is the Wigner-Poisson model
\citep{Wigner_30,Roos_60,Tatarskii_83,Markowich_90}.
In three dimensions, the Wigner equation for electrons can be written
\begin{equation}
\frac{\partial f}{\partial t} + {\bf v}\cdot \nabla f= -\frac{iem_e^3}{(2\pi)^3  \hbar^4}
\int\!\!\!\int d^3\lambda\, d^3v'
e^{im_e({\bf v}-{\bf v}')\cdot {\boldsymbol
\lambda}/\hbar}\bigg[\phi\bigg({\bf x }+\frac{\boldsymbol
\lambda}{2},t\bigg)-\phi\bigg({\bf x}-\frac{\boldsymbol
\lambda}{2},t\bigg)\bigg]f({\bf x},{\bf v}',t),
\end{equation}
which is coupled with the Poisson equation with immobile ions
\begin{equation}
\nabla^2 \phi =\frac{e}{\varepsilon_0}\left(\int f d^3v -n_0\right).
\end{equation}
One can show that the Wigner equation converges to the Vlasov equation in the formal limit $\hbar\rightarrow 0$,
\begin{equation}
\frac{\partial f}{\partial t} + {\bf v}\cdot \nabla f= -\frac{e}{m_e}\nabla\phi\cdot\frac{\partial f}{\partial {\bf v}}.
\end{equation}
It turns out that the Fourier technique in velocity space is well suited to solve the Wigner equation.
As an example we will study the 1D Wigner-Poisson system \citep{Manfredi_05}
\begin{equation}
  \frac{\partial f}{\partial t} + v \frac{\partial f}{\partial x}=
  -\frac{iem_e}{(2\pi) \hbar^2}\int\!\!\!\int
  e^{im_e(v-v'){\lambda}/\hbar}\bigg[\phi\bigg({x }+\frac{
  \lambda}{2},t\bigg)-\phi\bigg({x}-\frac{
  \lambda}{2},t\bigg)\bigg]f({x},{v}',t)d\lambda\, dv'
\end{equation}
\begin{equation}
  \frac{\partial^2\phi}{\partial x^2} =\frac{e}{\varepsilon_0}\left(\int f dv-n_0\right).
\end{equation}
As it stands, the Wigner is awkward to solve numerically. However,
introducing the Fourier transform pair in velocity space
\begin{equation}
  f(x,v,t)=\int_{-\infty}^{\infty} \widetilde{f}(x,\eta,t)\mathrm{e}^{-i \eta v}\,d \eta,
\end{equation}
\begin{equation}
    \widetilde{f}(x,\eta,t)=\frac{1}{2\pi}\int_{-\infty}^{\infty}f(x,v,t)e^{i\eta v}\,d v
\end{equation}
we obtain
\begin{equation}
 \frac{\partial \widetilde{f}}{\partial t}-i\frac{\partial^2 \widetilde{f}}{\partial x\partial \eta}-i\frac{e}{\hbar}\bigg[\phi\bigg(x+\frac{\hbar\eta}{2m_e}\bigg)-\phi\bigg(x-\frac{\hbar\eta}{2m_e}\bigg)\bigg]\widetilde{f}=0,
\label{F_Wigner}
\end{equation}
\begin{equation}
 \frac{\partial^2 \phi}{\partial x^2}= \frac{e}{\varepsilon_0}\big[2 \pi \widetilde{f}(x,\eta,t)_{\eta=0}-n_0\big],
\end{equation}
which is simpler to solve numerically. By using a pseudo-spectral method in space, the
spatial shifts by $\pm {\hbar\eta}/{2m_e}$ in Eq. (\ref{F_Wigner})
is converted to mulitplications by $\exp(\pm i k {\hbar\eta}/{2m_e})$, where $k$ is the wavenumber.
In $\eta$ space, we can apply the same absorbing boundary conditions,
\begin{equation}
  \widetilde{f}={\rm F}^{-1}[H(k){\rm F} \widetilde{f}]  \mbox{ at } \eta=\eta_{max}.
\end{equation}
as for the Vlasov equation. Hence the exisitng Vlasov codes are easily modified to simulate the
Wigner equation; see for example the work by \citet{Marklund_06} where the Wigner equation for
broadband electromagnetic radiation in a plasma was solved with the Fourier method as described here.
Finally we mention that other numerical methods for solving the Wigner equation exist in the litterature,
for example operator splitting methods \citep{Suh_91,Arnold_96}.

\section{Conclusions}
In this paper we have given a review of simulations of the Vlasov equation in higher dimensions.
In this method, the Vlasov equation is Fourier transformed in velocity space, and the resulting equation is solved numerically. We have discussed the main difficulties solving the Vlasov equation with a grid-based
solver, namely that in some problems, the particle distribution function becomes filamented in velocity space due to phase mixing. This can lead the recurrence of the initial condition on the numerical grid (the so-called recurrence phenomenon), which in turn leads to unphysical oscillations and instabilities in the simulations. By designing outflow boundary conditions in the Fourier transformed velocity space, the highest Fourier modes in velocity space are allowed to propagate over the boundary and to be removed from the simulations. In that way an effective dissipation is allowed in the Vlasov equation, and  the numerical recurrence phenomenon is strongly reduced. On the other hand, Fourier modes that have not reached the boundary in the Fourier transformed velocity space are not damped by the numerical method. In a sense, the method represents the minimum numerical viscosity possible one can introduce to the numerical simulation, which removes the recurrence phenomenon. The extension to multiple dimensions was also possible, by careful consideration of the magnetic field in the boundary conditions in the Fourier transformed velocity space. In this manner, the boundary conditions could be made strictly local in time and to only include boundary points in the Fourier transformed velocity space. The boundary conditions are highly absorbing and have been proved to be well-posed by energy estimates \citep{Eliasson_01,Eliasson_02,Eliasson_07}. The method may be an attractive alternative to existing methods for solving the Wigner equation, which describes the evolution systems of quantum particles. For the unmagnetized Wigner equation for charged particles, the method is directly applicable with only minor modifications. Future developments of the Fourier method could include relativistic effects and the extension to magnetized quantum plasmas.


\end{document}